\documentclass[twocolumn,citeautoscript,pra,superscriptaddress,longbibliography,aps,nofootinbib,10pt]{revtex4-2}

\usepackage{
  amsmath,
  amssymb,
  amsthm,
  bbm,
  bm,
  booktabs,
  braket,
  calc,
  dsfont,
  enumitem,
  float,
  graphicx,
  lipsum,
  mathrsfs,
  mathtools,
  tikz,
  upgreek,
  url,
  xparse
}
\graphicspath{{figures/}}
\usepackage{xcolor}

\usepackage[bookmarks=true,colorlinks=true,linkcolor=orange,citecolor=orange,urlcolor=orange,bookmarks]{hyperref}

\usepackage{iftex}
\ifPDFTeX
  \usepackage[utf8]{inputenc}
\fi

\usepackage[capitalise]{cleveref}

\usepackage[most]{tcolorbox}

\newcommand{\id}{\mathbbm{1}}
\newcommand{\tr}{\operatorname{Tr}}
\newcommand{\rrangle}{\rangle \! \rangle}
\newcommand{\llangle}{\langle \! \langle} 
\newcommand{\kket}[1]{|#1 \rrangle}
\newcommand{\bbra}[1]{\llangle #1 |}
\newcommand{\bbraket}[2]{\llangle #1 |#2\rrangle}
\newcommand{\kketbra}[2]{|#1 \rrangle \! \llangle #2|}
\newcommand{\Span}{\operatorname{span}}
\newcommand{\diag}{\operatorname{diag}}
\newcommand{\shortT}{{t_\varepsilon}}
\newcommand{\PT}{{\Upsilon_{k:0}}}
\newcommand{\shortPT}{{\Upsilon_{k:0}^{\shortT}}}

\newcommand{\PTchoi}{\hat{{\Upsilon}}_{k:0}}

\newcommand{\mPT}{\Upsilon_{k:0}^{\text{Mar}}}

\newcommand{\T}{\operatorname{T}}
\newcommand{\e}{\operatorname{e}}
\newcommand{\dd}{\operatorname{d}\!}
\newcommand{\Tr}{\operatorname{Tr}}

\newcommand{\cmps}[3]{\Ket{\Phi[#1,#2,#3]}}
\newcommand{\cpt}[3]{\Ket{\Upsilon_T[#1,#2,#3]}}
\newcommand{\cPT}{|\Upsilon_T \rangle}
\newcommand{\vac}{\ket{\Omega}}
\newcommand{\gen}{\mathbb H_{\rm SE}}
\newcommand{\Hgen}{\mathbb H_{\rm SE}}

\newcommand{\tint}{\int_0^T \mathrm{D}^{n}t}

\definecolor{jonas}{rgb}{0.9,0.17,0.31}

\newif\ifshowedits
\showeditstrue   % editing version
\ifshowedits

    \newcommand{\rem}[1]{{\color{red}[#1]}}

    \newcommand{\grem}[1]{{\color{red}(Greg): [#1]}}

    \newcommand{\jrem}[1]{{\color{red}(Jon{\'a}{\v s}): [#1]}}

    \newcommand{\crem}[1]{{\color{red}(Clara): [#1]}}

\else

    \newcommand{\rem}[1]{}

    \newcommand{\grem}[1]{}

    \newcommand{\jrem}[1]{}

    \newcommand{\crem}[1]{}

\fi

\newtheorem{theorem}{Theorem}%[section]

\newtheorem{definition}{Definition}%[section]
\newtheorem{result}{Result}%[section]
\newtheorem{proposition}{Proposition}

\makeatletter
\renewcommand\onecolumngrid{
\do@columngrid{one}{\@ne}
\def\set@footnotewidth{\onecolumngrid}
\def\footnoterule{\kern-6pt\hrule width 1.5in\kern6pt}
}
\renewcommand\twocolumngrid{
        \def\footnoterule{
        \dimen@\skip\footins\divide\dimen@\thr@@
        \kern-\dimen@\hrule width.5in\kern\dimen@}
        \do@columngrid{mlt}{\tw@}
}
\makeatother

\makeatletter
\newif\ifeqcontrib@this
\newif\ifeqcontrib@any
\eqcontrib@thisfalse
\eqcontrib@anyfalse

\newcommand{\eqcontrib}{\global\eqcontrib@thistrue\global\eqcontrib@anytrue}

\newcommand{\eqcontribmark}{\textsuperscript{\ensuremath{,\#}}}

\newcommand{\eqcontrib@maybe}{%
  \ifeqcontrib@this
    \global\eqcontrib@thisfalse
    \eqcontribmark
  \fi
}

\newcommand{\printEqContrib}{%
  \ifeqcontrib@any
    \begingroup
      \renewcommand{\thefootnote}{\ensuremath{\#}}%
      \footnotetext{These authors contributed equally to this work.}%
    \endgroup
  \fi
}

\def\doauthor#1#2#3{%
  \ignorespaces#1\unskip\@listcomma
  \begingroup
    #3%
  \@if@empty{#2}{\endgroup{}{}}{\endgroup{\comma@space}{}\frontmatter@footnote{#2}}%
  \eqcontrib@maybe
  \space \@listand
}%
\makeatother

\begin{document}
\title{An operational continuum limit of quantum combs}

\author{Clara Wassner\eqcontrib}
\email{c.wassner@fu-berlin.de}
\affiliation{Dahlem Center for Complex Quantum Systems, Freie Universit\"at Berlin, 14195 Berlin, Germany}

\author{Jon{\'a}{\v s} Fuksa\eqcontrib}
\email{jonas.fuksa@fu-berlin.de}
\affiliation{Dahlem Center for Complex Quantum Systems, Freie Universit\"at Berlin, 14195 Berlin, Germany}

\author{Jens Eisert}
\email{jense@zedat.fu-berlin.de}
\affiliation{Dahlem Center for Complex Quantum Systems, Freie Universit\"at Berlin, 14195 Berlin, Germany}
\affiliation{Helmholtz-Zentrum Berlin f{\"u}r Materialien und Energie, Berlin, Germany}

\author{Gregory A. L. White}
\email{gregory.white@fu-berlin.de}
\affiliation{Dahlem Center for Complex Quantum Systems, Freie Universit\"at Berlin, 14195 Berlin, Germany}

\begin{abstract}
Quantum combs are powerful conceptual tools for capturing multi-time processes in quantum information theory, constituting the most general quantum mechanical process. But, despite their causal nature, they lack a meaningful physical connection to time---and are, by and large, arguably incompatible with it without extra structure. The subclass of quantum combs which assumes an underlying process is described by the so-called process tensor framework, which has been successfully used to study and characterise non-Markovian open quantum systems. But, although process tensors are motivated by an underlying dynamics, it is not a priori clear how to connect them to a \textit{continuous} process tensor object mathematically---leaving an uncomfortable conceptual gap. 
In this work, we take a decisive step toward remedying this situation. 
We introduce a fully continuous process tensor framework by showing how the discrete multi-partite Choi matrix becomes a vector in bosonic Fock space, which is intrinsically and rigorously defined in the continuum. 
Equipped with this, we lay out the core structural elements of this framework and its properties. This translation allows for an information-theoretic treatment of multi-time correlations in the continuum via the analysis of their continuous matrix product state representatives. Our work closes a gap in the quantum information literature, and opens up the opportunity for the application of many-body physics insights to our understanding of quantum stochastic processes in the continuum.
\end{abstract}

\maketitle
\printEqContrib

\section{Introduction}\label{sec:introduction}

Fundamental to the history of our description of quantum mechanical systems is a history of capturing dynamics.
But modern quantum information often eschews the role of time, favouring instead a discretised picture where the wall clock, representing physical time, is conspicuously left absent. This reflects in part a change in priorities; rather than simply describing quantum systems, the emphasis amid the so-called `second quantum revolution' has typically been instead to \emph{manipulate} dynamics for the construction of some interesting state. In this idealisation, the temporal aspect is made implicit via the depth of some atomic circuit element and abstracted away from its physical implementation. But, as we contend in this work, the underlying laws of physics can and should be restored to the information-theoretic treatment of quantum dynamics.\par 

When viewed as a series of discrete steps, the evolution of a quantum system can be understood as a sequence of apparatuses that take states from one point in time to the next. This operational view of quantum theory emphasises the role of control in quantum mechanics, and suggests that experimenters making choices of measurements and collecting measurement outcomes is the most general way to view the dynamics of a quantum system.
Indeed, the study of multi-time processes using this perspective has steadily gathered more and more interest amidst the backdrop of maturing quantum technologies -- and in tandem with our growing theoretical understanding of closed and open quantum systems~\cite{strathearn_efficient_2018,Cygorek-2022, whiteWhatCanUnitary2025, white2022NonMarkovianQuantumProcess, whiteDemonstrationNonMarkovianProcess2020,aloisio-complexity,White-2025-Unifying-NM, Dowling-2024-Tree-PTs,Pollock2018-Process-Tensors,Milz-2021-GMEinTime, PhysRevLett.126.200401,Gribben2022usingenvironmentto}. \par

The appropriate framework from which one can correctly study finite multi-time processes under this lens has emerged in many different contexts over the years: in quantum combs, operator tensors, quantum strategies, process matrices, process tensors, influence matrices, and superdensity operators -- to name a few
\cite{Hardy_2012, PhysRevLett.101.060401,cotlerSuperdensityOperatorsSpacetime2018, BruknerNoCausalOrder,milzQuantumStochasticProcesses2021,whiteNonMarkovianQuantumProcess2022,Pollock2018-Process-Tensors,White-2025-Unifying-NM,PhysRevA.80.022339}.
Applications therein have ranged from finding optimal strategies in multi-round quantum games~\cite{Gutoski_2007}, 
through probing signatures of quantum chaos, to understanding non-Markovianity in open quantum systems \cite{Dowling_2024_chaos,White-2025-Unifying-NM}, as well as examining the structure of causality in quantum mechanics
\cite{BruknerNoCausalOrder}. 
A particularly useful property of all comb-like frameworks is that, via the generalised \emph{Choi-Jamio\l kowski isomorphism} (CJI), temporal correlations are mapped onto spatial ones, allowing temporal structure to be understood using the usual tools of quantum information~\cite{PhysRevA.80.022339,whiteWhatCanUnitary2025}.
In this work, we shall adopt the terminology `quantum combs' as a general designation for this class of objects. 
But our results correspond in particular to \emph{process tensors}~\cite{Pollock2018-Process-Tensors}, which occupy the context of discrete quantum stochastic processes and non-Markovian open quantum systems -- both theoretically, and in practice.
Here, a controllable quantum system is coupled to an inaccessible environment. 

\par 

Within this operational picture of dynamical processes, however, one encounters a contradiction in connecting to the underlying physical setting. Namely, we are studying dynamical processes using instruments which take no time at all.
This generates a substantial tension: a process tensor is a fixed object representing uncontrollable system-environment dynamics. But at every leg, the dynamics is paused and a perfect manipulation of the system takes place. The implication here is either that the underlying dynamics is not truly uncontrollable, or that transformations can take place arbitrarily quickly, and thus with infinitely high energy. 
At best, this is an approximation which holds in some settings -- at worst, it is an ill-conditioned treatment and can lead to contradictions when taken seriously. For example, many common measures of temporal correlations vanish in the continuum limit of decreasing time steps between system manipulations, as observed in 
Ref.~\cite{sonnerInfluenceFunctionalManybody2021} and analytically proven in our work. But this is merely an artefact of introducing arbitrarily strong probes at each time, and clearly  contradicts what one would want from a measure of physical dynamics. Thus, the discrete analysis of multi-time processes has shortcomings. So far, a tool for the information-theoretic study of quantum processes and quantum manipulations in a physically-operational setting -- i.e., under the mandatory condition of finite energy -- has not been available.

\begin{figure}[t]
    \centering
    \includegraphics[width=\linewidth]{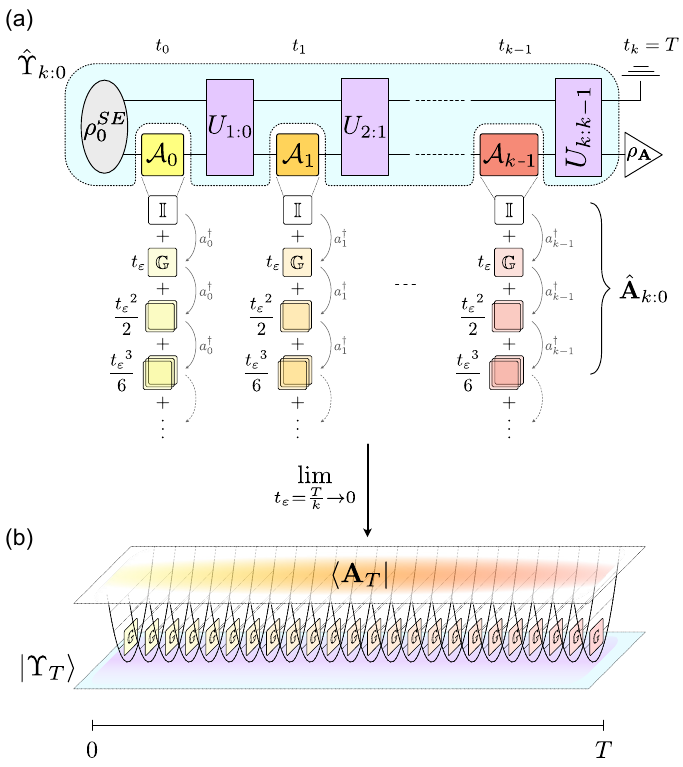}
    \caption{A high-level depiction of the results presented in this work. (a) A discrete process tensor $\PT$ is a multi-linear functional encoding the joint probabilities of all multi-time instruments $\mathbf{A}_{k:0}$ within the Choi matrix $\PTchoi$. When viewing $\PTchoi$ in a second-quantised form, these arbitrarily strong interventions can be interpreted as particles in a bosonic Fock space at higher and higher energy levels. (b) With this representation in place, the continuum limit of this object may then be taken, arriving at a bosonic Fock space vector $\cPT$ with at most one particle excitation (the instrument generator) at each time.}
    \label{fig:cpt-concept}
\end{figure}

In this work, we aim to close this gap by deriving a principled and well-defined \emph{continuum limit of the process tensor framework}.
The vectorised Choi matrix of the resulting object is not a spin system, but instead a vector in the continuum bosonic Fock space.
We depict this transformation illustratively in~\cref{fig:cpt-concept}: first, we show how process tensors, as they are traditionally defined, can be cast in a second-quantised form. In this picture, the Hilbert spaces representing the inputs and outputs of the system at each time are mapped onto a Fock space. Each order of an instrument in its power-series form can then be interpreted as the creation of a particle at higher and higher energy levels. Then, as one takes a continuum limit, only the zeroth and first energy levels survive~\cite{tingley2023notesfockspace}.
That is, once one drops the unphysical treatment that instruments might perform finite transformations in zero time, it follows immediately that the process tensor 
becomes a multi-linear functional on \emph{generators} of instruments, as opposed to their propagators. This result comes as a necessary consequence of introducing a wall clock into process tensor theory. 

Much like the generalised CJI exploits well-understood properties of quantum states to enlighten us about quantum processes in the discrete case, we furthermore believe that studying the correspondence between field theories and non-Markovian processes in the continuum (which is the natural extension of this correspondence) will lead to a better understanding of the latter by employing tools dedicated to the former. In this work, we primarily use the class of \emph{continuous matrix product states} (cMPS) to this end 
\cite{verstraeteContinuousMatrixProduct2010,osborneHolographicQuantumStates2010}, which is a powerful computational framework from which correlation structures in quantum field theories may be understood, and whose connection to continuous processes has already been hinted at in previous work \cite{hastingsConnectingEntanglementTime2015,parkTensorNetworkInfluence2024}.
Although our framework is not limited to this particular ansatz, the physical interpretability of cMPS processes in terms of SE dynamics grants it a special place in the theory.
The precise mathematical study of this connection is the subject of the companion work~\cite{fuksaDilationTheoremContinuum2026}.
With this formalism, we show how issues with discrete frameworks may be remedied. For example, we use this construction to introduce a family of well-defined continuum measures of non-Markovianity.
We anticipate that such a continuum treatment will furthermore have applications in tomography of non-Markovian noise~\cite{White-2025-Unifying-NM}, efficiency of simulation in non-Markovian open quantum systems~\cite{Link_2024,Cygorek-2022,parkTensorNetworkInfluence2024}, and a more rigorous treatment of measures of non-Markovianity~\cite{Pollock_2018_measure,zambonProcessTensorDistinguishability2024}. Each of these areas has benefitted immensely from employing the process tensor framework, but ultimately suffers in one way or another from ill-conditioned finiteness. 

Our work is organised as follows. In~\cref{sec:background} we first introduce our setting, providing the necessary background on both process tensors and the cMPS formalism.~\cref{sec:framework} represents the bulk of our work. Here we explicitly derive the continuum limit of the process tensor framework and arrive at an object which we call 
a \emph{continuous process tensor} (cPT); we furthermore illustrate its various representations and physical properties. We conclude the exposition of the framework by discussing generalised instruments and the spatiotemporal Born rule. The remainder of this work is then dedicated to working with the cPT framework and its applications. In~\cref{sec:examples} we present some worked examples of process tensors and their corresponding instruments. 
\cref{sec:NM} opens with a brief introduction to Choi-based non-Markovianity measures and a proof that they vanish in the continuum limit. We then leverage our framework to propose continuum measures of non-Markovianity and a structural organisation of the infinite hierarchy of multi-time correlations.
Lastly,~\cref{sec:apps} addresses the question: how does one work with cPTs in practice? To this end, we examine the complementary tasks of simulation and tomography. In the former, we propose a compression algorithm of cPTs -- in which truncation is built into the process tensor equations of motion. In the latter, we look at how cPTs can be robustly reconstructed from experimental data, which offers a better-conditioned means with which one can learn non-Markovian noise in real devices. We then conclude with some final remarks and outlook in~\cref{sec:discussion}.

\section{Background}\label{sec:background}
% !Tex root = ./paper.tex

Before deriving the continuum limit of what we shall call a \emph{quantum stochastic process}, we provide the necessary background. We will start by stressing that our terminology ``quantum stochastic process'' denotes the \emph{concept} of a dynamical open quantum system -- as opposed to a priori committing to a rigorous mathematical definition, per se. In the discrete case, one can treat process tensors as a definition of quantum stochastic processes~\cite{Milz_2021_QST_review}. In the continuous case, there is the mature mathematical field of the same name~\cite{accardiQuantumStochasticProcesses1982,hudson1984quantum} which is somewhat disparate from the language of process tensors, although connections have recently been formed~\cite{Nurdin_2021}. In our derivation, we assume an underlying generator which is well-behaved enough for quantum It\^{o} calculus to apply~\cite{hudson1984quantum}. This assumption allows us to identify that particular subclass of quantum stochastic processes in this way, but dealing with less regular behaviour is left as the subject of future work. In short: our cPTs should be viewed as a subclass of quantum stochastic processes, but not necessarily the entire genre.

We begin by establishing our notation and conventions, then review how discrete quantum processes are described by process tensors. 
Finally, we briefly recap bosonic non-relativistic field theory and 
recall the definition of continuous matrix product states, a special class of Fock space vectors that will feature prominently throughout our work. It should also be clear 
that throughout this manuscript we prioritise physical intuition over mathematical rigour per se.

\subsection{Notation and conventions}\label{ssec:notation}

We consider a physical system $\rm S$ with Hilbert space $\mathcal{H}_{\rm S}$ of dimension $d_{\rm S}$, which interacts with an environment $\rm E$ with Hilbert space $\mathcal{H}_{\rm E}$. 
Throughout this work, we restrict our attention to finite-dimensional systems with $d_{\rm S} < \infty$ composed of qubits.
The assumption of finite dimensional systems is fundamental for our framework but we restrict to qubit systems for convenience of presentation.
We impose no restriction on the dimensionality of the environment $d_{\rm E}$, allowing us to accommodate bosonic environments as well. 

We work in the Liouville representation, where density matrices $\rho \in \mathcal B(\mathcal H)$ are represented by vectors $\kket{\rho}$ and linear maps $\mathcal{A}: \mathcal B(\mathcal H)\to \mathcal B(\mathcal H)$  become matrices $\mathbb{A}$. 
To facilitate this identification, we use double bracket notation $\kket{\, \cdot \,}$ for Liouville space vectors throughout this work.

The evolution of the total system $\rm{SE}$ is characterised by an initial state
vector $\kket{\rho_0^{\rm{SE}}}$ and a (possibly time-dependent) generator $\gen(t)$, which is the matrix representation of the semigroup generator $\mathbb{G}^{\rm SE}(t)$ governing the time evolution of the total system.
For better readability we will often omit referencing explicitly the time-dependence on generators 
and write $\gen$ instead of $\gen(t)$.
The time-ordered exponential of a time-dependent generator is then denoted as $\mathcal T \mathrm{e}^{\int \mathrm{d}t \, \gen}$.
We denote an orthonormal basis of the system Liouville space by 
\begin{equation}
\{\kket{P_\mu}\}_{\mu = 0}^{d_{\rm S}^2-1},
\end{equation}
where $P_{\mu = 0}$ is conventionally identified with the identity, $P_{0} \equiv \id/\sqrt{d_{\rm S}}$. 
We will often use the shorthand $\kket{P_\mu} \coloneqq \kket{\mu}$. 
A summary of our notation and conventions is provided in  Appendix \ref{app:list_notation}.

\subsection{Process tensors and discrete quantum stochastic processes}\label{ssec:discrete process tensors}
No controllable quantum system is 
truly closed, and so inevitably the 
description of dynamics in a quantum setting must be one that accommodates stochastic behaviour due to an environment, possibly with memory.
In quantum information, the most general operational objects are often taken to be quantum channels, which are encapsulated as linear maps on quantum states with the additional property that they are \emph{completely positive, and trace-preserving} (CPTP)~\cite{wolf2012quantum}. 
But quantum channels encode only two-time correlations and, notably, cannot faithfully represent the situation where one has sequential channels which all share the same environment~\cite{Milz_2019_div}. Indeed, two chief issues are faced when attempting to generalise classical stochastic processes to the quantum setting: a loss of complete positivity when faced with initial system-environment correlations~\cite{pechukas_reduced_1994-1}, and the fact that observables at different times no longer commute~\cite{Milz_2021_QST_review}.

The resolution to this deficiency comes in the form of \emph{process tensors}~\cite{Pollock2018-Process-Tensors} (which, as we have mentioned, are equivalently stated in a series of alternate monikers). In words, a process tensor is a collection of possibly-correlated quantum channels defined with a causally-ordered structure. They map, therefore, not only states from one time to the next, but more generally multi-time instruments to multi-time outcomes. Process tensors were originally introduced to address some of the deficiencies of quantum channels in the non-Markovian setting, cf., for example, Ref.~\cite{Milz_2021_QST_review} for a historical overview and comprehensive tutorial.  \par 

To be precise, we consider an open quantum system $\rm S$ coupled to an environment $\rm  E$ via an arbitrary set of unitary maps $\{\mathcal{U}_{1:0},\cdots,\mathcal{U}_{k:k-1}\}$ and probed across a series of $k+1$ times $\{t_0,\cdots,t_k\}$. Now, in the most general scenario, an experimenter 
may intervene with the application of a set of (also possibly-correlated) CP maps $\mathbf{A}_{k-1:0} := \{\mathcal{A}_0,\cdots, \mathcal{A}_{k-1}\}$ after which a final state 
\begin{equation}
   \label{eq:pt-act}
   \rho_k(\mathbf{A}_{k-1:0}) = \Tr_{\rm E}\left[
   (\mathcal{U}_{k:k-1}\circ \mathcal{A}_{k-1} \circ \cdots \circ \mathcal{U}_{1:0}\circ\mathcal{A}_0)[\rho_0^{\rm SE}]\right]
\end{equation}
is obtained, conditioned on this choice of operations.
The process tensor $\PT$ is defined via \cref{eq:pt-act} as a multi-linear mapping
\begin{equation}
   \PT[\mathbf{A}_{k-1:0}] = \rho_k(\mathbf{A}_{k-1:0})
\end{equation}
from past controls $\mathbf{A}_{k-1:0}$ to future states.
It represents all of the uncontrollable $\rm SE$ dynamics within the open dynamics, taking in control operations as an argument. \par 
Process tensors also generalise the well-known state-process equivalence to the multi-time setting via a generalisation of the \emph{Choi-Jamio\l kowski isomorphism} (CJI). The Choi matrix $\hat{\mathcal{A}}\in\mathcal{B}(\mathcal{H}_{\mathfrak{o}}\otimes \mathcal{H}_{\mathfrak{i}})$ of an instrument $\mathcal{A}$ is defined with reference to its action on one half of an unnormalised maximally entangled state
as $\hat{\mathcal{A}} := (\mathcal{A}\otimes\id)[\Phi^+] = \sum_{i,j=1}^d \mathcal{A}[\ket{i}\!\bra{j}]\otimes\ket{i}\!\bra{j}$, with input $\mathfrak{i}$ and output $\mathfrak{o}$.
So, too, is the process tensor's Choi matrix $\hat{\Upsilon}_{k:0}$ defined by letting one 
half 
of a fresh Bell state participate in the dynamics 
at each step
\begin{equation}
   \label{eq:pt-choi-def}
   \hat{\Upsilon}_{k:0} = \Tr_E\left[\mathcal{U}_{k:k-1}[\Phi^+_{k-1}]\circ \cdots \circ{}\mathcal{U}_{1:0}[\Phi^+_{0}][\rho_0^{\rm SE}]\right].
\end{equation}
Here, $\circ$ should be understood to mean composition with respect to the environment, and its action on $\Phi^+_j\in \mathcal{B}(\mathcal{H}_{\mathfrak{o}_j}\otimes \mathcal{H}_{\mathfrak{i}_j})$ is only with respect to the $\mathfrak{o}_j$ subsystem. The process tensor Choi matrix is therefore an element of the space 
\begin{equation}
   \label{eq:pt-space}
   \mathcal{B}(\mathcal{H}_{\mathfrak{o}_k}\otimes\mathcal{H}_{\mathfrak{i}_k}\otimes\cdots\otimes\mathcal{H}_{\mathfrak{i}_1}\otimes\mathcal{H}_{\mathfrak{o}_0}),
\end{equation}
and, concretely, the labels $\mathfrak{o}_j$ and $\mathfrak{i}_{j+1}$ refer respectively to the output and input of the process at time $t_j$. \par

This multi-time CJI is extraordinarily useful. It allows one to place multi-time processes on the same footing as many-body states, and study temporal correlations by mapping them first onto spatial ones and then use standard information-theoretic tools. It also gives rise to a multi-time generalisation of Born's rule. In the single-time setting, Born's rule says that given a state $\rho$ and a measurement apparatus described by POVM $\{M^x\}_x$, the probability with which one obtains outcome $x$ is given by the overlap 
\begin{equation}
   \Pr(X=x\mid \{M^x\}_x) = \Tr[\rho M^x].
\end{equation}
Now, suppose in the dynamical setting that one applies a stochastic multi-time instrument $\mathbf{A}_{k:0} = \{\mathbf{A}_{k:0}^{(\mathbf{x}_k)}\}$ with outcome labels $\mathbf{x}_k = (x_k,\cdots ,  x_0)$. This encompasses, for example, measure-and-prepare channels at each time; deterministic unitary operations; or even repeated interactions with an auxiliary systed, followed by a projective measurement on that system. 
The probability with which one obtains a given $\mathbf{x}_k$ is given by the overlap
\begin{equation}
   \label{eq:pt-born}
   \Pr\left((X_k,\cdots,X_0)=\mathbf{x}_k\mid\mathbf{A}_{k:0}\right) = \Tr\left[\PTchoi \hat{\mathbf{A}}^{(\mathbf{x}_k)\T}_{k:0}\right]
\end{equation}
between the Choi matrix of the process tensor and the Choi matrix of the instrument,
which is exactly Born's rule extended to the multi-time setting. 

This identifies multi-time discrete quantum stochastic processes with operators on a Hilbert space (\cref{eq:pt-space}), but note that this is a strict inclusion of process tensors within this space; not all operators conversely represent physical processes. In particular, they must be \emph{completely positive} -- which is identical to positivity of the Choi matrix. They must also be \emph{causal}, which means that they must satisfy a set of no-signalling conditions which ensure that future interventions cannot affect past outcomes. Compactly, then, we have the following definition. 

\begin{definition}[Process tensor]
\label{def:discrete-pt}
   A process tensor $\PT$ defined on the time set $\{t_0,\cdots,t_k\}$ is a multi-linear mapping which is said to represent a physical process when its canonical Choi matrix 
   \[ \PTchoi \in\mathcal{B}(\mathcal{H}_{\mathfrak{o}_k}\otimes\mathcal{H}_{\mathfrak{i}_k}\otimes\cdots\otimes\mathcal{H}_{\mathfrak{i}_1}\otimes\mathcal{H}_{\mathfrak{o}_0}),\]
   satisfies
   \[\textbf{Complete positivity}\iff \PTchoi \succcurlyeq 0,\]
   and
   \[\textbf{Causality}\iff \Tr_{\mathfrak{o}_k}\left[\PTchoi\right] = \id_{\mathfrak{i}_k}\otimes \hat{\Upsilon}_{k-1:0}~~\forall~k.\]
\end{definition}

It should be stressed that, although there is an equivalence between states and process tensors, this analogy should not be taken too far. Multi-time processes are a strictly more general object than quantum states in that the `dual set' of experimental manipulations is larger. A quantum state $\rho_{\rm AB}$ can only be deterministically marginalised to $\rm B$ by performing a partial trace on $\rm A$ (measuring and forgetting). A \emph{process}, on the other hand, separated into some past($\mathsf{P}$)-future($\mathsf{F}$) partition $\Upsilon_{\mathsf{P}\mathsf{F}}$ can deterministically marginalise to $\Upsilon_{\mathsf{F}}^{\mathbf{A}_{\mathsf{P}}}$ by performing any deterministic instrument (such as a sequence of unitaries, for example) on the $\mathsf{P}$ slice. Under the CJI, this amounts to measurement of the Choi matrix with added post-selection. So, conditional states can only be produced with exponential overhead in sample complexity, whereas conditional processes can be produced deterministically. \par 
Indeed, this distinction is emblematic of the distinction between classical and quantum stochastic processes: classically, the choice not to measure is the same as the choice to measure and forget. Thus, there is no ambiguity in obtaining coarse-grained processes from fine-grained ones. Quantumly, however, this is of course not the case. The distinction between measure-and-forget as an operation from the identity operation is precisely what prompted a need to move from the master equation picture to process tensors in studying open quantum systems~\cite{Milz_2021_QST_review, milzKolmogorovExtensionTheorem2020}.

\subsection{Continuous matrix product states}\label{ssec:cMPS}

\emph{Continuous matrix product states}  (cMPS) \cite{osborneHolographicQuantumStates2010,verstraeteContinuousMatrixProduct2010,haegemanCalculusContinuousMatrix2013} are a subclass of (not necessarily normalised) states on a non-relativistic, one dimensional field theory. They can be represented and manipulated via computations on a holographic `virtual' space, which is possibly much smaller than the complete physical space. We will provide a minimalist introduction here, for further details see, e.g., Refs.~\cite{haegemanCalculusContinuousMatrix2013,tilloy2021notes}. 
First, we give the relevant facts about the involved Hilbert space.
Since in this work, we consider a field theory with $q$ different bosonic particle species,
the relevant Hilbert space is the bosonic Fock space over the single-particle Hilbert space $L^2([0,T])$ of all square-integrable functions on an interval $[0,T]$.
We will eventually interpret this interval as the 
time interval over which our processes are defined. 
The Fock space reads
\begin{equation}
\Gamma_T^{(q)}\Bigl(L^2([0,T])\Bigr) = \bigoplus_{\mathbf{n}}^\infty \mathcal{H}_{\mathbf{n}} 
=:
\Gamma_T^{(q)} ,
\label{eq:Fock_space_cmMS}
\end{equation}
where $\mathbf{n} = (n_1,\cdots,n_q)$ is a 
vector of particle numbers and $\mathcal{H}_{\mathbf{n}}$ denotes the fixed particle number sector, which is the symmetrised tensor product space $\mathcal{H}_{\mathbf{n}} = \bigotimes_{\nu=1}^q \text{Sym}^{n_\nu}(L^2([0,T]))$.
We denote the vacuum state of the field theory by $\ket{\Omega}$. 
The field operators $\psi_\nu(t)$ for the different particle species $\nu \in [q]$ obey the canonical bosonic commutation relations
\begin{equation}
\begin{split}
    [\psi_\nu(t), \psi^\dagger_{\nu'}(t')]& = \delta_{\nu,\nu'}\delta(t-t'), \\
    \quad [\psi_\nu(t), \psi_{\nu'}(t')] &= 0.
\end{split}
\end{equation}
These operators act on the Fock space by annihilating a particle of species $\nu$ at time $t$, and satisfy $\psi_\nu(t)\ket{\Omega} = 0$ for all $\nu$ and $t$.
A general vector $\ket{f}$ in the Fock space $\Gamma_T^{(q)}$ is often described by specifying for all $n \in \mathbb N_0$  its expansion coefficients $\{g_{\nu_1,\cdots, \nu_n}(t_1, \dots, t_n): \nu_j\in[q], j\in [n]\}$, in the Fock basis of (non-normalizable) particle number states. Those are square-integrable functions on $[0,T]$, which are symmetric under the simultaneous permutation of $t_1, \dots, t_n$ and $\nu_1, \dots, \nu_n$, and they are often called \textit{particle amplitudes}. The inner product between two vectors $ \ket{g}, \ket{f} \in \Gamma_T^{(q)}$ reads
\begin{multline}
\label{eq:inner_prod_Fock}
   \braket{g|f} = \\ \sum_{n=0}^\infty \sum_{\nu_1,\cdots,\nu_n=1}^q \int_0^T \mathrm{D}^{n}t \bigl(g_{\nu_1,\cdots, \nu_n}^\ast f_{\nu_1,\cdots, \nu_n}\bigr)(t_1,\cdots,t_n),
\end{multline}
where we adopted the notation $\int_0^T \mathrm{D}^{n}t \coloneqq \int_{0\leq t_1 \leq \cdots \leq t_n \leq T} \mathrm{d}t_1 \cdots\mathrm{d}t_n$. 
This inner product induces a norm on the Fock space which will be the relevant norm in our work. Concretely, for a vector $\ket{f} \in \Gamma_T^{(q)}$
\begin{equation}
    \| \ket{f} \| \coloneqq \sqrt{\braket{g|g}}.
\end{equation}
Distances between Fock space vectors will be measured with respect to this norm. 
Now we turn to the definition of the cMPS.
\begin{definition}[Continuous matrix product states]
   \label{def:cMPS}
   Given a $D$-dimensional Hilbert space $\mathcal{H}_{\rm aux}$, a cMPS is a vector $\ket{\Phi} \in \Gamma_T^{(q)}(L^2([0,T]))$ parametrised by the collection of matrix-valued `bulk' functions $Q(t), \{R_\nu(t)\} \in \mathcal{B}(\mathcal{H}_{\rm aux})$ for all $t \in [0,T]$ and $\nu \in [q]$, as well as the constant `boundary' matrix $B\in\mathcal{B}(\mathcal{H}_{\rm aux})$.
   The parametrisation takes the following form
\begin{multline}
   \label{eq:cMPS_definition}
      \cmps{Q}{\{R_\nu\}_{\nu \in [q]}}{B} := \\ \Tr_{\rm aux}\Bigl[B\,\mathcal{T}\exp\Bigl(\int_0^T \mathrm{d}t \,Q(t) \otimes \id + \sum_{\nu=1}^{q}R_\nu(t) \otimes \psi^\dagger_\nu(t)\Bigr)\Bigr]\ket{\Omega}.
\end{multline}
\end{definition}
We see that the class of cMPSs is parameterised by families of matrices on a $D$-dimensional auxiliary space. 
We will use the abbreviation $\cmps{Q}{\{R_\nu\}_{\nu \in [q]}}{B} \coloneqq \cmps{Q}{\{R_\nu\}}{B}$ if it is clear from the context how many particle species are being considered.
The values of the particle amplitudes of a cMPS $\ket{g} = \cmps{Q}{\{R_\nu\}_{\nu \in [q]}}{B} \in \Gamma^{(q)}_T $ are given by
\begin{multline}\label{eq:cMPS_Fock_space_expansion}
    g_{\nu_1,\cdots, \nu_n}(t_1,\cdots,t_n) = \\ \Tr_{\rm aux}\Bigl[B\,M_Q(T,t_n)R_{\nu_n}(t_n)\cdots R_{\nu_1}(t_1)M_Q(t_1,0)\Bigr],
\end{multline}
where $M_Q(t,s):= \mathcal{T}\exp\bigl(\int_s^t \mathrm{d}\tau \,Q(\tau)\bigr)$.
Note that $D$ can also be infinite, in which case the operators $Q(t)$, $\{R_\nu(t)\}$, and $B \in \mathcal{B}(\mathcal{H}_{\rm aux})$ must be such that the Fock space expansion in \cref{eq:cMPS_Fock_space_expansion} yields a vector with finite norm.
The class of cMPS is particularly useful because expectation values of field operators for a cMPS and overlaps of cMPSs are entirely expressible in terms of the parameter matrices of the involved cMPSs. 
For our work the overlap formula 
\begin{align}
\label{eq:cMPS_overlap}
&\braket{\Phi[Q',\{R'_\nu\}, B']|\Phi[Q,\{R_\nu\}, B]} =  \notag \\
&\text{Tr}_{\rm aux\otimes aux'}\,\Bigl[(B\otimes B'^\ast)\, \mathcal{T}\mathrm{e}^{\int_0^T \mathrm{d}t\,  Q\otimes \id+ \id \otimes Q'^\ast+ \sum_\nu R_\nu\otimes R'^\ast_\nu }  \Bigr]
\end{align}
for two general cMPSs will be particularly important.
Here and in the following, explicit reference to potential time-dependence of $\{R_\nu\}, \{R'_\nu\}$ and $Q, Q'$ is omitted for better readability. 

As their name suggests, cMPSs can be viewed as the continuum limit of \emph{matrix product states}  (MPS). 
Consider an MPS defined on a discrete lattice with $N = T/\epsilon$ sites and lattice spacing $\epsilon$, where each site hosts $q$ bosonic modes. 
Such an MPS can be written as
\begin{equation}
 \Tr_{\rm aux}\Bigl[B \prod_{j=1}^{N} \exp\Bigl(Q_j \otimes \id + \sum_{\nu=1}^q R_{\nu,j} \otimes a^\dagger_{\nu,j}\Bigr)\Bigr]\ket{0},
\end{equation}
where $a^\dagger_{\nu,j}$ are creation operators for the bosonic modes at site $j$.
The continuum limit is obtained by taking $N \to \infty$ (equivalently $\epsilon \to 0$) while keeping $T$ fixed, with the rescaling
\begin{equation}
Q_j = \epsilon Q(j\epsilon), \quad R_{\nu,j} = \sqrt{\epsilon} R_\nu(j\epsilon).
\label{eq:scaling_discrete_MPS_to_cMPS}
\end{equation}
Under this rescaling, the discrete product converges to the time-ordered exponential in \cref{def:cMPS}, and the discrete creation operators become field operators in the standard way as $a^\dagger_{\nu,j}/\sqrt{\epsilon}\mapsto\psi^\dagger_\nu(j\epsilon)$.
This scaling of the discrete parameter matrices in \cref{eq:scaling_discrete_MPS_to_cMPS}  ensures that in the limit the cMPS has finite particle density. 

Another informative interpretation of cMPS is viewing the field theory as a continuous meter which records the outcome of a continuous measurement performed on the system $\mathcal{H}_{\rm aux}$ \cite{osborneHolographicQuantumStates2010}. 
In this picture, the auxiliary space $\mathcal{H}_{\rm aux}$ represents the quantum system being measured, the matrices $\{R_\nu(t)\}$ encode the measurement operators coupling the system to the field at time $t$, the matrix $Q(t)$ describes the coherent evolution of the auxiliary system, and the boundary matrix $B$ contains the initial state of the auxiliary system and on which final state the auxiliary system is projected at the end. 
The Fock space vector $\ket{\Phi}$ then encodes the complete measurement record. 
It is this picture of cMPS which will eventually help us to construct the representation of continuous instruments in our framework. In analogy to this picture, we will conceptually refer to $Q$ as the \emph{drift matrix} of a cMPS, and $\{R_\nu(t)\}$ as the \emph{jump operators}.

We lastly remark that, much like their discrete counterparts, cMPS have an inherent gauge freedom built into the virtual space. Specifically, for any smoothly-defined position-dependent and invertible gauge matrix $g(t)$ on the interval $[0,T]$, we can perform the transformations
\begin{equation}
\label{eq:cmps-gauges}
    \begin{split}
        Q(t)&\mapsto g(t)^{-1}Q(t)g(t) + g(t)^{-1}\frac{\dd g}{\dd t}(t),\\
        R(t)&\mapsto g^{-1}(t)R(t)g(t),\\
        B&\mapsto g(0)^{-1}B g(T),
    \end{split}
\end{equation}
and the physical field theory vector will remain unchanged. This freedom in representation can be used for both conceptual clarity and computational stability. We refer the interested reader to~\cref{app:gauges} for further information on the different standard cMPS gauges.

\section{The continuous process tensor framework}\label{sec:framework}

% !Tex root = ./paper.tex

We wish to describe the continuum limit of discrete process tensors  defined on the time interval $[0, T]$. As we shall see, the natural continuum limit of process tensors is that of a one-dimensional, non-relativistic bosonic field theory over the space interval $[0,T]$.  In this sense, we will have a strict correspondence between non-trivial generators of instruments and particle excitations from the vacuum. In particular, the vacuum corresponds identically to the `do-nothing' operation, and so each $n$-particle sector of the field theory corresponds to $n$-time correlation functions. We will call this object a continuous process tensor (cPT). 
Although our derivation arrives at a cMPS representation of the cPT, with our work we wish to stress that viewing the continuum limit of process tensors from this field-theoretic perspective is not merely a useful analogy, but the appropriate operational lens of continuous process tensors. 
Our framework is centred around constructing a cMPS and exploiting known characteristics to both shed insight and develop methods for continuous process tensors.
 
\subsection{The continuum limit of process tensors}
\label{ssec:processes as cMPS}

In this section, we will motivate and state the key points of taking the continuum limit. We have explicitly left in the details of steps which communicate the physical insight of this framework, with further mathematical technicalities available in Appendix~\ref{app:derivation} and in our companion paper \cite{fuksaDilationTheoremContinuum2026}. At a high level: we (i) write the process tensor in its MPS form, (ii) convert it to a particular number basis of a bosonic Fock space, and finally (iii) take the continuum limit to derive its form as a cMPS. \par
The second step is a necessary component in taking the limit, and we shall see that it is physically well-principled: if one takes a process tensor at face value, then as an object it permits arbitrary manipulations in zero time. Naively taking $\Delta t\to 0$ with respect to the Choi matrix results in the creation of a Bell pair every $\text{d}t$ and has an ill-defined continuum limit. The resolution is to view instruments as objects which perform transformations of a system over a finite time window. The infinite energy levels in the Fock space associated to each time step can then be constructed to accordingly represent each order in the power expansion of an instrument. 
In this form, we see that only the \emph{generator} of the instruments survive the continuum limit -- and none of its higher orders -- which is enough to keep the continuous process tensor ``UV finite''. That is, an essential conclusion to introducing a wall clock into the process tensor framework is the requirement that the total energy (equivalently, excitations in the field theory) of any instrument probing the system is to remain finite. 
\par 
To start with, consider a $k$-step process tensor $\PT$ constructed via a set of system-environment unitaries $\{\mathcal{U}_{j:j-1}^{\rm SE}\}_{j=1}^k$ from whom we have the corresponding (vectorised) Hamiltonian generators $\{\Hgen^{(j)}\coloneqq -i(H_j\otimes\id - \id \otimes H_j^{\T}\}$ and for which---without loss of generality---we take the corresponding evolution times to each be $\shortT$. Note that although the assumption of an underlying generator to our dynamics keeps our framework more physically grounded, it comes at the expense of full generality in the classification of quantum stochastic processes. Nevertheless, it is only our intent to have an operational framework of such processes in the continuum. 
One could more generally think of a continuous process tensor as the CP functional underlying an infinite collection of consistent finite process tensors~\cite{milzKolmogorovExtensionTheorem2020} (which need not be governed by a stochastic differential equation, for example), but it is not entirely clear then how to extract physical meaning from such an object. Our starting point, therefore, shall be the Choi matrix $\hat{\Upsilon}_{k:0}^{\shortT}$, constructed such that each $\mathcal{U}_{j:j-1}(\shortT) = \exp(\shortT\Hgen^{(j)}).$
We set the number of steps to $k=\frac{T}{\shortT}$ and then construct the object conceptually represented by $\lim_{\shortT\to 0}[\hat \Upsilon_{k:0}^{\shortT}]$. 
To do this, we will first construct a second quantised version of $\hat{\Upsilon}_{k:0}^{\shortT}$ and show that it is identical to $\PTchoi$ in its preservation of the generalised Born rule (\cref{eq:pt-born}). 
It is this action as a linear mapping which we shall use to guide our derivation of this representation.

It is clearer to see the matrix-product nature of the process tensor by working instead with the vectorisation of \cref{eq:pt-choi-def} by employing the Liouville space representation. 
Here, $\mathcal{U}$ acts as a matrix on the (row-)vectorised $\rm SE$ state and our instruments' Choi matrices $\hat{\mathcal{A}}$ will be cast in the normalised Pauli basis $\{\kket{P_\mu}\}_{\mu = 0}^{d_{\rm S}^2-1}=:  \{\kket{\mu}\}_\mu$ of the system's Liouville space such that 
\[\bbra{\mu}\hat{\mathcal{A}}\kket{\mu'} = \Tr[\mathbb A(P_{\mu}\otimes P_{\mu'}^*)]\equiv \Tr[\mathbb A\, \mathbb{P}_{\mu, \mu'}],\]
where $\mathbb A$ denotes the vectorisation of $\mathcal A$.
This choice defines a vectorised process tensor $\kket{\Upsilon^{\shortT}_{k:0}}$ in an orthonormal basis, 
\begin{equation}
\label{eq:pt-mps}
\begin{split}
    \kket{\Upsilon^{\shortT}_{k:0}} = \sum_{\bm{\mu},\bm{\mu}'}\bbra{\bm{\mu},\bm{\mu}'}\Upsilon^{\shortT}_{k:0}\rangle\!\rangle \,\kket{\bm{\mu},\bm{\mu}'}
\end{split}
\end{equation}
where the $\bm{\mu},\bm{\mu}'$ coefficients take their values from the open quantum evolution
\begin{equation}
    \label{eq:pt-coeff}
    \begin{split}
  \bbra{\bm{\mu},\bm{\mu}'}\Upsilon^{\shortT}_{k:0}\rangle\!\rangle &=    \bbra{\id_{\rm SE}} \mathbb{P}_{\mu_k,\mu_k'}\e^{\shortT \Hgen^{(k)}}\cdots \\ 
 &\quad \qquad\cdots \mathbb{P}_{\mu_{1},\mu_{1}'}\e^{\shortT\Hgen^{(1)}}\mathbb{P}_{\mu_0,\mu_0'}\kket{\rho_0^{\rm SE}}.
     \end{split}
\end{equation}
This form illustrates the structure of $\kket{\Upsilon^{\shortT}_{k:0}}$ as an MPS: \cref{eq:pt-coeff} defines the matrix-product form. There are now two spaces to speak of: the \emph{virtual space}, which encodes the uncontrollable system-environment evolution; and the \emph{physical} indices which index controllable operations that an experimenter may perform on the virtual space.
One also sees here that the process tensor represents an affine sum over an informationally complete set of `event' histories, and that its action on any instrument sequence $\mathbf{A}_{k:0}$ is simply a linear expansion over these histories (which, like in the current choice of basis, need not necessarily be physical). 

At this point, let us characterise our (vectorised) instruments instead by generators such that $\mathbb A = \e^{\shortT \mathbb G}$. Note that this does not necessitate them to form a dynamical semigroup, or even for $\mathbb G$ to be uniquely defined at this point. Such a $\mathbb{G}$ always exists on the support of $\mathbb{A}$. At this stage, this `generator' and the `time' $\shortT$ need not correspond to the underlying physics of the instrument, but are an artifice of the proof which will adopt meaning in the continuum limit. 
In the limit, all instruments \emph{must} come equipped with a notion of time to be physical.
In particular, this means that each $\mathbb{A}_j$ can be expanded into a power series in this representation 
\begin{equation}
    \label{eq:power-exp}
    \mathbb A_j = \sum_{\ell=0}^\infty \frac{\shortT^\ell}{\ell!}\mathbb G_j^\ell = \sum_{\ell=0}^\infty \sum_{\mu\mu'}\frac{\shortT^\ell}{\ell!}(\mathbb G_j^{\ell})^{\mu , \mu'}\mathbb{P}_{\mu,\mu'},
\end{equation}
where $(\mathbb G_j^{\ell})^{\mu , \mu'}= \tr[\mathbb G_j^{\ell} \mathbb P_{\mu,\mu'}]$.
We will use this suggestive form to construct our Fock space representation of the process tensor. \par 
Let $\mathcal{H}_j:=\Span\{\kket{\mu,\mu'}\}_{\mu,\mu'}$, and note that
\begin{equation}
    \mathcal{H}_j\cong \mathcal{B}(\mathcal{H}_{\mathfrak{o}_j}\otimes \mathcal{H}_{\mathfrak{i}_{j+1}}).
\end{equation}
We would like to build a bosonic Fock space out of the local Hilbert space, which is where the physical indices live, at any given time $t_j$.  Within this Fock space, we have the freedom to associate the vacuum with something physically interpretable. Or, rather, to choose our vacuum in such a way that we argue is well-principled for processes. 
Specifically, we would like the vacuum to constitute the ``do nothing'' operation in the virtual space at that associated time $t_j$.
There are multiple reasons for this. Perhaps most compelling is that it emphasises the genuine \emph{process} nature of our object. The conventional Choi representation is an (unnormalised) quantum state, and manipulations thereof are state-like. One has to carve out the special subset of operations which carry physical meaning -- such as the identification of a Bell state as doing nothing. 
Our goal is that this is made inherent in the framework. Particles will therefore represent excitations of the system with respect to the `do-nothing' operation, which has now been imbued with a special status. Their meaning will be made mathematically explicit but for now should be understood as non-trivial manipulations of the system at a given time.
Consequently, low-excitation sectors represent marginals of higher particle sectors in the Kolmogorov sense, which is essential for defining a sensible notion of any stochastic process. 

Concretely, the identity operation corresponds to the zeroth element of our Liouville Pauli basis, and hence we define a ``process vacuum'' by $\kket{\text{vac}}_j := \kket{\mu=0,\mu'=0}_j$.
The Hilbert space $\mathcal{H}_j$ at time $t_j$ then correspondingly has the decomposition 
\begin{equation}
    \begin{split}
        \mathcal{H}_{j} &= \mathbb{C}\kket{\text{vac}}_j\oplus \mathcal{H}_j^{\nu},\quad\text{where}~\\
        \mathcal{H}_j^{\nu} &= \Span\{\kket{\nu}~:~\nu\neq(0,0)\}.
    \end{split}
\end{equation}
Note that we have combined the $(\mu,\mu')$ indices into a single $\nu$ which indicates the $d_{\rm S}^4-1$ traceless Pauli operators on the system Liouville space. 
From this identification, we can build a bosonic Fock space $\Gamma(\mathcal{H}_j^\nu)$ out of the \emph{excitation subspace} $\mathcal{H}_j^\nu$ via the standard construction
\begin{equation}
    \Gamma(\mathcal{H}_j^\nu) := \bigoplus_{\bm n=0}^\infty  \text{Sym}^{\bm n}(\mathcal{H}_j^\nu),
\end{equation}
where $\bm n = (n_{(0,1)},n_{(1,0)},\cdots,n_{\,\cramped{(d_{\rm S}^2,d_{\rm S}^2)}})
=:(n_\nu)_\nu$.
We have that $\text{dim }\mathcal H_j^\nu = d_{\rm S}^4-1$ and
each $\kket{\nu}\in \mathcal{H}^\nu_j$ hence defines a bosonic \emph{species} $\nu$ with associated creation and annihilation operators $a_{\nu,j}^\dagger$ and $a_{\nu,j}$. 
For $\bm n=\bm{0}\equiv(0,0,\cdots,0)$ we have the \emph{vacuum} subspace $\text{Sym}^{\bm 0}(\mathcal{H}^\nu_j) = \mathbb{C}$; meanwhile the $n \equiv  \sum_\nu n_\nu \geq1$ are the $n$-particle sectors. \par

We construct our vacuum more rigorously in~\cref{app:derivation}, but informally speaking: we can identify this vacuum as being the subspace of $\Gamma(\mathcal{H}_j)$ which corresponds to $\Gamma(\mathbb{C}\kket{\rm vac})$. The particles created by $a_{\nu,j}^\dagger$ should therefore be understood as excitations with respect to the creation of an arbitrary number of $(0,0)$-particles.
In other words, the action of the linear functional $\bbra{0,0}_j$ -- which is exactly the identity operation in the virtual space -- shall correspond to no particle at all. Through an abuse of notation, we say that
\begin{equation}
    \text{Sym}^{\bm 0}(\mathcal{H}_j^\nu)  \simeq  \Gamma(\mathbb{C}\kket{\rm vac})  \simeq \ket{\rm vac}_j.
\end{equation}
\par 

We show in \cref{app:derivation} that this is a genuine bosonic theory for which particles correspond identically to excitations relative to $\Gamma(\mathbb{C}\kket{\text{vac}}_j)$ as contained in $\Gamma(\mathcal{H}_j)$. We will therefore freely use the terminology that 
\begin{equation}
        \text{no particle at time }t_j \equiv \bbra{\id_j}\shortPT\rangle\!\rangle.
\end{equation}
With this, we have simply made the delineation that performing any kind of manipulation of the system in a process amounts to the creation of particles. To attach precise meaning now to the particles in this Fock space, we will re-write our discrete process tensor from Eq.~\eqref{eq:pt-mps} in this Fock basis. This follows by mapping the basis vectors of the local Hilbert space to elements of the Fock space. Although there are many ways to do this, our choice will momentarily become clear.
In particular, elements of the local Hilbert space $\kket{\nu}\in \mathcal{H}_j^\nu$ can be lifted to a vector in the Fock space $\Gamma(\mathcal{H}_j^\nu)$ via the `exponential' mapping~\cite{parthasarathy2014quantumstochasticcalculusquantum}, defined as
\begin{equation}
    \kket{\nu}\mapsto \ket{e(\nu)}:=\exp(a_\nu^\dagger)\ket{\text{vac}} =\bigoplus_{n=0}^\infty \frac{(a_\nu^\dagger)^{ n}\ket{\text{vac}}}{\sqrt{n!}}.
\end{equation}
 This has the property that $\langle e(\nu)|e(\nu')\rangle = \exp(\langle\!\langle \nu|\nu'\rangle\!\rangle)$.

We can extend this construction in a canonical way to include operators on some virtual space $\mathcal{H}$. Then for any scalar $c$ and operator $X \in \mathcal{B}(\mathcal{H})$ we have
\begin{equation}
    \begin{split}
    \exp(a_\nu^\dagger\otimes cX)[\ket{\text{vac}}_j\otimes \id] &= \bigoplus_{n=0}^\infty\frac{(a_\nu^\dagger)^{ n}\ket{\text{vac}}_j}{\sqrt{n!}}\otimes c^nX^n \\
        &\in~\Gamma(\mathcal{H}_j^\nu)\otimes\mathcal{B}(\mathcal{H}).
    \end{split}
\end{equation}
For a second (not necessarily identical) Hilbert space $\mathcal H'$ and operator $B \in \mathcal B(\mathcal H')$ we define an inner product on the Fock spaces as
\begin{equation}
\label{eq:exp-inner}
    \begin{split}
        &\bra{\text{vac}}_j\exp(a_{\nu}\otimes X)\exp(a_{\nu}^\dagger\otimes Y)\ket{\text{vac}}_j, \\
        &= \exp(\langle\!\langle \nu|\nu'\rangle\!\rangle X\otimes Y).
    \end{split}
\end{equation}
The use of this exponential mapping now becomes clear: the $\kket{\nu}\mapsto \ket{e(\nu)}$ association allows us to embed the exponentiation of an operator into the virtual space carried by its coefficients. Readers familiar with quantum combs may notice that~\cref{eq:exp-inner} resembles the form of a \emph{link product} with respect to physical and virtual spaces~\cite{milzQuantumStochasticProcesses2021}.\par
To see this in a schematic example (we will examine instruments in full generality in ~\cref{ssec:born rule}), let us now return to the action of the instrument $\mathcal{A}$ on the system at time $t_j$, and re-write its application in the process tensor using our second quantised language. Recall that we take $\mathbb A= \e^{\shortT \mathbb G}$. 
By the informational completeness of $\{\mathbb{P}_\nu\}$ on traceless operators, we know that 
\begin{equation}
\mathbb G = \sum_{\nu}\Tr[\mathbb G\mathbb{P}_\nu] \mathbb{P}_\nu:=\sum_{\nu}c_\nu \mathbb{P}_\nu. 
\end{equation}
If we take as fixed the physical-virtual tensor $\exp(\sum_\nu a_\nu^\dagger \otimes \mathbb{P}_\nu)\ket{\text{vac}}$ and construct a dual object $\bra{\text{vac}}\exp(\sum_{\nu'}c_{\nu'}a_{\nu'})$ then their inner product yields 
\begin{equation}
    \begin{split}
        &\bra{\text{vac}}\exp\left(\sqrt{\shortT}\sum_{\nu'}c_{\nu'}a_\nu\right)\exp\left(\sum_\nu a_\nu^\dagger \otimes \sqrt{\shortT}\mathbb{P}_\nu\right)\ket{\text{vac}}\\
        &= \exp\left(\shortT\sum_{\nu}c_\nu \mathbb{P}_\nu\right)\\
        &= \exp(\shortT \mathbb G) = \mathbb A.
    \end{split}
\end{equation}
As this construction can be easily generalised to correlated instruments by allowing $c_\nu$ to be matrix valued for fixed Fock space vector\[\exp\left(\sum_\nu a_\nu^\dagger \otimes \sqrt{\shortT}\mathbb{P}_\nu\right)\ket{\text{vac}}_j,\] we can always find a linear functional such that the contraction of the two builds an arbitrary instrument on the level of the virtual space at exactly time $t_j$. 

We are now ready to write $\kket{\Upsilon^{\shortT}_{k:0}}$ in its Fock space representation, which we refer to as the \emph{second-quantised process tensor} $\ket{\Upsilon^{\shortT}_{k:0}}\in\bigotimes_{j=1}^k \Gamma(\mathcal{H}_j^\nu)$ and its product-instrument dual $\bra{\mathbf{A}^{\shortT}_{k:0}}$. Letting $\ket{\Omega}_k=\bigotimes_j\ket{\text{vac}}_j$, we have 
\begin{widetext}
\begin{align}
    \label{eq:pt-2nd}
        &\ket{\Upsilon^{\shortT}_{k:0}}~~~~~~~= \bbra{\id_{\rm SE}}\prod_{j=1}^{k+1} \left[\exp\left(\id\otimes \shortT \Hgen^{(j)} + \sum_{\nu_{j}=1}^{d_{\rm S}^4-1}(a_{\nu_{j}}^\dagger \otimes \sqrt{\shortT}\mathbb{P}_{\nu_{j}})\right)\right]\ket{\Omega}_k\kket{\rho_0^{\rm SE}},\\  \label{eq:discrete_coherent_control}
        &\bra{\mathbf{A}^{\shortT}_{k:0}}~~~~~~~= \bra{\Omega}_k\prod_{j=0}^k\exp\left(\sum_{\nu_{j}=1}^{d_{\rm S}^4-1}\Tr[\sqrt{\shortT} \mathbb G_j \mathbb{P}_{\nu_{j}}] a_{\nu_{j}}\right),\qquad\text{such that,}\\
        &\langle \mathbf{A}^{\shortT}_{k:0}\ket{\Upsilon^{\shortT}_{k:0}} = \bbra{\id_{\rm SE}}\left[\prod_{j=1}^{k+1} \underbrace{\exp(\shortT\Hgen^{(j)})}_{\rm dynamics}\underbrace{\mathbb A_{j-1}}_{\rm instrument}\right]\kket{\rho_0^{\rm SE}}\qquad\text{the appropriate inner product, as desired.}
\end{align}
\end{widetext}
Examining this last equation, we can ascribe some physical meaning to our second quantisation. Specifically, a process tensor may be understood as a bosonic Fock space vector, for which the vacuum at any lattice site represents the ``do-nothing'', or identity operation. The bosonic species populate an informationally-complete representation of the basis elements of the Liouville space on the system, and the $n$-particle sector at a specific time $t_j$ consists of all of the non-trivial excitations of some arbitrary generator $\mathbb G_j$ at order $\mathcal{O}(\shortT^n)$. Note, however, that as it is written, this $\bra{\mathbf{A}_{k:0}^{\shortT}}$ is in its minimal form to demonstrate the structure of the process tensor. In full generality, it can be written with an interaction term on a controllable auxiliary space. This will be explored further in~\cref{ssec:born rule}.
Equation~\eqref{eq:pt-2nd} is still in MPS form, but with infinite-dimensional bosonic physical indices. Concretely,
\begin{equation}
    \begin{split}
    &\ket{\Upsilon^{\shortT}_{k:0}} = \sum_{s=0}^{\infty}\sum_{\substack{\bm{\nu}:|\bm{\nu}|=s}}\left(\bbra{L}\prod_{j=1}^k M_j^{s,\bm{\nu}}\kket{R}\right) a_{\nu_1}^\dagger\cdots a_{\nu_s}^\dagger \ket{\Omega}_k,
        \end{split}
\end{equation}
   where
\begin{equation}
    \begin{split}
        &~~\bbra{L} ~~= \bbra{\id_{\rm SE}}, \\
        &~~\kket{R} ~~\!= \kket{\rho_0^{\rm SE}},\\
        &~~M_j^0~~~= \exp(\shortT\Hgen^{(j)}),\\
        &~~M_{j}^{s>0,\bm{\nu}}= \exp(\shortT\Hgen^{(j)})\frac{\sqrt{\shortT^s}}{\sqrt{s!}}\prod_{i=1}^s \mathbb{P}_{\nu_i}.
    \end{split}
\end{equation}
We have thus arrived at the exact form of the pre-continuum limit MPS given in \cref{eq:scaling_discrete_MPS_to_cMPS}. From here, we can use the well-known cMPS architecture and take the natural continuum limit to construct the cMPS representation of a process tensor. We first fix a time interval $[0,T]$ on which the process is defined. Next, with $\shortT$ as the spacing, let $k=\frac{T}{\shortT}$ be the number of steps. It is well-known that the bosonic Fock space is only as big as the fermionic one in the continuum limit: as $\shortT\to 0$, there will only be two particle sectors that survive. These are the zero particle and the single particle sector at each position $j$
according to
\begin{equation}
    \begin{split}
        M_j^0 &= \id + \shortT \Hgen^{(j)},\\
        M_j^\nu &= \sqrt{\shortT}\mathbb{P}_\nu.
    \end{split}
\end{equation}
Thus, letting $\psi^\dagger_\nu(j\shortT) = a_{\nu_j}^\dagger / \sqrt{\shortT}$ be our continuum field operators, we finally have the following result:
\begin{result}
    [Field theory representation of a continuous process]\label{res:cPT}
    A continuous process tensor $\cPT$ in our chosen basis can be understood as a vector in $\Gamma_{[0, T]}^{(d_{\rm S}^4 - 1)}$ of the form
    \begin{equation}
        \label{eq:pt-cmps}
        \begin{split}
            &\cpt{\Hgen(t)}{\{\mathbb{P}_\nu\}}{B} :=\\
            &\hspace{1cm}\Tr\left(B\,\mathcal{T}\e^{\int_0^T \dd t \Hgen(t)\otimes \id + \sum_{\nu=1}^{d_{\rm S}^4-1}\mathbb{P}_\nu \otimes \psi^\dagger_\nu(t)}\right)\ket{\Omega}
        \end{split}
    \end{equation}
where $B:=\kket{\rho_0^{\rm SE}}\!\bbra{\id_{\rm SE}}$. The vacuum drift term is precisely the uncontrollable system-environment interaction; and the field operators $\{\psi_\nu^\dagger(t)\}$ create the process tensor legs $\{\mathbb{P}_\nu\}$ dilutely in jump form at each instant of time. 
\end{result}
The relationship to Eq.~\eqref{eq:pt-2nd} should be explicitly understood as
\begin{equation}
    \cpt{\Hgen(t)}{\{\mathbb{P}_\nu\}}{B} = \lim_{k\left(=\frac{T}{\shortT}\right)\to \infty} \ket{\Upsilon^{\shortT}_{k:0}}.
\end{equation}
\cref{res:cPT} is the starting point of our results in this work, and is the conceptual insight from which we develop our framework.

We conclude this section with a few remarks. One might notice that replacing the Hamiltonian drift term in~\cref{eq:pt-cmps} with a Lindbladian one will also yield a physically completely reasonable process. Indeed, the steps to arrive at this point could have been conducted with an open system-environment evolution instead of a closed one. 
Since the dilation of Lindblad evolution -- either to constant Hamiltonians on an infinite-dimensional~\cite{Burgarth_2022_dilations}, or time-dependent Hamiltonians on $(d_{\rm S}d_{\rm E})^2$-dimensional auxiliary spaces~\cite{Burgath_2015_TD_dilations} -- is both well-understood and well-behaved, we can define cPTs purely from a Hamiltonian perspective, or equivalently interchange the drift matrix with Lindblad dynamics without loss of generality.

\par We have derived a genuine continuum form of process tensors. It is exact, and moreover extends the state-process equivalence to show how non-Markovian processes should be operationally understood as bosonic Fock space vectors on a continuous interval. The legs of the Choi matrix now become dilute particle excitations amidst a vacuum which represents the always-on system-environment interaction. 
By virtue of introducing time into this framework, we see that the action of any physical instruments can only be meaningfully understood as the infrequent activation of their generators.
Indeed, as we shall momentarily see, the recovery of discrete process tensors is achieved through a choice of delta function instruments, which have infinite norm. 
In the rest of this section, we will abstract away from the cMPS representation and study general vectors in $\Gamma^{(d_{\rm S}^4 - 1)}_{[0,T]}$ as representatives of continuous processes.
\par

\subsection{Getting back to finite process tensors}\label{ssec:discrete marginals}
Now that we have derived a continuum representation $\cPT$ of a quantum stochastic process, we will say a few words about how to proceed in the reverse direction and recover finite process tensors in their Hilbert space form. When considering the wavefunctions of the cPT~\cref{eq:pt-cmps} using~\cref{eq:cMPS_Fock_space_expansion}, we notice that they can be seen as matrix elements of all the $k$-step process tensors, marginals of the cPT, for $k=0, \dots, \infty$, with the positioning of the legs at all possible times on the given interval. Therefore, in order to retrieve $\PTchoi$ from $\cPT$, we simply need to project \cref{eq:pt-cmps} onto its $k+1$-particle sector with field operators at the relevant times. 

Concretely, for a given set of times $\bm{t} := (t_0,\cdots , t_{k})$ with $0\leq t_0 < t_1 < \cdots < t_{k-1} < t_k\leq T$, let us define the operators $\{\xi_{\nu_i}(t_i)\}_{\nu_i}$ corresponding to the $i$th time, where $i = 0,1,\cdots,k$, by
\begin{equation}
    \xi_{\nu_i}(t_i) = 
    \begin{cases}
        \id & \text{if } \nu_i = 0,\\
        \psi_{\nu_i}(t_i) & \text{if } \nu_i \in \{1,\cdots,d_{\rm S}^4-1\}.
    \end{cases}
\end{equation}
This recombines the `do-nothing' operation with the non-trivial excitations, from which we hence re-institute the $\nu \rightarrow (\mu,\mu')$ indexing via the definition $\mathbb{P}_{\nu} \rightarrow \mathbb{P}_{\mu,\mu'}  = P_{\mu}\otimes P_{\mu'}^\ast$. 
With these operators, we can define the tensor elements $(\Upsilon_{k:0})_{\bm{\mu},\bm{\mu'}}(\bm{t})$ from the finite-particle wavefunctions of the $\cPT$ Fock space vector
\begin{equation}\label{eq:discrete marginal}
    \begin{split}
    (\Upsilon_{k:0})_{\bm{\mu}, \bm{\mu'}}(\bm{t})&\coloneqq \langle\Omega|\prod_{j=0}^k \xi_{\mu_j,\mu_j'}(t_j)\cPT\\
        =\bbra{\id_{\rm SE}}&M_{\Hgen}(T,t_{k})\mathbb{P}_{\mu_{k},\mu_{k}'} M_{\Hgen}(t_k,t_{\cramped{k-1}})\cdots \\
         & \qquad \qquad \cdots\mathbb{P}_{\mu_0,\mu_0'}M_{\Hgen}(t_0,0)\kket{\rho_0^{\rm SE}}.
    \end{split}
\end{equation}
As in~\cref{eq:cMPS_Fock_space_expansion}, $M_\mathbb{H}(t,s):= \mathcal{T}\exp\bigl(\int_s^t \mathrm{d}\tau \,\mathbb{H}(\tau)\bigr)$.
This exactly recovers the form of the process tensor coefficients from \cref{eq:pt-coeff}, and hence is precisely the object $\PT$ defined on $\{t_0,\cdots , t_k\}\subseteq[0,T]$. Note that if $t_0 > 0$, then $\kket{\rho_0^{\rm SE}}\mapsto M_{\Hgen}(t_0,0)\kket{\rho_0^{\rm SE}}$. And further, owing to the trace-preservation of $M_{\Hgen}(t_k,T)$, we have that $\bbra{\id_{\rm SE}}\mapsto \bbra{\id_{\rm SE}}M_{\Hgen}(t_k,T) = \bbra{\id_{\rm SE}}$.

Note that the object $\Upsilon_{k:0}$ defined above does not strictly speaking correspond to a process tensor as defined in \cref{ssec:discrete process tensors}, since it is missing an output leg.
In fact, it is equivalent to a process tensor with the output leg traced out. 
In the literature, such an object is sometimes called a \emph{tester}.
However, this is a purely technical distinction, which does not have physical consequences, due to the causality constraint.

\Cref{eq:discrete marginal} illustrates how the cPT contains all of the information required to construct any element of the entire family of process tensors over the interval $[0,T]$. 
It is instructive to note that in order to recover these discrete process tensors, we needed to compute the overlap with $\langle\Omega|\prod_{j=0}^k \xi_{\nu_j}(t_j)$. Such vectors are not normalisable, as their norm is a product of delta functions. In accordance with our earlier arguments about the natural unphysicality of discrete process tensors, we can see from this fact that obtaining them necessitates instruments with unbounded norm, or -- in a sense -- the application of ``infinite energy'' manipulations of the system. 
We formalise this intuition using the Gelfand triple construction in the companion work~\cite{fuksaDilationTheoremContinuum2026}.

Finally, let us remark that we might want to get the finite process tensors in an interaction basis other than $\mathbb{P}_\nu$.
For instance, we will be interested in the basis $\kket{P_\alpha}\!\bbra{P_\alpha'} = \sum_\nu c^\nu_{\alpha, \alpha'} \mathbb{P}_\nu$.
As can be readily seen from~\cref{eq:discrete marginal}, this can be achieved by rotating the $\xi_\nu$ operators as
\begin{equation}\label{eq:basis change}
    \tilde{\xi}_{\alpha, \alpha'}(t) \coloneqq \sum_\nu (c^{-1})^\nu_{\alpha, \alpha'} \xi_\nu(t),
\end{equation}
where $c^{-1}$ is the inverse of $c$ as a matrix from $\nu$ to the multi-index $(\alpha, \alpha')$.
If the basis transformation is unitary and leaves the identity invariant, it is a type of a Bogoliubov transformation, defining a new set of field operators that can be used to formulate the whole theory in the new basis.

\subsection{Continuous processes as Fock space vectors}
\label{ssec:physicality}
Although we have used the well-known cMPS continuum limit to arrive at our main result, we wish to emphasise that it need not be siloed in this form.
In this section, we will show this by abstracting away from the cMPS representation and by doing so we will obtain the class of Fock space vectors with $d_{\rm S}^4 - 1$ bosonic species that represent a quantum stochastic process on $[0, T]$.
Mathematically, this can be interpreted as closing the set of cPTs represented by a cMPS under the Fock space norm, as elaborated on in the companion work~\cite{fuksaDilationTheoremContinuum2026}.
This generalisation paves the way towards using techniques from quantum field theory other than cMPSs to study quantum stochastic processes, and makes the theory mathematically well behaved.
Additionally, let us recall that we do not claim that Fock space vectors describe \emph{all} quantum stochastic processes, as they are studied by the mathematical literature, but rather that they define a physically motivated subset of them, which is relevant from a practical standpoint.
We leave the rigorous mathematical treatment of the connection between the bosonic Fock space and quantum stochastic processes as mathematical objects to future work.

\subsubsection*{Causality and positivity for cPTs}\label{ssec:physical conditions}
Recall from~\cref{def:discrete-pt} that the correspondence between (unnormalised) quantum states and process tensors can be made via the basis-independent conditions that (a) a process must not be able to signal from future to past (\emph{causality}), and (b) its action as an operator on the complete set of valid instruments should produce only non-negative numbers (\emph{complete positivity}). Briefly, as a reminder, these conditions manifest in the Choi matrix via
\begin{enumerate}[label=(\roman*)]
    \item $\Tr_{\mathfrak{o}_k}[\PTchoi] = \id_{\mathfrak{i}_k}\otimes \hat{\Upsilon}_{k-1:0}~\forall~k$ (causality).
    \item $\PTchoi\succcurlyeq 0$ (positivity).
\end{enumerate}
The first condition can be understood as an affine condition that the projection of $\mathfrak{o}_k$ onto the identity, along with $\mathfrak{i}_k$ onto any non-identity Pauli $P_{\mathfrak{i}_k}$ must be identically zero. i.e., that $\Tr[(\id_{\mathfrak{o}_k}\otimes P_{\mathfrak{i}_k})\PTchoi] = 0$ for $P_{\mathfrak{i}_k} \neq d_{\rm S}^{-1/2} \id_{\mathfrak{i}_k}$.
We will now examine how those conditions generalise to the continuum. 
Note that the causality condition serves as a normalisation condition for Choi matrices of process tensors.
This condition is distinct from the state normalisation condition: Choi matrices are not normalised from the quantum state perspective.

We will state the conditions in a representation-independent manner, as well as for a cMPS parametrisation. 
In the case of generic Fock space vectors, we will designate that we have $d_{\rm S}^4 - 1$ bosonic fields, defined on the interval $[0, T]$, with the pre-set interpretation that the particles correspond to interventions in the traceless Pauli basis -- meaning that finite particle amplitudes define discrete process tensors as in~\cref{eq:discrete marginal}.

\textbf{Causality.}
The physical meaning of causality should be taken to be that, given an instrument applied across a future interval $[s,T]$, the process itself should be unchanged across the past interval $[0,s]$.
We will derive the causality conditions on $\ket{\Psi_T} \in \Gamma_{[0, T]}^{d_{\rm S}^4 -1}$ from its finite particle sectors which, as above, are taken to be related to finite process tensors in the Liouville Pauli basis. 
Consider, first, the  basis transformation 
\begin{equation}
\label{eq:basis_transformation_alphas}
    \kket{P_\alpha}\!\bbra{P_{\alpha'}} = \sum_{\nu\in[d_{\rm S}^4]} c^\nu_{\alpha, \alpha'} \mathbb{P}_\nu
\end{equation}
of the jump operators from $P_{\mu}\otimes P_{\mu'}^\ast$ to $\kket{P_{\alpha}}\!\bbra{P_{\alpha'}}$,
and the corresponding transformation $\xi_\nu(t) \to \tilde{\xi}_{\alpha, \alpha'}$ in~\cref{eq:basis change}. This associates a given label $\alpha'_j$ with the space $\mathfrak{o}_j$ and $\alpha_j$ with $\mathfrak{i}_{j+1}$, allowing us to separate out the causally distinct spaces. The analogous $\Tr_{\mathfrak{o}_k}(\cdot)$ is implemented by construction by the boundary trace at $T$ in the cPT.
Consider, then, the full set of $(k+1)$-point finite-particle amplitudes on the ordered set $\{t_0,\cdots, t_k\}.$ The causality conditions for the associated $k$-step discrete marginals of $\ket{\Psi_T}$ now read:
\begin{equation}
    \underbrace{\bra{\Omega} \prod_{i}^k \tilde{\xi}_{\alpha_i,\alpha_i'}(t_i)}_{\substack{\text{finite-particle}\\\text{projection}}} \ket{\Psi_T} = 0, 
    \quad \forall~\bm{\alpha},\bm{\alpha'} : \alpha_k\neq 0,
\end{equation}
along with the normalisation condition $\braket{\Omega | \Psi_T} = 1$.
However, as the vector $\bra{\Omega}\prod_{i=1}^k \tilde{\xi}_{\alpha_i, \alpha'_i}(t_i)$ is not normalisable, we have to use a smeared version, integrating the above against a test function.
To write causality as a condition directly on the field theory vector and not just on its marginals, we will need the projector onto the subspace with no particles in the interval $[t,T]$,
\begin{equation*}
    \begin{split}
    \Pi_{t} & \coloneqq \sum_{n=0}^\infty\sum_{\nu_1,\cdots,\nu_n}\int_0^t{\rm{D}}^nt \\
   &\qquad \psi_{\nu_n}^\dagger(t_n)\cdots \psi_{\nu_1}^\dagger(t_1) \ket{\Omega}\!\bra{\Omega}\psi_{\nu_1}(t_1) \cdots \psi_{\nu_n}(t_n).
    \end{split}
\end{equation*}
This operator conditions on the cases where there is no intervention after time $t$.
We are now in the position to write down the causality condition for cPTs.
\begin{definition}[Causality -- continuum]\label{def:causality}
    A vector $\ket{\Psi_T} \in \Gamma_{[0,T]}^{d_{\rm S}^4 - 1}$ is \emph{causal} if for all $f\in L_2([0,T])$ and $\alpha \neq 0, \alpha'$
    \begin{align}
        \braket{\Omega | \Psi_T} &= 1\label{eq:causality}, \\
        \int {\rm{d}} t \ f(t) \Pi_{t} \tilde{\xi}_{\alpha,\alpha'}(t) \ket{\Psi_T} &= 0.
        \label{eq:causality2}
    \end{align}
    A cMPS $\ket{\Psi_T[Q, \{R_\nu\}, B]}$ parametrises a causal process if
    \begin{align}
    \tr\left[\mathcal{T}\left(\e^{\int_0^T \mathrm d s Q(s)}\right) B\right] &= 1,\\
     B \mathcal{T}\left(\e^{\int_t^T \mathrm d s  Q(s)}\right) \tilde{R}_{\alpha,\alpha'} &= 0 \quad \forall  t, \alpha \neq 0, \alpha'.
    \end{align}
\end{definition}

\textbf{Positivity.}
The positivity can also be defined by considering the smeared discrete marginals of a process.
\begin{definition}[Complete positivity -- continuum]\label{def:positivity}
    For any function $f\ge0 \in L^2(\Delta_k)$ with $\int_0^T {\rm D}^k\bm{t}|f(\bm{t})| = 1$ consider
    \begin{equation}
        \Psi_{\bm{\nu}}^f \coloneqq \int_0^T {\rm D}^k \bm{t} \ f(\bm{t}) \bra{\Omega} \prod_{j=1}^k \xi_{\nu_i}(t_i) \ket{\Psi_T}.
    \end{equation}
    The vector $\ket{\Psi_T}$ is completely positive if and only if for all $f$ and $k$ the matrix $(\Psi^f_{\bm{\mu},\bm{\mu'}})_{\bm{\mu},\bm{\mu'}} \equiv (\Psi_{\bm{\nu}}^f)_{\bm{\nu}}$ is positive semi-definite.
    For a cMPS $\ket{\Psi_T[Q, \{R_\nu\}, B]}$ this implies that for all $k=0,\dots,\infty$ and all $0\le t_1<\dots<t_k\le T$, the matrix from $\bm{\mu'}$ to $\bm{\mu}$
    \begin{equation}
        \tr\left[B M_Q(T, t_k) R_{(\mu_k, \mu'_k)} \dots R_{(\mu_1, \mu_1')} M_Q(t_1, 0) \right]
    \end{equation}
    is positive semi-definite, where we identify $\nu_i \to (\mu_i, \mu'_i)$ and set $M_Q(t',t) \coloneqq \mathcal{T}\exp[\int_t^{t'} {\rm d}s Q(s)]$.
\end{definition}

There are also two necessary, but insufficient conditions implied by positivity on a cPT in a cMPS form.
First, note that imposing positivity on the discrete marginals implies that the $R$ matrices are linearly independent~\cite{Milz_2018_restricted}.
Second, since the marginals need to be Hermitian, we have the requirement that 
\begin{equation}
\begin{split}
    \hat{\Psi}_T[Q, &\{R_{\mu, \mu'}\}_{(\mu, \mu') \neq (0, 0)}, B]^\dag \\
    &= \hat{\Psi}_T[Q^*, \{R_{\mu', \mu}^*\}_{(\mu, \mu') \neq (0, 0)}, B^*],
\end{split}
\end{equation}
noting the reversed index on the $R$ matrices on the RHS.
For this to be satisfied, we require that there exists a gauge transformation $g$ with action $G$ of the form given in~\cref{eq:cmps-gauges} -- i.e., a differentiable and everywhere-invertible matrix-valued function $g: [0, T] \rightarrow \mathbb{C}^{\chi \times \chi}$ -- such that
\begin{equation}\label{eq:hermitian choi state}
\begin{split}
    Q^\ast(t) &= G[Q(t)],\\
    R^\ast_{\mu,\mu'}(t) &=  G[R_{\mu',\mu}],\\
    B^\ast &= G[B].
\end{split}
\end{equation}
The role of the gauge transformation can be seen from the cMPS we obtained in~\cref{sec:framework}, where $Q = \mathbb{H}_{\rm SE}$ is a Liouvillian time generator of SE dynamics and $R_{\mu, \mu'} = \mathbb{P}_\mu \otimes \mathbb{P}_{\mu'}^*$.
This cMPS of course satisfies the Hermiticity condition, and requires $g$ to swap the forward and backward components of the Liouville system space: the condition states $\mathbb{P}^*_{\mu, \mu'} = P^*_\mu \otimes P_{\mu'} = g^{-1} \mathbb{P}_{\mu', \mu} g$, implying $g = \text{SWAP}$.

Finally, we arrive at a definition of a continuous process tensor in terms of generic Fock space vectors, which generalises the one made in~\cref{def:discrete-pt}.
\begin{definition}[Continuous process tensor (cPT)]\label{def:cpt}
    A vector $\ket{\Psi_T} \in \Gamma_{[0, T]}^{d_{\rm S}^4 - 1}$ is a cPT if it is \underline{causal} according to~\cref{def:causality} and \underline{completely positive} according to~\cref{def:positivity}. We denote the set of all cPTs by $\mathsf{cPT}$.
\end{definition}
This completes the ansatz-free formulation of continuum quantum stochastic processes, and thus establishes a correspondence between Fock space vectors on an interval, and continuous process tensors.

\subsection{Process-canonical representation of cPTs}\label{ssec:PCR}

In~\cref{ssec:processes as cMPS} we have derived cPTs as a cMPS vector in a Fock space, where the cMPS parameters had a clear relation to the underlying $\rm SE$ dynamics.
The connection to the dynamics motivates the following definition.
\begin{definition}[Process-canonical representation (PCR)]\label{def:process-canonical rep}
    A \emph{process-canonical representation} of a continuous process of a $d_{\rm S}$ dimensional quantum system is a particluar cMPS parametrisation $\cmps{Q}{\{R_\nu\}}{B} \in \Gamma^{d_{\rm S}^4 - 1}_{[0,T]}$ whose auxiliary space corresponds to the Liouville space of some $\rm SE$ dynamics.
    Concretely, this means that the drift matrix $Q(t) = i\left(H_{\rm SE}(t) \otimes \id_{\rm SE} - \id_{\rm SE} \otimes H_{\rm SE}^*(t)\right)$ is a vectorised Liouvillian, where $H_{\rm SE}(t)$ is a Hamiltonian on a $d_{\rm S} \times d_{\rm E}$ dimensional quantum system, the jump matrices $R_\nu = \mathbb{P}_\nu \otimes \id_{\rm E}$ for $\nu \in [d_{\rm S}^4 - 1]$ are the non-identity Pauli matrices on the system and the boundary matrix $B = \kket{\rho_0^{\rm SE}}\!\bbra{\id}$ for $\rho_0^{\rm SE}$ a pure quantum state.
    Moreover, we impose that $\|H_{\rm SE}(t)\|_\infty<\infty$ for all $t\in[0,T]$.
\end{definition}
\noindent
This definition brings up a natural question.
What is the connection between all cPTs as defined in~\cref{ssec:physicality} and those that have a PCR?
In other words, what is the connection between positive causal vectors in the Fock space $\Gamma^{d_{\rm S}^4 - 1}_{[0, T]}$ and the Schrödinger equation?
In the companion work~\cite{fuksaDilationTheoremContinuum2026} we find that in fact any cPT can be approximated arbitrarily well by a PCR.
We prove the following.
\begin{theorem}[PCR approximation theorem~\cite{fuksaDilationTheoremContinuum2026}]\label{thrm:pcr density}
For any $\cPT \in \Gamma^{d_{\rm S}^4 - 1}_{[0,T]}$ which is positive and causal and any $\epsilon>0$, we can find a PCR $\cmps{\mathbb{H}_{\rm SE}}{\{\mathbb{P}_\nu\otimes \id_{\rm E}\}_\nu}{\kket{\rho_{\rm SE}}\!\bbra{\id_{\rm SE}}}$, such that
\begin{equation}
    \|\cPT - \cmps{\mathbb{H}_{\rm SE}}{\{\mathbb{P}_\nu \otimes \id_{\rm E}\}}{\kket{\rho_{\rm SE}}\!\bbra{\id_{\rm SE}}}\| \le \epsilon,
\end{equation}
where $\|\cdot\|$ is the norm on $\Gamma^{d_{\rm S}^4 - 1}_{[0,T]}$.
\end{theorem}
\noindent
This theorem can be understood as a dilation theorem for cPTs: it connects the abstract mathematical concept $\cPT$ to quantum dynamics, which we can, e.g., simulate on a quantum computer, or otherwise implement in the lab, with an arbitrarily small approximation error.
Furthermore, it shows that the physicality conditions obtained in~\cref{ssec:physicality} by analogy with the discrete case, are complete.

\subsection{Instruments and the generalised Born rule}
\label{ssec:born rule}

After defining $\mathsf{cPT}$ and finding their canonical representation, we now turn to the definition of instruments and their representation in our framework.
Instruments are maps from the space of Fock vectors to probability density functions over some outcome set $\mathcal{X}$.
In particular, the probability density of obtaining $x\in\mathcal{X}$ is given by
\begin{equation}
    p(x | \mathcal{I}) = \langle \mathcal{I}(x) \cPT,
\end{equation}
such that the probability of the event $E\subset \mathcal{X}$ is
\begin{equation}
    p(E| \mathcal{I}) = \int_E {\rm d}x \langle\mathcal{I}(x) \cPT,
    \label{eq:gen_Born_rule}
\end{equation}
where ${\rm d}x$ is a reference measure on $\mathcal{X}$ (e.g. the Lesbegue measure for $\mathcal{X}$ an Euclidean space). We note that the integral becomes a (potentially weighted) sum in the case when $\mathcal{X}$ is a discrete set.
The Riesz representation theorem hence provides to each $E \subset \mathcal{X}$ a dual vector $\bra{\mathcal{I}(E)}$, which we call the \emph{effects} of the instrument $\mathcal{I}$.
Note that while the effects are always elements of the dual space, this is not necessarily the case for $\bra{\mathcal{I}(x)}$.
This object should be understood as the map from cPTs to probability densities when contracted with a cPT, not necessarily as a dual vector.

So an instrument effect is a map $\langle\mathcal{I}(E)| : \mathsf{cPT}\to [0,1]$ telling us the probability of event $E$ given some cPT. Relatedly, one can also \emph{condition} on events occurring to obtain either a spatially or temporally reduced (and possibly subnormalised) cPT. We briefly discuss here how to define reduced cPTs of subsystems. Among other physical instances, this is desirable in order to have a robust notion of positivity. Specifically, suppose we have a tensor product system $\rm{S}\otimes \rm{S}'$ on which a cPT $\cPT\in\mathsf{cPT}_{\rm SS'}$ is defined. Then, given some instrument effect $\langle  \mathcal{I}(E)|:\mathsf{cPT}_{ \rm{S}}\to [0,1]$ it is natural to ask what the resultant cPT is on $\rm{S}'$. In this vein, one would like to define the operator which plays the role of ``$\langle \mathcal{I}(E)|_{\rm{S}}\otimes\id_{\rm{S'}}$''. We call this operator $c_E : \mathsf{cPT}_{\rm{S}\rm{S}'}\to\mathsf{cPT}_{\rm{S}'}$ and define it as follows. First, let $\mathsf{cPT}_{\rm SS'}$ be defined with field operators $\psi_{\nu,\nu'}^\dagger$ corresponding to the Liouville basis $\{\mathbb{P}_\nu^{(\rm{S})}\otimes\mathbb{P}_{\nu '}^{(\rm{S}')}\}$, and denote by $\{\tilde \psi_{\nu'}^\dagger\}$ the field operators of $\mathsf{cPT}_{\rm S'}$. We can resolve the identity $\mathscr{I}_{\rm S'}$ on $\rm S'$ with respect to its Liouville space by summing up and matching the two sets of field operators. Explicitly, if we let 
\begin{equation}
    \mathscr{I}_{\rm{S}'}(\cdot) \coloneqq \!\!\!\!\! \sum_{\substack{m=0\\\bm{\nu'}\in [d_{\rm{S}'}^4-1]^m}}^{\infty}\int \mathrm{D}^m\tau \prod_{j=1}^m \tilde \psi_{\nu'_j}^\dagger(\tau_j) (\cdot)  \psi_{0,\nu'_{j}}(\tau_j),
\end{equation}
and, letting $\{\Psi_{\bm{\nu}}^{(n)}(\bm t)\}$ be the finite-particle wavefunctions of $\langle \mathcal{I}(E)|$, then we have
\begin{equation}\label{eq:cond-op}
    c_{E} \coloneqq \!\!\!\!\!\sum_{\substack{n=0\\\bm{\nu}\in[d_S^4-1]^n}}^{\infty}\!\!\!\!\!\!\!\! \int \mathrm{D}^n t \; \Psi^{(n)}_{\bm{\nu}}(\bm{t})\mathscr{I}_{\rm{S}'}[|\Omega_{\rm{S}'}\rangle\!\langle \Omega_{\rm{S}\rm{S}'}|]\prod_{k=1}^n\psi_{\nu_k,0}(t_k).
\end{equation}

With a definition of conditioning in place, we will now define an appropriate notion of combining instruments on two systems $\rm S$ and $\rm S'$.
Consider the respective effects $\bra{\mathcal{I}(E)}$ and $\bra{\mathcal{I}'(E')}$ with wavefunctions $\{\Psi_{\bm{\nu}}^{(n)}\}$ and $\{{\Psi'}^{(n')}_{\bm{\nu'}}\}$.
The joint effect on $\rm SS'$ is given by
\begin{equation}
\begin{split}
    \bra{\mathcal{I}(E) \otimes \mathcal{I}'(E')} \coloneqq \sum_{n,n'=0}^\infty \sum_{\substack{\bm{\nu}\in[d_{\rm S}^4 - 1]^n \\ \bm{\nu'}\in[d_{\rm S'}^4 - 1]^{n'}}} \! \int_0^T {\rm D}^n\bm{t} \int_0^T {\rm D}^{n'}\bm{t'} \\ \Psi^{(n)}_{\bm{\nu}}(\bm{t}){\Psi'}^{(n')}_{\bm{\nu'}}(\bm{t'})\bra{\Omega}\prod_{k=1}^n \psi_{\nu_k, 0}(t_k) \prod_{k'=1}^{n'} \psi_{0, \nu'_{k'}}(t'_{k'}).
\end{split}
\end{equation}
Although we adapt the tensor product notation for combining quantum systems, note that the combined effect is \emph{not} the tensor product of the individual ones.
It is rather consistent with the way dynamics generators combine: given generators $\mathbb{A}_{\rm S}$ and $\mathbb{A}'_{\rm S'}$ on $\rm S$ and $\rm S'$ the total generator on $\rm SS'$ is $\mathbb{A}_{\rm S}\otimes \id_{\rm S'} + \id_{\rm S} \otimes \mathbb{A}'_{\rm S'}$, such that only the $(\nu,0)$ and $(0,\nu')$ components in the Pauli basis survive.
The two definitions are compatible in the sense that $\langle \mathcal{I}(E)\otimes \mathcal{I}'(E')|\Upsilon_T\rangle = \langle \mathcal{I}'(E')|c_{E} |\Upsilon_T\rangle$ for all cPTs $\ket{\Upsilon_T}$ on $\rm SS'$.
Consequently, we can interpret $|\Upsilon_T^{(E)}\rangle = c_{E}|\Upsilon_T\rangle$ as being proportional to the (not necessarily causal) cPT on $\rm{S}'$ conditioned on outcome $E$; to renormalise one simply needs to divide by the probability $p(E|\mathcal{I})$. 

The operation of conditioning is essential for enforcing the robustness of positivity under extension of the quantum system to a larger one.  We are now in the position to define the space of physical continuous instruments.

\begin{definition}[Continuous instruments]
\label{def:instrument}
    Suppose an outcome set $\mathcal{X}$ with a reference measure ${\rm d} x$.
    A \emph{continuous instrument} on a $d_{\rm S}$-dimensional quantum system is a map $\bra{\mathcal{I}(\cdot)}$ from cPTs to probability density functions over $\mathcal{X}$.
    For any measurable subset $E$ of $\mathcal{X}$, this defines an \emph{effect} $\bra{\mathcal{I}(E)} \in \big(\Gamma^{d_{\rm S}^4 - 1}_{[0, T]}\big)^*$, such that for all $\ket{\Phi} \in \Gamma^{d_{\rm S}^4 - 1}_{[0,T]}$
    \begin{equation}
        \braket{\mathcal{I}(E)|\Phi} = \int_E {\rm d}x \braket{\mathcal{I}(x)|\Phi}.
    \end{equation}
    Analogous to cPTs, instruments satisfy the respective complete positivity and causality conditions:
    For all $E\subseteq \mathcal{X}$ and all $\cPT\in\mathsf{cPT}_{\rm S}$ we have $0\leq \langle\mathcal{I}(E)|\Upsilon_T\rangle \leq 1 = \langle \mathcal{I}(\mathcal{X}) \cPT$, and moreover, for any $\cPT\in\mathsf{cPT}_{\rm SS'}$ under any extension $\rm{S}'$ we have both that $c_E|\Upsilon_T\rangle$ is proportional to a cPT as per~\cref{def:cpt}, and that $c_{\mathcal{X}}|\Upsilon_T\rangle$ is exactly a cPT.
\end{definition}
\noindent
Imposing that $c_E \cPT$ and $c_{\mathcal{X}} \cPT$ are proportional and equal to cPTs respectively provides instruments with equivalent notions of complete positivity and causality (\cref{def:positivity,def:causality}).
Lastly, the definition also implies that the instrument integrated over all outcomes $\bra{\mathcal{I}(\mathcal{X})}$ is deterministic across all processes.

We have thus defined the continuous instruments in terms of their action on the set $\mathsf{cPT}$.
We now propose their direct characterisation.
In the discrete setting, given an instrument $\hat{\mathbf{A}}_{k:0}^{(\mathbf{x})} \in \mathcal{B}(\mathcal{H}_{\mathfrak{i}_k}\otimes \mathcal{H}_{\mathfrak{o}_{k-1}} \otimes \dots \otimes \mathcal{H}_{\mathfrak{i}_1} \otimes \mathcal{H}_{\mathfrak{o}_0})$, we can find a corresponding process tensor in the following two steps.
First, one must sum over the possible experimental outcomes $\mathbf{x}$ such that the process preserves the trace.
Second, we exchange the roles $\mathcal{H}_{\mathfrak{o}_j} \mapsto \mathcal{H}_{\mathfrak{i}_{j+1}}$ for all $j=0, \dots, k-1$, while mapping $\mathcal{H}_{\mathfrak{i}_j} \mapsto \mathcal{H}_{\mathfrak{o}_{j}}$ for all $j=1, \dots, k$ (since outputs of a process feed into inputs of instruments, and vice versa).
The outcome of this procedure is a process tensor with a system input leg.
In the continuum, the first step becomes to evaluate $\bra{\mathcal{I}(\mathcal{X})}$.
The second step becomes a bit more involved, as in the cPT formulation we are effectively labelling the elements of each $\mathcal{B}(\mathcal{H}_{\mathfrak{i}_{j+1}} \otimes \mathcal{H}_{\mathfrak{o}_{j}})$ with a single index $\nu_j$.
We are therefore required to split and reconnect the indices.
To do this in the discrete case, the $\bm{\nu}, \alpha, \alpha'$ indexed elements of the process tensor corresponding to $\hat{\mathbf{A}}_{k:0}$ are computed by
\begin{equation}
\begin{split}
    (\hat{\Upsilon}(&\mathbf{A})_{k-1:0})_{\alpha', \bm{\nu}, \alpha} =\\
    &\sum_{\bm{\nu'}} (\hat{\mathbf{A}}_{k:0})_{\bm{\nu'}} \bbra{\alpha'}\mathbb{P}_{\nu_{k}'}\mathbb{P}_{\nu_{k-1}}\mathbb{P}_{\nu_{k-1}'}\dots \mathbb{P}_{\nu_1}\mathbb{P}_{\nu_1'}\kket{\alpha},
    \end{split}
\end{equation}
where $\alpha$ is its input leg and $\alpha'$ is its final output leg.
This effect can be achieved in the continuum via the operator
\begin{equation}\label{eq:T operator}
    \mathbb{T}_T \coloneqq \mathcal{T}\e^{\int_0^T \sum_\nu \mathbb{P}_\nu \otimes \psi^\dag_\nu(t)[\cdot] + \sum_{\nu'}\mathbb{P}_{\nu'} \otimes [\cdot] \psi_{\nu'}(t) \mathrm dt}\left[\ket{\Omega}\!\bra{\Omega}\right],
\end{equation}
such that the corresponding cPT indexed by input and output indices $\alpha, \alpha'$ becomes $\bbra{\alpha'}\mathbb{T}\kket{\alpha} \ket{\mathcal{I}(\mathcal{X})}$.
Intuitively, the way $\mathbb{T}$ works is that it places $\mathbb{P}_{\nu'}$ at every intervention in $\ket{\mathcal{I}(\mathcal{X})}$, matrix multiplying it with $\mathbb{P}_\nu$ wherever we place a particle from the left.
The operator $\mathbb{T}$ is a continuous matrix operator (cMPO), as recently introduced in \cite{tjoaContinuousMatrixProduct2025} and summarised in \cref{app:cmpo}, where a more detailed explanation of how $\mathbb{T}$ works can be found.

By analogy with the discrete case, we conjecture that requiring for all quantum states $\rho_{\rm S}$ the Fock vectors $\bbra{\id}\mathbb{T}\kket{\rho_{\rm S}}\ket{\mathcal{I}(E)}$ are completely positive for all $E\subset \mathcal{X}$ and $\bbra{\id}\mathbb{T}\kket{\rho_{\rm S}}\ket{\mathcal{I}(\mathcal{X})}$ is a cPT according to~\cref{def:cpt}, imposes that $\bra{\mathcal{I}(x)}$ is a valid instrument.
We leave the rigorous proof of this statement to future work.

\subsection{Choi duality for cPTs}\label{ssec:choi duality}

The Choi isomorphism, stating the equivalence between quantum channels and quantum states, extends to the multitime setting: any discrete process tensor $\Upsilon_{k:0}$ has an associated Choi matrix $\hat\Upsilon_{k:0}$, which is a non-normalised state on $\mathbb{C}^{d_{\rm S}^{2k}}$, $2k$ qudits with local dimension $d_{\rm S}$.
Mathematically, this relation arises because of the fundamental isomorphism $\mathcal{B}(\mathbb{C}^d) \equiv \mathbb{C}^d \otimes \left(\mathbb{C}^d\right)^*$ between the bounded linear operators on a finite dimensional Hilbert space and the doubled Hilbert space.
But this relation does not generalise to the infinite dimensional setting because $\Gamma^{d_{\rm S}^4 - 1}_{[0,T]}$ does not factorise.
Therefore, cPTs can be thought of as vectorised Choi matrices of the process, but obtaining their operator representation is non-trivial.

However, as the Choi isomorphism has proven exceedingly useful in the finite dimensional setting, in the companion work \cite{fuksaDilationTheoremContinuum2026} we prove a version of the Choi duality suitable to the continuum.
Here, we give an overview of the key elements and refer to the companion work for details.

In~\cref{ssec:born rule} we have shown that the probability associated to measuring an outcome observing an event $E \subseteq \mathcal{X}$ is computed by~\cref{eq:gen_Born_rule}.
The Choi duality introduces a pair of linear maps: $\mathcal{P}$ mapping cPTs to Hilbert-Schmidt operators on $\Gamma^{d_{\rm S}^2 - 1}_{[0, T]}$ and $\mathcal{V}$ mapping instrument effects to an \emph{integration kernel} on $\Gamma^{d_{\rm S}^2 - 1}_{[0,T]}$.
An integration kernel is a (potentially distribution valued) function $A^{n;n'}_{\bm{\mu};\bm{\mu'}}(\bm{t};\bm{t'})$ that defines a sesquilinear form $f_A$ on $\Gamma^{d_{\rm S}^2 - 1}_{[0, T]}$ via
\begin{equation}
\begin{split}
    f_A(\ket{\phi}, \ket{\tilde \phi}) \coloneqq \sum_{n,n'=0}^\infty &\sum_{\bm{\mu}\in[d_{\rm S}^2 - 1]^n}\sum_{\bm{\mu'} \in [d_{\rm S}^2 - 1]^{n'}}\!\!\!\! \int {\rm D}^n \bm{t} \int{\rm D}^{n'} \bm{t'} \\
    &\left(\phi_{\bm{\mu}}^{(n)}(\bm{t})\right)^*
    A^{(n;n')}_{\bm{\mu};\bm{\mu'}}(\bm{t};\bm{t'}) \tilde{\phi}_{\bm{\mu'}}^{(n')}(\bm{t'}).
\end{split}
\end{equation}
The action of the map $\mathcal{P}$ can be understood by the mapping
\begin{equation}\label{eq:choi map}
\begin{split}
    \vac &\mapsto \ket{\tilde{\Omega}}\\
    \psi_{(\mu, 0)}(t) &\mapsto \phi_{\mu}(t) [\cdot]\\
    \psi_{(0, \mu')}(t) &\mapsto [\cdot] \phi_{\mu'}^\dag(t)\\
    \psi_{(\mu, \mu')}(t) &\mapsto 0, \quad \mu,\mu'\neq0,
\end{split}
\end{equation}
ignoring all the ``cross species", while the mapping $\mathcal{V}$ can be intuitively understood as the mapping
\begin{equation}
\begin{split}
    \vac &\mapsto \ket{\tilde{\Omega}}\\
    \psi_{(\mu, 0)}(t) &\mapsto \phi_{\mu}(t) [\cdot]\\
    \psi_{(0, \mu')}(t) &\mapsto [\cdot] \phi_{\mu'}^\dag(t)\\
    \psi_{(\mu, \mu')}(t) &\mapsto \phi_{\mu}(t) [\cdot]\phi_{\mu'}^\dag(t), \quad \mu,\mu'\neq0,
\end{split}
\end{equation}
which retains all the information about the input at the expense of the output not being a bounded operator (or even a densely defined unbounded operator) in general.
Here, we label by $\phi_\mu(t)$ and $\ket{\tilde{\Omega}}$ the field operators and the vacuum in $\Gamma^{d_{\rm S}^2 - 1}_{[0,T]}$, respectively.
These two maps map cPTs and instrument effects to their Choi duals via $\mathcal{P}: \cPT \mapsto \hat\Upsilon_T$ and $\mathcal{V}: \ket{\mathcal{I}(E)} \mapsto \check{\mathcal{I}}(E)$ respectively.
The duality is embodied by the Born rule equivalence
\begin{equation}
    \langle \check{\mathcal{I}}(E), \hat\Upsilon_T \rangle_\text{ker} = \braket{\mathcal{I}(E)|\Upsilon_T},
\end{equation}
where $\langle\cdot,\cdot\rangle_\text{ker}$ is a sesquilinear function on integration kernels, which reduces to the Hilbert-Schmidt inner product $\tr[\check{\mathcal{I}}(E)^\dag \hat{\Upsilon}_T]$ whenever $\check{\mathcal{I}}(E)$ happens to be a Hilbert-Schmidt operator.
It is important to note that this duality does not hold on all $\Gamma^{d_{\rm S}^4 - 1}_{[0,T]}$ and requires $\cPT$ to be a cPT.

\subsubsection*{Kraus operator-sum representation}

Consider the Choi dual $\hat \Upsilon_T$ of a PCR with Hamiltonian $H_{\rm SE}$ and initial state $\rho_0^{\rm SE}$.
Defining first the operator $V_T'\in\mathcal{B}(\mathcal{H}_{\rm SE}\otimes \Gamma_T^{(d_{\rm S}^2-1)})$ as 
\begin{equation}
    V_T' := \mathcal{T}\exp\left(\int_0^T\dd t[H_{\rm SE}(t)\otimes\id + \sum_{\mu=1}^{d_{\rm S}^2-1} P_\mu\otimes \phi_{\mu}^\dagger(t)]\right), 
\end{equation}
applying~\cref{eq:choi map} the Choi dual becomes 
\begin{equation}
    \label{eq:cmpo choi state}
    \hat{\Upsilon}_T\left[H_{\rm SE}(t), \{P_\mu\}, \rho_0^{\rm SE}\right] = \Tr_{\rm aux}[V_T'(|\tilde{\Omega}\rangle\!\langle\tilde{\Omega}|\otimes \rho_0^{\rm SE})V_T'^\dagger].
\end{equation}
The operator $\hat{\Upsilon}_T$ is a cMPO~\cite{tjoaContinuousMatrixProduct2025}. For a more explicit writing of $\hat{\Upsilon}_T$, refer to~\cref{app:cmpo}. 
Decomposing the initial state as $\rho_{\rm SE} = \sum_i \lambda_i \ket{i}\!\bra{i}$, we can write the Choi dual as
\begin{equation}\label{eq:kraus form}
    \hat\Upsilon_T = \sum_{i,j=1}^{d_{\rm S}d_{\rm E}} \lambda_i \ket{u_{i,j}}\!\bra{u_{i,j}},
\end{equation}
where we define the cMPS
\begin{equation}
    \ket{u_{i,j}} = \ket{\Phi_T[H_{\rm SE}, \{P_\mu\}_\mu, \ket{i}\!\bra{j}]}.
\end{equation}
Now, Choi duality contraction becomes
\begin{equation}
    p(E|\mathcal{I}) = \sum_{i,j=1}^{d_{\rm S}d_{\rm E}} \lambda_i\braket{u_{i,j}|\check{\mathcal{I}}_T(E)|u_{i,j}}_\text{ker},
\end{equation}
where $\braket{\cdot|\check{\mathcal{I}}_T(E)|\cdot}_\text{ker}$ is the sesquilinear form defined by the integral kernel $\check{\mathcal{I}}_T(E)$.
This decomposition can be understood as the Kraus representation of the generalised Born rule~\cref{eq:gen_Born_rule}.

\section{Worked examples}\label{sec:examples}
% !Tex root = ./paper.tex

After the derivation of continuous process tensors $\cPT \in  \mathsf{cPT}$ in terms of Fock space vectors and the abstract discussion of its physical properties, we turn in this section to concrete examples. First we consider the cPT associated to the archetypical spin-boson model, and present its explicit form. This showcases that our framework also incorporates environments of infinite dimension. In the second part, we present explicit expressions of instruments for the operational scenarios of coherent evolution of the physical system, continuous measurements and a feedback-control setting.

\subsection{Infinite dimensional environment: Spin-boson model}\label{sec:spin-boson_ex}
The spin-boson model describes a quantum spin system $\rm S$ which is coupled to an environment $\rm E$ comprised of bosonic modes. 
It is extensively studied in the open quantum system literature. See, e.g., Refs.~\cite{le2008entanglement,LeHurrBook2010,strathearn_efficient_2018,Dowling-2024-Tree-PTs,Link_2024} for some background and examples.

Under standard assumptions used -- such as a Gaussian environment, a linear coupling between $\rm S$ and $\rm E$, and no initial correlations -- the discrete process tensors deriving from this model can be explicitly evaluated \cite{Pollock2018-Process-Tensors}.
We will use this in the following to find the particle amplitude of the cPT field theory vector. 
We remark that although the spin-boson cPT can be thought of as a cMPS with infinite bond dimension it is not strictly speaking in process-canonical representation $\ket{\Upsilon_T^{SB}}$ according to \cref{def:process-canonical rep} because of the unbounded Hamiltonian.

For clarity of presentation, we restrict here to a two level system which interacts with an environment of $M$ bosonic modes. 
The environmental modes evolve under the free Hamiltonian $\sum_m \omega_m b^\dagger_m b_m$, where $b_m$ is the annihilation operator of the $m$-th environment mode.
Choosing the rotating frame basis for the environment (interaction picture), any linearly coupled system and environment can be described by the time-dependent Hamiltonian
\begin{align}
    \label{eq:spin_boson_Hamiltonian}
    H_{{\rm SE}}(t) &= \hat s \otimes  \sum_m  g_{m}(b_m e^{-\operatorname{i}\omega_m t} + \text{h.c.}) \nonumber \\
    &=: \sum_{s = 0}^1  \ket{s}\!\bra{s} \otimes B_{s}(t)
\end{align}
where $\{g_m\}$ are real-valued coupling constants and $\{\ket{s}\}_s$ is the eigenbasis of the operator  $\hat s \in  \mathcal B(\mathbb C^2)$.
The corresponding generator in the Liouville space, with $\ket{s}\!\! \bra{s'}\to \kket{s,s'}$, is 
\begin{equation}
    \label{eq:SB_Liouvillian}
    \gen = - \operatorname{i}\sum_{s,r}  \kket{s,r}\!\bbra{s,r} \otimes \Bigl(B_{s}(t)\otimes \id_{\rm E} - \id_{\rm E}\otimes  B^*_{r}(t)\Bigr).
\end{equation}
The initial state $\rho_0^{\rm SE} = \rho_0^{\rm S}\otimes \rho_0^{\rm E}$ at $t_0$ is uncorrelated and $\rho_0^{\rm E}$ is a bosonic Gaussian state. 
It 
has been
shown in Ref.~\cite{Pollock2018-Process-Tensors} that the Choi matrix of the $k$-step process tensor defined on $\bm{t}=(t_0,t_1,\cdots, t_{k})$ whose last output has been traced out, can be explicitly calculated. Its elements in the $\hat s$ eigenbasis, with $\bm{s},\bm{s}',\bm{r},\bm{r}' \in \{0,1\}^{k}$, are
\begin{align}
\label{eq:SB_model_marginals}
(&\tr_{\mathfrak{o}_{k}}[\Upsilon_{k:0}])_{\bm{s},\bm{s}',\bm{r},\bm{r}'}(\bm{t})=\nonumber\\
&\delta_{s_0, s_1'}\delta_{r_0,r_1'}\dotsi\delta_{r_{k-2},r'_{k-1}} \delta_{r_{k-1},s_{k-1}}\bbra{s'_0,r'_0}\rho_0^{\rm S}\rangle\!\rangle\times\\
&\times\exp\bigl[-\tfrac 12\sum_{i\geq j =0}^{k-1} W_{\cramped{s_i,r_i,s_j,r_j}}(t_i,t_{i+1},t_j,t_{j+1}) \bigr] \nonumber
,
\end{align}
where
\begin{multline}
\label{eq:W_SB_model}
 W_{\cramped{s_i,r_i,s_j,r_j}}(t_i,t_{i+1},t_j,t_{j+1}) :=  \\\zeta_{i,j}^{(s_i,s_j)}+ 
   (\zeta_{i,j}^{(r_i,r_j)})^\ast-(\zeta_{i,j}^{(s_i,r_j)})^\ast-\zeta_{i,j}^{(r_i,s_j)}.
\end{multline}
This operator captures the relevant two-point correlations of the bosonic environment which can be seen most clearly in the definition of each $\zeta$. For $i \neq j$ we have
\begin{align}
\label{eq:zeta_SB_1}
    \zeta_{i,j}^{(u,v)}:= \int_{t_{i}}^{t_{i+1}}\mathrm{d}x\int_{t_{j}}^{t_{j+1}}\mathrm{d}y \tr_{\rm E}\left[B_{u}(x)B_{v}(y)\rho_{0}^{\rm E}\right] ,
\end{align}
and for $i =j$
\begin{align}
\label{eq:zeta_SB_2}
    \zeta_{i,i}^{(u,v)}:= \int_{t_{i}}^{t_{i+1}}\mathrm{d}x\int_{t_{i}}^{x}\mathrm{d}y\tr_{\rm E}\left[B_{u}(x)B_{v}(y)\rho_{0}^{\rm E}\right].
\end{align}

In \cref{ssec:discrete marginals} we saw that marginals of a cPT in the Pauli basis are given by $ (\tr_{\mathfrak{o}_{k}}[\Upsilon_{k:0}])_{\bm{\mu},\bm{\mu'}}(\bm{t})= \langle\Omega|\prod_{j=0}^k \xi_{\mu_j,\mu_j'}(t_j)\cPT$. 
Therefore we consider the basis transformation $\kketbra{s_i,r_i}{s'_i,r_i'} \mapsto P_{\mu_i}\otimes P^\ast_{\mu_i'}$ and express \cref{eq:SB_model_marginals} in the Pauli basis labeled by $\bm{\mu},\bm{\mu}'$.
We now want to relate those explicit expressions with the $k$-particle amplitudes of the cPT $ \ket{\Upsilon_T^{SB}}$ in process-canonical representation associated to the model from \cref{eq:spin_boson_Hamiltonian}. For all index values $\bm{\mu},\bm{\mu'}$ such that  $(\mu_i,\mu_i')\neq (0,0)$ for any $i$, this is a straightforward association 
  \begin{equation}
    \langle\Omega|\prod_{j=0}^k \psi_{\mu_j,\mu_j'}(t_j)\ket{\Upsilon_T^{SB}} =  (\tr_{\mathfrak{o}_{k}}[\Upsilon_{k:0}])_{\bm{\mu},\bm{\mu}'}(\bm{t}) .
  \end{equation}
With respect to the $(\mu,\mu')=(0,0)$ identity contributions, we remark that if for some $j$: $(\mu_j,\mu_j')=(0,0)$, then the respective marginal is equal to the $(k-1)^{\rm th}$-step marginal of $ \ket{\Upsilon_T^{SB}}$ with the intervention time $t_j$ projected onto the identity. 
This follows directly from the form of \cref{eq:zeta_SB_1} and \cref{eq:zeta_SB_2}. 
As expected, the identity intervention is really equivalent with doing nothing. 
Consequently, we can write down the cPT in its expanded form
\begin{widetext}
\begin{multline}
        \ket{\Upsilon_T^{SB}}  = 
        \sum_{n=0}^\infty \sum_{\substack{\bm{\mu},\bm{\mu}' \\ (\mu_j,\mu_j')\neq (0,0)}}\tint \sum_{\substack{s_0,\cdots ,s_n \\ r_0,\cdots, r_n}}
          \delta_{r_{n},s_{n}}   \bbra{s_0,r_0}\rho_0^{\rm S}\rangle\!\rangle
\exp^{-\tfrac 12\sum_{i\geq j =0}^{n} W_{\cramped{s_i,r_i,s_j,r_j}}(t_i,t_{i+1},t_j,t_{j+1})}\Big|_{t_0\equiv0} \\ \braket{s_n|P_{\mu_n}|s_{n\!-\!1}}\!\braket{s_{n\!-\!1}|P_{\mu_{n\!-\!1}}|s_{n\!-\!2}}\cdots \braket{r_{n\!-\!2}|P_{\mu'_{n\!-\!1}}|r_{n\!-\!1}}\!\braket{r_{n\!-\!1}|P_{\mu'_n}|r_{n}}\psi_{\mu_n,\mu'_{n}}^\dagger(t_n)\cdots \psi_{\mu_1,\mu'_1}^\dagger(t_1) \ket{\Omega}.
\end{multline}
\end{widetext}

\subsection{Control family}\label{ssec:control_family}

Instruments compute the probabilities of experimental outcomes via the generalised Born rule (see \cref{ssec:born rule}). In what follows, we identify the instruments  corresponding to common operational scenarios, including deterministic coherent control, continuous measurements, and a simple feedback-based control protocol.
We derive their form by examining the overlaps with cPTs in the PCR.
In the companion work~\cite{fuksaDilationTheoremContinuum2026}, using~\cref{thrm:pcr density}, we discuss how this generalises to the action on all cPTs.

\subsubsection{Deterministic control}

As can be seen from the derivation in \cref{ssec:processes as cMPS}, and more concretely from \cref{eq:discrete_coherent_control}, the Fock space vector corresponding to the coherent control of the system under some traceless generator  $\mathbb G(t),\, t \in [0,T]$ is a cMPS $\cmps{0}{\{c_\nu(t)\}}{1}$ with bond-dimension $D =1$, where $c_\nu(t) = \tr[\mathbb G(t) \mathbb P_\nu]^*$. 
In fact, this vector is an unnormalised coherent field theory state. 
Via the overlap formula for two cMPSs in \cref{eq:cMPS_overlap}, one can then directly see that the overlap of a cPT in process-canonical representation with the instrument gives the desired expression 
\begin{multline}
   \langle \Phi[0,\{c_\nu(t)\},1]\cpt{\gen(t)}{\{\mathbb P_\nu\}}{\kket{\rho_0^{\rm SE}}\bbra{\id_{\rm SE}}} = \\  \text{Tr}\,\Bigl[\kket{\rho_0^{\rm SE}}\bbra{\id_{\rm SE}} \,  \mathcal T\mathrm e^{\int_0^T \mathrm dt\,  \gen(t) + \id_{\rm E} \otimes \mathbb G(t) }  \Bigr].
\end{multline} 
Note that due to the causality conditions the expression above evaluates to $1$, as we are coherently controlling the system without measuring it.
Therefore, coherent control is an example of a deterministic instrument: its contraction with any cPT returns unity.
As such, this overlap is therefore not very interesting.
However, note that if one makes the replacement $\bbra{\id_{\rm SE}} \mapsto \bbra{B}$, where $B$ is an observable or a POVM element, effectively redefining the meaning of the cPT, the expression would evaluate the corresponding conditional expectation value or probability at time $T$.

Note that the ``do nothing" operation is a special case of coherent control with the zero Hamiltonian.
We have found in~\cref{sec:framework} that it corresponds to the vacuum state of the field theory
\begin{equation}
    \ket{\mathcal{I} = \id} =  \ket{\Omega} = \cmps{0}{0}{1},
\end{equation}
which is consistent with the derivation above.
We can lift the assumption of tracelessness on the generator by setting $Q = \tr[\mathbb G]^*$ for the drift parameter of the instrument cMPS above such that the instrument cMPS is then 
\begin{equation}
    |\mathcal{I}(\mathbb{G})\rangle = |\Phi[\tr[\mathbb G]^*, \{c_{\nu}\}_{\nu},1] \rangle.
\end{equation}
The overlap formula for two cMPSs in \cref{eq:cMPS_overlap} then exactly gives the desired expression when contracting this instrument with a cPT. 

More generally, to implement \emph{correlated} coherent instruments, operational access to an auxiliary quantum system $\rm A$ with Hilbert space $\mathcal{H}_{\rm A}$ of dimension $d_{\rm A}$ can be given. 
This auxiliary system can be coupled to $\rm S$ via a joint coherent evolution generated by $\mathbb L_{\rm {AS}}$. 
We obtain the corresponding Fock space vector from the decomposition
\begin{equation}
 \mathbb L_{\rm{AS}}= \sum_{\nu}\underbrace{\tr_{\rm S} [\mathbb L_{\rm{AS}}\mathbb P_\nu]}_{c_{\nu}^*} \otimes \mathbb P_\nu, 
\end{equation}
which we can interpret as an expansion of the system generator in the Pauli basis with the coefficients $c_{\nu}\coloneqq \tr_{\rm S} [\mathbb{L}_{\rm AS} \mathbb{P}_\nu]^*$ now being matrices of dimension $d_{\rm A}^2 \times d_{\rm A}^2$. 
Consequently, the instrument for deterministic control with quantum memory is a cMPS of bond dimension $d_{\rm A}^2$ 
\begin{equation}
 \ket{\mathcal{I}[\mathbb (\mathbb{L}_{\rm A })^*,\{c_{\nu}\}_{\nu},\kketbra{(\rho_0^{\rm A})^*}{\id_{\rm A}}]}.  
    \label{eq:inst_coherent_control_A}
\end{equation}
A potential ambient time evolution of the auxiliary system is captured by $\mathbb L_{\rm A }$ and the auxiliary system is initialised in the vectorised state $\kket{\rho_0^{\rm A}}$. 
The contraction of the instrument with a continuous process tensor then gives, via the cMPS overlap formula in \cref{eq:cMPS_overlap}, the desired expression 
\begin{multline}
   \langle \mathcal{I}[(\mathbb L_{\rm A })^*,\{c_{\nu}\}_{\nu},\kketbra{(\rho_0^{\rm A})^*}{\id_{\rm A}}] \cPT = \\ \bbra{ \id_{\rm{ASE}}}  \, \mathcal T \e^{\int_0^T \rm d t \mathbb L_{\rm{SE}} + \mathbb L_{\rm{AS}}+ \mathbb L_{\rm{A}} }\kket{\rho_0^{\rm{SE}}}\kket{\rho_0^{\rm A}}
\end{multline}
for the action of the instrument on the cPT.

\subsubsection{Continuous measurements: jump and diffusive cases}

In this section, we present the effects of instruments corresponding to continuous measurements, sometimes referred to as weak quantum measurements. 
We refer the reader to the literature for an introduction to this well-established topic \cite{wisemanQuantumMeasurementControl2014, guilminParametersEstimationFitting2024a,rosal2025deterministicequationsfeedbackcontrol,gardiner2004quantum}.
The following brief exposition serves to establish notation and make our expressions for the respective instruments comprehensible. 
It follows the excellent exposition in
Ref.~\cite{guilminParametersEstimationFitting2024a}.

The evolution of a quantum state $\rho$ subject to continuous measurement of $k$ linear operators $\{L_j\}_{j \in [k]}$ on $\mathcal{H}_{\rm S}$ is described by a linear stochastic master equation  
\begin{multline}
\mathrm{d} \bm{\rho} = \sum_{j \in [k]} \bigl(L_j\bm{\rho} L_j^\dagger - \tfrac{1}{2} (L^\dagger_jL_j\bm{\rho} + \bm{\rho} L^\dagger_jL_j)\bigr) \mathrm{d} t \\ + \mathcal{M}(\bm{\rho}, L_j)\mathrm{d} \bm{S}_j
\label{eq:cont_measurement_SME} 
\end{multline}
with stochastic increments $\mathrm{d} \bm{S}_j$ corresponding to stochastic processes and $\mathcal{M}$ being a superoperator acting on the stochastic density matrix $\bm{\rho}$ (not necessarily linear) that depends on the measured operator $L_j$.
The stochastic process $\mathrm{d} \bm{S}_j$ represents the measurement signal of the operator $L_j$. 
For simplicity, we omit from \cref{eq:cont_measurement_SME} any coherent evolution of the system, take the operators $\{L_j\}$ to be time-independent, and assume perfect efficiency of the continuous measurement. 
Our derivation can be easily adapted to remove these assumptions.

There are two regimes of continuous measurements that are commonly studied, distinguished by their possible measurement outcomes as determined by the specific form of the stochastic increments $\mathrm{d} \bm{S}_j$. 
In the ``quantum jump" regime, the measurement outcomes form a finite and discrete set. 
The second regime is the diffusive measurement regime, where the possible outcomes form a continuous set. 
In the following, we specify these two regimes and give the form of the respective instruments on a high level. A more detailed derivation is given in \cref{app:continuous_instruments}.

\textbf{Quantum jump measurement.}
The quantum jump measurement corresponds \cref{eq:cont_measurement_SME} to $\mathcal{M}(\bm{\rho}, L_j) = L\rho L^\dagger - \rho$ and the stochastic increments $\mathrm{d} \bm{S}_j = \mathrm{d} \bm{N}_j -\mathrm d t$, containing Poissonian point processes with rate 1 which satisfy
\begin{equation}
\label{eq:jump_meas_increment}
\begin{split}
    \mathrm{d} \bm{N}_j &= (\mathrm{d} \bm{N}_j)^2, \\
    \mathbb{E}[\mathrm{d} \bm{N}_j] &= \mathrm{d} t,
    \end{split}
\end{equation}
where $\mathrm{d} t$ is the infinitesimal time increment.
This means that most of the time we obtain the outcome ``0", whereas with infinitesimal probability we record a ``click" corresponding to one of the outcomes $j \in [k]$.
Physically, such a POVM can be realised by coupling the system to a quantum field serving as a meter and performing a projective measurement on the meter.
The relationship between weak measurements and the cMPS formalism has been studied previously \cite{osborneHolographicQuantumStates2010}.
We use the formulation from this reference as the starting point to derive the field theory formulation for the instrument of a jump measurement.
We find (see~\cref{app:continuous_instruments}) that the probability density of observing one jump $x \in [k]$ at time $t \in [0,T]$ is
\begin{equation}
    p(x, t) =  \sum_{\nu = 0}^{d_S^4 - 1} \chi^x_{\nu} \bra{\Phi[c_0, \, \{c_{\nu'}\}_{\nu'\in[d_{\rm S}^4-1]}, \, 1]} \xi_{\nu}(t) \cPT,
\end{equation}
where $c_{\nu}$ and $\chi^x_{\nu}$ are given by
\begin{equation}
\begin{split}
    c_\nu &\coloneqq - \tfrac{1}{2} \tr\Bigl[\mathbb{P}_\nu\bigl(\sum_{j\in[k]} L_j^\dag L_j \otimes \id + \id \otimes L_j^T L_j^*\bigr)\Bigr]\\
    \chi^x_\nu &\coloneqq \tr\left[L_x \otimes L_x^* \mathbb{P}_\nu\right].
    \end{split}
    \label{eq:coefficients_jump_instrument}
\end{equation}
This allows us to directly deduce that the effects can be computed by summing and integrating the object 
\begin{multline}
    \ket{\mathcal{I}(\bm{x},\bm{t})} =\\  \sum_{\nu_1,\cdots, \nu_n = 0}^{d_S^4-1} \chi^{x_1}_{\nu_1} \dots \chi^{x_n}_{\nu_n} \prod_{j\in[n]} \xi^\dagger_{\nu_j}(t_j)\cmps{c_0}{\{c_{\nu'}\}_{\nu'}}{1}.
\label{eq:mulitple_jump_instrument}
\end{multline}
over (subsets of) $n \in \mathbb{N}_0^+, \bm{x}\in[k]^n$ and (subsets of) time-ordered $\bm{t} \in [0, T]^n$.
The object~\cref{eq:mulitple_jump_instrument} is no longer a cMPS, but rather a superposition of particle-added cMPSs and it has distribution valued wavefunctions.
It hence does not lie in the Fock space.
However, the effects \emph{are} actually in the Fock space (see the companion work~\cite{fuksaDilationTheoremContinuum2026} for details).
For instance, the effect associated to obtaining two clicks in the time interval $[0,T]$ given by
\begin{equation}
    \bra{\mathcal{I}(\text{two clicks})} =
               \sum_{x, x' \in [k]} \int_0^T \mathrm{d} t \int_0^t \mathrm{d} t' \langle \mathcal{I}\left((x, x'), (t, t')\right) |
\end{equation}
is a valid element of the (dual) Fock space, such that the associated probability can be computed as
\begin{equation}
    \mathrm{Pr}[\,\text{two clicks}\, ] = \langle \mathcal{I}(\text{two clicks}) \cPT.
\end{equation}

\textbf{Diffusive measurement.}
A diffusive measurement is characterised by stochastic Wiener (Itô-) increments $\mathrm{d} \bm{S}_j(t) = \mathrm{d} \bm{Z}_j(t)$ \cite{wisemanCompleteParameterizationInvariance2001}. The corresponding linear stochastic master equation has $\mathcal{M}(\bm{\rho}, L_j) = \sum_j L_j\rho + \rho L^\dagger_j$. We restrict here for simplicity to the case of complex, independent Wiener processes which satisfy 
\begin{equation}
\begin{split}
    \mathbb{E}[\mathrm{d} \bm{Z}_j] &= 0 ,\\
     \mathrm{d} \bm{Z}_j \mathrm{d} \bm{Z}_{j'} &= \mathrm{d}t \, \delta_{j,j'}.
     \end{split}
\end{equation}
This corresponds to the physical scenario where the system is coupled to $k$ quantum fields which are then subject to heterodyne detection.  
The outcome of such a diffusive measurement are $k$ continuous functions $Y_j:[0,T] \to \mathbb{C}$. Due to the inherent randomness in the measurement they are stochastic quantities. Their increments in the infinitesimal time interval $[t, t+\mathrm{d}t)$ satisfy
\begin{equation}
   \mathrm{d} \bm{Y}_j  = \tr[L_j\bm{\rho}(t)]\mathrm{d}t  + \mathrm{d} \bm{Z}_j.
\label{eq:diff_currents}
\end{equation} 
We associate to the Wiener processes their derivatives in a distributional sense $\bm J_i= \mathrm d \bm Z_j/\mathrm d t$. In the case of $k$ diffusively measured operators $\{L_j\}$, we collectively denote them by $\vec{J}$ \cite{wisemanCompleteParameterizationInvariance2001}.
In \cref{app:continuous_instruments} we show that the instrument effects corresponding to this diffusive measurement are integrals with respect to the complex Wiener measures of 
\begin{align}
\label{eq:diff_measu_state}
\ket{\mathcal{I}(\vec{J})} = \cmps{c_0 + d_0+\chi_0^{\vec{J}}}{\{c_{\nu} +d_\nu+ \chi^{\vec{J}}_{\nu}\}_{\nu}}{1},
\end{align}
where $c_\nu$ is as in \cref{eq:coefficients_jump_instrument} and
\begin{equation}
    \begin{split}
        \chi^{\vec{J}}_\nu(t) &\coloneqq \!\! \sum_{j\in[k]}\!\!\tr[\mathbb{P}_\nu  \bigl(J_j(t) L_j(t) \otimes \id + \id \otimes J_j(t)L^\ast_{j}(t)\bigr)]^*, \\
        d_\nu &\coloneqq \sum_{j\in[k]} \tr[\mathbb P\nu (L_j\otimes L_j^\ast)]^\ast .
    \end{split}
\end{equation}
The object~\cref{eq:diff_measu_state} is a non-normalisable coherent field theory state with white noise functionals as wave functions.
The stochastic integrals in its formal expansion are understood in an Itô sense.
For details, see~\cref{app:continuous_instruments}.

\subsubsection{Feedback-based control}\label{ssec:control}
With a formulation of continuous measurement in the cPT framework in place, we will now discuss the form of instruments corresponding to quantum control in two settings of interest.
As the first setting, suppose we measure the POVM $\{M^x\}_x$ at time $t \in [0, T]$ and then at time $t'>t$ we apply a unitary control operation $U^x$ according to the outcome $x$.
We are interested in the $x$ conditioned state $\rho^{(x)}$ of the system at time $T$ and the probability of obtaining the outcome $x$.
As for PCRs this setting can be fully described in terms of the discrete process tensor $\braket{\Omega | \xi_{\mu'}(t')\xi_\mu(t)\xi_{\mu''}(T) | \Upsilon_T }$ (see \cref{ssec:discrete marginals}), all the results follow from the discrete theory.
Due to the causality condition in \cref{def:causality}, the probability of obtaining the outcome $x$ will be given by 
\begin{equation}
    p(x) = \sum_{\mu'} M^x_\mu \braket{\Omega | \xi_{\mu'}(t') | \Upsilon_T},
\end{equation}
where $M^x_\mu = \tr\left[\kket{\id_{\rm S}}\!\bbra{M^x} \mathbb{P}_\mu\right]$ and the corresponding state will have Pauli components
\begin{equation}
    \begin{split}
        \rho^{(x)}_\alpha &= \tr\left[P_\alpha \rho_x \right]\\
                  &= \frac1{p(x)}\sum_{\mu, \mu', \mu''} c_{0,\alpha}^{\mu''} U^x_{\mu'} M^x_{\mu} \braket{\Omega | \xi_{\mu''}(T)\xi_{\mu'}(t')\xi_{\mu}(t) | \Upsilon_T },
    \end{split}
\end{equation}
where $c^\mu_{\alpha, \alpha'} \coloneqq \tr\left[\mathbb{P}_\mu \kket{\alpha}\!\bbra{\alpha'}\right]$ and $U^x_\mu \coloneqq \tr\left[\mathbb{P}_\mu U^x\right]$.
Note that the associated effect is distribution valued and hence does not belong to our theory.
    This can be mitigated for well behaved cPTs by the means of a Gelfand triple construction, as we elaborate in the companion paper \cite{fuksaDilationTheoremContinuum2026}.

We will now, as the second setting, use the intuition obtained from the discrete time example in the following, inherently continuous-time scenario.
Suppose we are performing a quantum jump measurement on the system and any time we obtain the outcome $x$, we apply the control operation $U^x$ a time $\epsilon$ later.
For simplicity, in this case we are only interested in the probabilities of obtaining the outcome $\bm{x}$.
The generalisation to obtaining the density matrix at time $T$ given the outcome $\bm{x}$ follows just like in the previous case.
We obtain
\begin{equation}
p(\bm{x}) = \sum_{\bm{\mu}} \prod_{i=1}^{|\bm{x}|} U^{x_i}_{\mu_i} \bra{\mathcal{I}(\bm{x},\bm{t})} \prod_{j=1}^{|\bm{x}|} \xi_{\mu_j}(t_j+\epsilon) \cPT,
\end{equation}
where $\bra{\mathcal{I}(\bm{x}, \bm{t})}$ is defined in \cref{eq:mulitple_jump_instrument}.

\section{Results on non-Markovianity in the continuum}\label{sec:NM}
With our framework 
on continuous process tensors now established, 
we would like to be able to apply it to determine insights on the operational effects of a given system-environment interaction. One 
of the great advantages of the process tensor viewpoint is that direct knowledge of the environment is not necessary to say anything about the system: one can adopt an instrumental viewpoint and ask the question ``to what extent does the presence of 
a quantum environment generate temporal correlations in any \emph{measurable} sense?''~\cite{Pollock_2018_measure,whiteWhatCanUnitary2025}. Measurable correlations mean that the stochastic process is therefore non-Markovian. Intuitively, this means that there exist at least two instruments which, when applied across some fixed window in the past, lead to conditionally different dynamics on the system in the future. Ideally, one would like to quantify exactly how complex such temporal correlations can become for a given non-Markovian system. In this section, we thus turn to both qualitatively and quantitatively studying non-Markovianity using our continuous process tensor framework.

As we shall see, for a variety of reasons this question in the continuum limit is fraught. But we present two sets of results. In the first subsection, we provide background to the problem and prove that typical measures of non-Markovianity 
using discrete process tensors identically vanish when the continuum limit is taken. We then proceed to motivate and derive entropic measures which quantify non-Markovianity for continuous processes. In the second subsection, we turn more to the question of -- not how \emph{large} the non-Markovianity of a given dynamics is -- but how \emph{complex} it can be~\cite{aloisio-complexity}. Specifically, we look at processes from the lens: of up to which 
$n$ does one need to measure joint 
$n$-point multi-time correlation functions in 
order to have all $k$-point correlation 
functions 
for $k$ generically larger than $n$. 

\subsection{Quantifying non-Markovianity}\label{ssec:nm measures}
In the quantum setting, the question of how to appropriately measure the level of non-Markovianity is a rich topic which has historically been addressed from many different perspectives~\cite{MarkovCirac,PhysRevA.83.052128,Pollock_2018_measure,rivas2014quantum,PhysRevLett.103.210401,ObservationNonMarkovian}. At a high level, most measures fall into one of two schools: the breakdown of CP-divisibility of the dynamical maps governing a system dynamics; and the presence of multi-time correlations which are detectable via interventions. The former, it has been more recently elucidated~\cite{Milz_2019_div} is actually a measure of the non-Lindbladness of dynamics, and a value of zero does \emph{not} imply Markovianity, since higher-point temporal correlations may exist. This distinction can also be elucidated with explicit physical models~\cite{Arenz_2015_decoupling}. 
The latter category is defined by examining correlations in the Choi state of a discrete process tensor. \par Although these latter measures are arguably more well-posed, it is the former ones which are well-defined in the continuum. To our knowledge, a measure which incorporates the strengths of both perspectives does not exist in the literature.
In this section, we propose a family of non-Markovianity measures which are valid for continuous-time dynamics. 
We will briefly review these discrete process tensor measures before showing that they vanish in the continuum limit. Then, employing our cPT framework, we derive entropic measures of the environment's past-future influence at arbitrary time which come equipped with operational meaning. 

\subsubsection{Background of Choi-based non-Markovian measures}
In the discrete time setting, process tensors define a strict notion of Markovianity that agrees in the classical limit with that of classical stochastic processes. 
This is because process tensors allow one to incorporate \textit{causal breaks} at the level of the system and infer therefore whether any information has persisted via the environment. I.e., one can look for inequalities 
that can discern whether the probabilities of observing an outcome $y$ in the future $(\mathfrak{o}_{k} \cdots \mathfrak{i}_{j+1})$ process differ, given that we observed two distinct outcomes $x$ and $x'$ in the past at $(\mathfrak{o}_{j} \cdots \mathfrak{o}_{0})$.
Importantly, we must break up the instrument at $j$ into a past and a future component -- across $\mathfrak{o}_j$ and $\mathfrak{i}_{j+1}$ -- this is the aforementioned causal break. It ensures that the future depends on the past if and only if the process is non-Markovian. This is because, when the instrument at $j$ is broken no information passes from the past to the future via the system. 
\par

It is possible to guarantee that there exist no instruments for which an environment-mediated past-future dependence can be found. In particular, a discrete quantum stochastic process is said to be \emph{Markovian} if, and only if, its Choi state can be written exactly as a product of the Choi states of channels at each time step~\cite{Pollock2018-Process-Tensors}. That is, 
\begin{equation}
    \label{eq:dpt-mark}
     \mPT := \bigotimes_{j=1}^k \hat{\mathcal{E}}_{j:j-1}\otimes \rho_0.
\end{equation}
Denoting the (non-convex) space of all $k$-step Markovian processes by $\mathcal{M}_k$, measures of non-Markovianity typically correspond to quasi-distances from this set. In particular, a popular measure of non-Markovianity, introduced in Ref.~\cite{Pollock_2018_measure}, is that given any CP-contractive divergence $\mathcal{D}(\rho\mid\!\mid\sigma)$, the measure of non-Markovianity induced by $\mathcal{D}$ is
\begin{equation}
    \label{eq:dnm-meas}
    \mathcal{N}_{\mathcal{D}}(\PTchoi) = \min_{\mPT\in\mathcal{M}_k}\mathcal{D}(\PTchoi \mid\!\mid \mPT).
\end{equation}
Typically, $\mathcal{D}$ is chosen to be the quantum relative entropy $S(\rho\mid\!\mid\sigma)\coloneqq \Tr[\rho\log\rho - \rho\log\sigma]$, for which the unique minimum in Eq.~\eqref{eq:dnm-meas} is conveniently given by the tensor product of the marginals of~$\PTchoi$, i.e.,
\begin{equation}
    \label{eq:re-dnm}
    \mathcal{N}_{S}(\PTchoi) = S\left(\PTchoi \big|\!\big| \bigotimes_{j=1}^k\hat{\mathcal{E}}_{j:j-1}\otimes \rho_0\right).
\end{equation}
This corresponds to the generalised quantum mutual information of the Choi state, and incorporates both classical and quantum temporal correlations. Operationally, this can be understood as an asymptotic limit of the success of hypothesis testing between a process tensor's Choi state and the best Markov model, employing measurements on those Choi states. 
For a single cut, it can be related to the entanglement entropy across a bond of the (normalised) vectorised Choi state. In this context, the non-Markovianity can be interpreted as the cost of simulating the environment, sometimes referred to as ``temporal entanglement''~\cite{Abanin_2021-IF}.

\subsubsection{Zeno-type vanishing of temporal correlations}
More recently, it has in fact been pointed out that there are some conceptual issues with~\cref{eq:dpt-mark}. In particular, the fact that $\mathcal{D}$ is a \emph{state} divergence means that it satisfies the data processing inequality with respect to channels acting on the Choi state. However, the set of operations which one can perform on processes is larger than that of states.
Indeed, it has been shown by way of example that $\mathcal{D}(\Upsilon_1\mid\!\mid\Upsilon_2)$ can actually increase with the application of a superprocess to each comb. 
This leads one to the conclusion that since a Choi divergence can increase under superprocesses, then the hypothesis testing problem between $\PT$ and the closest Markovian process could be strictly easier. Therefore, Eq.~\eqref{eq:dnm-meas} should only be seen as a lower bound of non-Markovianity. 

Here, we will show that this particular issue is what obfuscates a naive continuum limit of the above measure. One natural starting point for a continuum measure would be the idea that: given a process tensor $\PT$ constructed from some dynamics on an interval, compute the quantity
\begin{equation}
\label{eq:dnm-inf}
    \lim_{k\left(=\frac{T}{\shortT}\right)\to\infty}\mathcal{N}_{S}(\PT).
\end{equation}
However, as was already observed numerically in Ref.~\cite{sonnerInfluenceFunctionalManybody2021}, this quantity vanishes in the continuum limit. Here we shall prove analytically the following result.
\begin{proposition}[Vanishing of discrete NM measures]
    Given any family of discrete process tensors $\{\PT\}_k$ defined on a fixed interval $[0,T]$ with respect to an underlying system-environment Hamiltonian $H_{\rm SE}(t)$ with bounded operator norm $\|H_{\rm SE}(t)\|_\infty < \infty$ for all $t\in [0,T]$, the value of the limiting measure given in~\cref{eq:dnm-inf} is identically zero.
    \label{prop:vanishing_NM_measu}
\end{proposition}

First, consider the iterated construction of the Choi state from~\cref{eq:pt-choi-def}, where we instead explicitly keep the environment bond open. That is, introduce $\hat{\Upsilon}_{k:0}^{\rm E}$ where
\begin{equation}
    \hat{\Upsilon}_{k:0}^{\rm E} = \mathcal{U}_{k}[\Phi_{k-1}^+] \circ \cdots \circ \mathcal{U}_{1}[\Phi^+_{0}][\rho_0^{\rm SE}].
\end{equation}
It follows that $\PTchoi=\Tr_{\rm E}[\hat{\Upsilon}_{k:0}^{\rm E}]$.
We will take the initial state to be pure for simplicity, this does not change our argument. Now, let each $\mathcal{U}_{j}\equiv u_j(\cdot)u_j^\dagger \equiv\mathcal{U}_j(\shortT)$, where $u_j=\mathcal{T}\exp\left[-i\int_{t_{j-1}}^{t_{j-1}+\shortT}H_{\rm SE}(t)\dd t\right]$. Then each $\PTchoi^E$ is related to the last by $\hat{\Upsilon}_{k:0}^{\rm E} = \mathcal{U}_k [\Phi_{k-1}^+] \circ \hat{\Upsilon}_{k-1:0}^{\rm E}$. Importantly, each $\hat{\Upsilon}_{k:0}^{\rm E}$ is a pure state. Using the so-called chain-rule decomposition of the quantum mutual information~\cite{khatri2024principlesquantumcommunicationtheory}, we can re-write~\cref{eq:re-dnm} as
\begin{equation}
    \mathcal{N}_{S}(\PTchoi) = \sum_{j=2}^k S(\hat{\Upsilon}_{j:0}\mid\!\mid \hat{\mathcal{E}}_j\otimes\hat{\Upsilon}_{j-1:0}),
\end{equation}
which is the iterated sum of each \emph{new} step's mutual information with the rest of the process. It follows by the data-processing inequality then that 
\begin{equation}
    \mathcal{N}_{S}(\PTchoi) \leq \sum_{j=2}^k S(\hat{\Upsilon}^{\rm E}_{j:0}\mid\!\mid \hat{\mathcal{E}}_j\otimes\hat{\Upsilon}_{j-1:0}^{\rm E}).
\end{equation}
We will use this form to show that each of these terms goes to zero faster than linearly in $\shortT$, and so their sum also goes to zero. 
Let us now examine each term in the summand. Since $\hat{\Upsilon}_{j:0}^{\rm E}$ is pure, then the mutual information $S(\hat{\Upsilon}^{\rm E}_{j:0}\mid\!\mid \hat{\mathcal{E}}_j\otimes\hat{\Upsilon}_{j-1:0}^{\rm E}) = 2S(\hat{\mathcal{E}}_j)$, where the latter term is the von Neumann entropy of $\hat{\mathcal{E}}_j$: $S(\rho) \coloneqq-\Tr[\rho\log\rho]$. 
Recall that $\hat{\mathcal{E}}_j$ is constructed by taking the Bell state $\Phi^+$, which represents the input and output spaces $\mathfrak{i}_j$ and $\mathfrak{o}_j$ and applying $\mathcal{U}_j(\shortT)$ across the latter half and the environment:
\begin{equation}
\label{eq:nm-growth}
    \hat{\mathcal{E}}_j = \Tr_{\mathrm{E}, j-1:0}\left[\mathcal{U}_{j}^{\mathfrak{o}_j {\rm E}}(\shortT)[\Phi^+\otimes \hat{\Upsilon}^{\rm E}_{j-1:0}]\right].
\end{equation}
Clearly, we have $S(\hat{\mathcal{E}}_j) \overset{\shortT\to 0}{\longrightarrow} 0$.
As the next step we will show that the entropy in fact vanishes faster than linearly.
To this end, we will use the Fannes-Audanert inequality \cite{audenaertSharpContinuityEstimate2007}, which upper bounds the von-Neumann entropy of a $d$-dimensional quantum state $\rho$ by its trace distance $T(\rho, \sigma) \coloneqq \frac12 \|\rho - \sigma\|_1$ to a pure state $\sigma$ via
\begin{equation}\label{eq:fannes-audanert}
    S(\rho) \le T \log(d-1) - T \log T - (1-T)\log(1-T).
\end{equation}
The base of the logarithm has to be the same as the one used in the von-Neumann entropy $S(\rho)$. 
Consider now setting $\rho(t) = \Tr_{\mathrm{E}, j-1:0}\left[\mathcal{U}_{j}^{\mathfrak{o}_j {\rm E}}(t)[\Phi^+\otimes \hat{\Upsilon}^{\rm E}_{j-1:0}]\right]
$ and $\sigma(t) = \widetilde{\mathcal{U}}_j(t) [\rho(0)]$, where $\widetilde{\mathcal{U}}_j(t) \coloneqq \tilde{u}_j(t) (\cdot) \tilde{u}_j^\dag(t)$ with $\tilde{u}_j(t) \coloneqq \mathcal{T}\exp\left[-i \int_{t_{j-1}}^{t_{j-1}+t}  \Tr_{{\rm E}, j-1:0}[H_{\rm SE} \hat{\Upsilon}_{j-1:0}^{\rm E}]\right]$.
By expanding $\mathcal{U}_j(t)$ and $\widetilde{\mathcal{U}}_j(t)$ in their power series in $t$ one finds, employing the condition $\|H_{\rm SE}(t)\|_\infty < \infty$, that
\begin{equation}
    T(\rho(t), \sigma(t)) = \mathcal{O}(t^2\|H_{\rm SE}\|_\infty^2).
\end{equation}
Now we can compute the limit of each term in~\cref{eq:fannes-audanert}, and obtain
\begin{equation}
    \tfrac{S(\rho(t))}{t} \overset{t \to 0}{\longrightarrow} 0.
\end{equation}
Consequently, for any $\eta > 0$ there exists a time $t_\eta$ such that for all $j$ and all $t \leq t_\eta$ it holds that $\tfrac{2 S(\hat{\mathcal{E}}_j)}{\shortT} < \eta $. 
Then we can write for any $\shortT < t_\eta$
\begin{equation}
 \mathcal{N}_S(\PTchoi) \leq    \sum_{j=2}^{k}\shortT \frac{2S(\hat{\mathcal{E}}_j)}{\shortT} < (k-1) \shortT \eta = T \eta.
\end{equation}
where in the last step we set $(k-1) \shortT \equiv T$. Since the above has to hold for any $\eta >0$, and because $0 <\mathcal{N}_S(\PTchoi)$, we conclude that in the limit $k \to \infty$, $\shortT \to 0$, where $T = (k-1) \shortT$ is kept constant, we get the claimed result
\begin{equation}   \lim_{\substack{\shortT\to0\\k\to\infty}}\mathcal{N}_S(\PTchoi^{\shortT}) = 0.
\end{equation}

\par 

Conceptually, this is troubling, and highlights another angle from which one must be careful in working with discrete process tensors. Allowing for the possibility of more interventions should \emph{increase} the discrimination power of an experimenter, not the contrary. The issue here is that when treated purely as a state, the creation of Bell pairs acts purely as a causal break, which in turns leads to a Zeno-type freezing of the system-environment dynamics in the large step number limit. In particular, this example illustrates the absence of the coarse-graining superprocess, in which an experimenter may choose to perform the do-nothing operation and recover a fewer-step process marginal over the same time window. While the appropriate distinguishability measure for two \emph{fixed} combs is resolved in the consideration of generalised comb divergences, see \cref{ssec:distance} for a discussion of them in the continuum limit, to our knowledge there is no known efficiently-computable measure of non-Markovianity.

\subsubsection{Deriving robust continuum measures of quantum non-Markovianity}
Although we do not aim to resolve this question of Choi divergence versus comb divergence in its full generality in the present work, let us now turn to the specific problem of properly defining and measuring quantum non-Markovianity in the continuum. In this discussion we restrict ourselves to the PCR such that we have readily-accessible singular values of the process in addition to a clear system-environment split.
In analogy to MPS in the discrete case, one might initially think that this corresponds to quantifying the `bond' entropy $\sigma(t)$ in the cMPS, which we study in \cref{ssec:compressibility}. However, because the continuum limit forces together all excitations at a single site, the relevant entropic contribution is from $t$ to $t+\text{d}t$, rather than \emph{at} $t$. 
In the discrete picture, this amounts to performing an SVD at the $\mathfrak{i}_j\mid\mathfrak{o}_j$ cut instead of $\mathfrak{o}_j\mid\mathfrak{i}_{j+1}$. 
Thus, even though the entropy of the cMPS singular values can be rigorously understood in the context of simulation cost for some non-Markovian process, this measure can only supply the information propagation by $\rm S$ and $\rm E$ \emph{simultaneously}, making it a poor measure. \par 

Instead of making a cut at an instantaneous moment and asking about the remaining past-future correlations, we condition on it.
We can think of the cMPS as generating a record field (the process tensor legs) plus an internal bond (memory). In the weak limit, the record is mostly vacuum and occasionally has excitations. Those excitations are the pieces that carry distinguishable ``outcome" information, so they are natural carriers of causal correlations. Our strategy will be to compute the quantum mutual information -- at a fixed point $t$ -- between the past $\mathsf{P}_t:=[0,t)$ and the future $\mathsf{F}_t:=(t,T]$ field theory state, \emph{conditioned} on the system $\mathsf{S}_t$ at present. Importantly, the ``system at present'' includes both the jump operators on the field theory state \emph{and} the instantaneous input and output of the system on the virtual space. 
This defines the past-future dependence at a given point in time. We will then use this quantity to define an integrated measure of non-Markovianity over the whole interval. \par

Formally, given a cPT $\cPT$ in a cMPS form, let us define the splitting $\mathcal{H}_{\mathsf{P}_t}\otimes\mathcal{H}_{\mathsf{B}_t}\otimes \mathcal{H}_{\mathsf{F}_t}$, where $\mathcal{H}_{\mathsf{B}_t}$ is the virtual space at time $t$, and $\mathcal{H}_{\mathsf{P}_t}$ and $\mathcal{H}_{\mathsf{F}_t}$ are the Fock spaces of the field theory state defined on the intervals mentioned in the previous paragraph. Note that from here on we will let the $t$-dependence be implicit. This is the continuum equivalent to opening up a bond in MPS to define a vector on the physical and virtual spaces. 
Consider the decomposition of $\mathcal{H}_{\mathsf{B}}$ into the system-environment split $\mathcal{H}_{\mathsf{S}}\otimes\mathcal{H}_{\mathsf{E}}$, which will be necessary to perform the appropriate conditioning.
There is a unitary $\hat{\mathcal{U}}(t_1,t_2)$ whose action 
\begin{equation}
    \label{eq:pt-bond-open}
    \hat{\mathcal{U}}(t,0)\left(\kket{\rho_0^{\rm SE}}\ket{\Omega}\right)\left(\bbra{\id_{\rm SE}}\bra{\Omega}\right)\hat{\mathcal{U}}^\dagger (t,T)
\end{equation}
creates the unnormalised field-virtual state $\rho_{\mathsf{P}\mathsf{S}\mathsf{E}\mathsf{F}}$ on $\mathsf{PSEF}$ at time $t$. Roughly speaking, we must construct the state $\rho_{\mathsf{P}\mathsf{S}\mathsf{E}\mathsf{F}}$ and compute the conditional quantum mutual information (CMI) which quantifies non-Markovianity in the continuum limit:
\begin{equation}
    \label{eq:cnm-meas}
    \begin{split}
        \mathcal{N}_{\Upsilon_T}(t) := I(\mathsf{P}_t:\mathsf{F}_t\mid \mathsf{S}_t),
    \end{split}
\end{equation}
where the CMI is $I(A:C\mid B) = S(AB) + S(BC) - S(ABC) - S(B)$. However, it is infeasible to compute Eq.~\eqref{eq:pt-bond-open} directly. 
Instead, we can operate with the $\cPT$ in its \emph{central canonical form}, with drift and jump parameters denoted by $Q^{(c)}(t)$ and $\{R^{(c)}_\nu(t)\}$. This gauge is defined in analogue to the mixed canonical gauge in discrete MPS theory~\cite{RevModPhys.93.045003}, and we refer readers to~\cref{ssec:compressibility} and~\cref{app:gauges} for explicit definitions and details. The important feature of this gauge is that there is a readily computable $D\times D$ matrix whose singular values are exactly the Schmidt coefficients of $\cPT$ with respect to a cut made at time $t$. In this gauge, one can compute the instantaneous Schmidt decomposition of $\cPT$:
\begin{equation}
    \cPT = \sum_i \sigma_i(t)\ket{\Psi_i}_{\mathsf{P}}\otimes\ket{\Psi_i'}_{\mathsf{F}},
\end{equation}
where $\{\ket{\Psi_i}_{\mathsf{P}}\}_i$ and $\{\ket{\Psi_i'}_{\mathsf{F}}\}_i$ are the respective past and future orthonormal Schmidt vectors, and $\{\sigma_i(t)\}_i$ the Schmidt coefficients. Despite the fact that each $\ket{\Psi_i}_{\mathsf{P}}$ and $\ket{\Psi_i'}_{\mathsf{F}}$ live in infinite dimensional spaces, the relevant correlations are stored in only a finite number of Schmidt coefficients. Let us now interpret this diagonal matrix $\Sigma_{ii}(t)=\sigma_i(t)$ as a linear map
from right-bond space $\mathsf{B}_R$ to left-bond space $\mathsf{B}_L$, and where we also make the identification of $\mathsf{B}_{L/R}\cong {\rm S}_{L/R}\otimes {\rm E}_{L/R}$, with $i\equiv (i_{\rm S},i_{\rm E})$ to distinguish the system and environment. Using this fact, and absorbing the $\sqrt{\sigma_i(t)}$ into $\ket{\Psi_i}_{\mathsf{P}}$ and $\ket{\Psi_i'}_{\mathsf{F}}$, then we may define the opened-bond pure state $\ket{\zeta}_{\mathsf{PSEF}}$ as
\begin{equation}
    \sum_{i_{\rm S},i_{\rm E},j_{\rm S},j_{\rm E}}\ket{\widetilde{\Psi}_{i_{\rm S},i_{\rm E}}}_{\mathsf{P}}\otimes \ket{i_{\rm S},i_{\rm E}}_{\mathsf{B}_L}\otimes \ket{j_{\rm S},j_{\rm E}}_{\mathsf{B}_R}\otimes\ket{\widetilde{\Psi}'_{j_{\rm S},j_{\rm E}}}_{\mathsf{F}}.
\end{equation}
This vector defines an instantaneous quantum comb where $\mathsf{B}_L$ and $\mathsf{B}_R$ are now the respective outputs and inputs of S and E at $t$. It is related to the cPT by a contraction of the identity $\delta_{(i_{\rm S},i_{\rm E})(j_{\rm S},j_{\rm E})}$. Given that we want to measure information which propagates through the environment, we restore \emph{this} component of the bond, and define the comb $\ket{\zeta}_{\mathsf{PSF}}$ by contracting in the environment-only identity $\delta_{i_E, j_E}$, yielding
\begin{equation}
    \ket{\zeta}_{\mathsf{PSF}} = \sum_{i_{\rm S},i_{\rm E},j_{\rm S}}\ket{\widetilde{\Psi}_{i_{\rm S},i_{\rm E}}}_{\mathsf{P}}\otimes \ket{i_{\rm S}}_{\mathsf{S}_L}\otimes \ket{j_{\rm S}}_{\mathsf{S}_R}\ket{\widetilde{\Psi}'_{j_{\rm S},i_{\rm E}}}_{\mathsf{F}}.
\end{equation}

From this, we now have $\rho_{\mathsf{PSF}} = |\zeta\rangle\!\langle \zeta|$ expressed in a finite-dimensional coefficient matrix with respect to an orthonormal basis. At this point, we must pause and emphasise a conceptual distinction in process-based correlation measures in contrast to state-based ones. Quantum states have canonically defined marginals: for a bipartite state $\rho_{A,B}$ there is only one deterministic measurement that can be performed to produce a state on system $B$. In the process picture, this is no longer the case. Indeed, \emph{all} deterministic control on $\mathsf{P}$ will produce a valid conditional marginal on $\mathsf{SF}$. Owing to causality, the marginal on $\mathsf{P}$ is always the same, however. It becomes unavoidable then to have the correlations defined in~\cref{eq:cnm-meas} be further \emph{conditioned} on a definition of a marginal over the state on $\mathsf{P}$ in terms of a specific choice of a deterministic operation.

\begin{figure*}[t!]
    \centering
    \includegraphics[width=\linewidth]{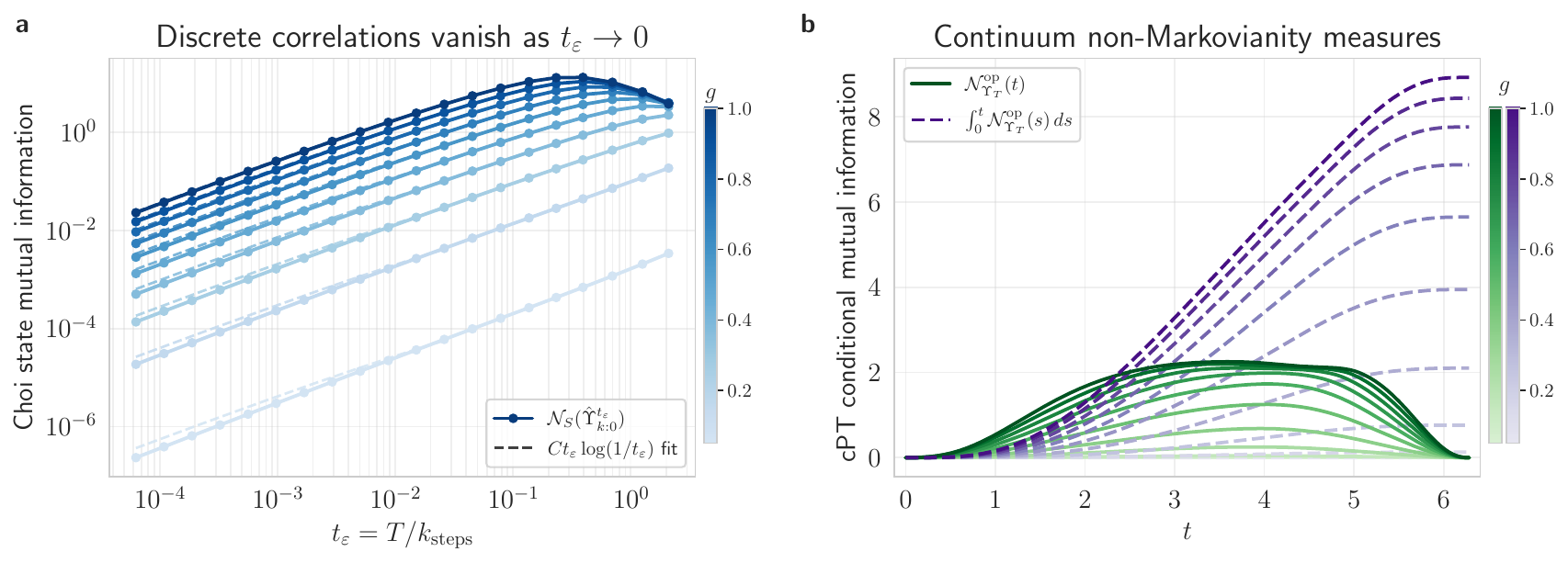}
    \caption{Illustration of continuum non-Markovianity results for a truncated Jaynes-Cummings model. \textbf{a} We compute the total quantum mutual information of the discrete process tensor Choi state $\mathcal{N}_S(\hat{\Upsilon}^{\shortT}_{k:0})$ for different coupling strengths as a function of Trotter step. From this, one sees that the measures identically vanish in the continuum limit at a rate of $\mathcal{O}(\shortT \log(1/\shortT))$. \textbf{b} For the same model, we compute our proposed continuum non-Markovianity measure $\mathcal{N}_{\rm op}$ in both its instantaneous (solid, green lines) and integrated (dashed, purple lines) forms, demonstrating that it remains non-trivial at all times and scales as expected with coupling strength. }
    \label{fig:nm_measures}
\end{figure*}

Once this marginalisation is chosen, $\rho_{\mathsf{PS}}$ can be straightforwardly computed and, from it, the necessary entropies to compute~\cref{eq:cnm-meas}.
For the others, we identify three distinct but principled choices one can make for the construction of $\rho_{\mathsf{SF}},$ which induce three possible measures of non-Markovianity. We define these as
\begin{enumerate}[label=(\roman*)]
        \item $\mathcal{N}_{\rm id},~$ which is computed via the projection of $\mathsf{P}$ onto the vacuum state $\ket{\Omega}_{\mathsf{P}}$, and amounts to no past intervention at all:
        \begin{equation}
            \rho_{\mathsf{SF}}^{\rm (id)} = \bra{\Omega}_{\mathsf{P}}\ket{\zeta_{\mathsf{PSF}}}.
        \end{equation}
        We consider this to be the choice which is most `natural' for quantum stochastic processes.
        \item $\mathcal{N}_{\rm tr},~$ which computes the future marginals by taking the partial trace of the Choi state $\hat{\zeta}$ over the interval $\mathsf{P}$. This is maximally invasive, and in a sense unphysical, but corresponds to a complete discarding of the past:
        \begin{equation}
            \rho_{\mathsf{SF}}^{\rm (tr)} = \Tr_{\mathsf{P}}[\hat{\zeta}_{\mathsf{PSF}}].
        \end{equation}
        Although this quantity does not reflect a physically implementable operation in the continuum -- since it averages over all finite particle sectors -- it is the most close in spirit to~\cref{eq:dnm-meas}, in the sense of a causal break between past and future. Note that as a consequence, this is only computable when the rank of the Choi state is small (see~\cref{ssec:choi duality}).
        \item $\mathcal{N}_{\rm op},~$ which, analogous to temporal entanglement in the finite case, measures operator correlations via environment-mediated entanglement in the \emph{vectorised} Choi state. That is to say, it comes from taking the partial trace 
        \begin{equation}
            \rho_{\mathsf{SF}}^{\rm (op)} = \Tr_{\mathsf{P}}[|\zeta_{\mathsf{PSF}}\rangle\!\langle\zeta_{\mathsf{PSF}}|].
        \end{equation}
        This quantity is slightly different in that it is not related to any physical operation on the system over $\mathsf{P}$. It corresponds, therefore, to compressibility, but does not carry an operational interpretation. Note additionally that this quantity requires $\rho_{\mathsf{PS}}$ to be computed in a similarly alternative way.
    \end{enumerate}
Each of these quantities constitute genuine and non-trivial measures of quantum non-Markovianity in the continuum and may be straightforwardly generalised from bipartite mutual information measures to the multipartite, which would encode the existence of higher point correlations between multiple different \emph{regions} of the process. Lastly, since they are defined for a chosen time $t$, there might seem some arbitrariness to the measure. For a more complete notion of non-Markovianity over an interval, one might instead consider the integrated quantities relative to a given cPT
\begin{equation}
    \int_0^T\mathcal{N}_{\Upsilon_T}^{(\cdot)}(t)~\dd t.
\end{equation}
Both are equally valid, equally computable, with the distinction of being an instantaneous entropic rate verses an integrated one. In analogy to discrete measures, the former would be equivalent to Schmidt entropy at a single cut, and the latter equivalent to the full mutual information of the Choi state. 

We conclude this section with some brief numerical demonstrations of our results. Concretely, we consider a minimal non-Markovian interaction: a damped Jaynes-Cummings Hamiltonian with the environment truncated down to a single mode. These dynamics are described by a Lindbladian acting on two spins: $\mathcal{L}_{\rm SE}(\rho) = -i[H_{\rm SE},\rho] + L\rho L^\dagger - \frac12 \{L^\dagger L,\rho\}$, with
\begin{equation}
\begin{split}
    H_{\rm SE} &= \frac{\omega_{\rm S}}{2} \sigma_Z^{({\rm S})} + \frac{\omega_{\rm E}}{2}  \sigma_Z^{({\rm E})} + g(\sigma_+^{({\rm S})} \sigma_-^{({\rm E})} + \sigma_-^{({\rm S})} \sigma_+^{({\rm E})}),\\
    &\text{and}~~L = \sqrt{\kappa}\id^{(\rm S)}\otimes \sigma_-^{(\rm E)}.
\end{split}
\end{equation}
This corresponds to an excitation-exchange model where the parameter $g$ governs the coupling strength between system and environment, and $\kappa$ is the environmental damping factor. Setting $\kappa=0.3$ and varying $g$, we compute a collection of discrete $\hat{\Upsilon}_{k:0}^{\shortT}$ and continuous $\cPT$ process tensors on the interval $[0,T]$ where $T=2\pi$. The discrete process tensors are computed for a range of step sizes. We then evaluate their respective non-Markovianity measures $\mathcal{N}_S(\hat{\Upsilon}_{k:0}^{\shortT})$ and $\mathcal{N}_{\rm op}$, which are displayed in~\cref{fig:nm_measures}. Panel \textbf{a} illustrates how the discrete measures all identically tend towards zero in the $\shortT\to 0$ limit, which we additionally match to the expected $\mathcal{O}(\shortT \log(1/\shortT))$ scaling. Meanwhile, Panel \textbf{b} shows how our introduced continuum measures constitute a non-trivial quantifier of the non-Markovianity. Collectively, these expound the arbitrariness of studying physical models with a typical Trotter approach, and demonstrate a robust replacement. It is our hope and intention that these proposed measures will allow for a more principled study of non-Markovian models in future. Moreover, on the theoretical side, we are motivated to extend the classification of non-Markovianity measures from PCRs to all cPTs. Along these lines, for instance, it will be important to have representation-agnostic ways of finding a system-environment decomposition. Such instances will also constitute an interesting case study in the sense of: do there exist cPTs which are not exact PCRs but which still even have a sensible and finite notion of non-Markovianity?

\subsection{A hierarchy of genuine multi-time correlations}\label{ssec:wick}
% !Tex root = ./paper.tex

Identifying the continuous process tensor as a Fock space vector opens up possibilities for transferring existing results from the field theory literature to quantum processes.
A particularly relevant result for quantum processes is the analysis in Ref.~\cite{hubenerWicksTheoremMatrix2013}, where the authors prove a version of ``Wick's theorem'' for randomly generated cMPS. Higher order correlation functions are computed from lower order ones by using structural properties of the cMPS rather than algebraic properties of correlation functions. 
More concretely, they identify a property (which they call the ``$p$-number'') that allows them to categorise field theory states in terms of the size of the particle sectors that completely determine the state. In the following, we adapt this result to continuous process tensors which are PCRs and examine the implications for quantum stochastic processes. 
We view this exposition as a first step toward constructing a hierarchy of quantum processes based on the ``$p$-number'', which should motivate further investigation of this property in future work.

To motivate the definition of the $p$-number of a continuous process tensor, we consider objects of the form
\begin{align}
     C^{(n)}_{\bm{\nu}}(\bm{t} \,) &\coloneqq  \bra{\Omega}\prod_{j=1}^n \psi_{\nu_j}(t_j)\cPT,     \label{eq:finite_particle_marginal_cPT}
\end{align}
where $\bm{t} = (t_1,\cdots, t_n) \in \mathbb{R}^n$ is a list of times such that $0 \leq t_1 \leq t_2 \leq \cdots \leq t_n \leq T$ and $\bm{\nu} = (\nu_1, \cdots, \nu_n )$ is a list of particle species.  For $\cPT$ a PCR, the functions~\cref{eq:finite_particle_marginal_cPT} are point-wise well defined.
From a field theory perspective, these are the $n$-particle wavefunctions of the vector $\cPT$. 
From the process tensor perspective, $C^{(n)}_{\bm{\nu}}\!(\bm{t})$ 
are entries in the finite $n$-point marginal of the continuous 
process $\Upsilon_T$.

From now on, we work in this subsection under the assumption that the cPT is a PCR with a finite dimensional environment and time-independent, Liouvillian generator $\gen(t) \equiv \gen$. 
We denote by $\gen= \sum_{k=0}^{d^2_{\rm SE}-1} \lambda_k \kketbra{k}{k}$ the (possibly non-unique) eigendecomposition of $\gen$. 
Note that in general the corresponding eigenbases are non-orthogonal.
Since $\gen$ is a Liouvillian, we know that
$\kket{\id_{\rm SE}}$ must be among its eigenvectors corresponding to eigenvalue zero, and we identify $\kket{0} \equiv \kket{\id_{\rm SE}}$.
Then one finds that
\begin{multline}
C^{(n)}_{\bm{\nu}}(\bm{\tau} \,) =  \sum_{k_1,\cdots, k_n =0}^{d^2_{\rm SE}-1}c^{(n)}_{\bm{\nu}}(k_1,\cdots,k_{n}) \times \\ \mathrm{e}^{\tau_n \lambda_{k_n}}\cdots \mathrm{e}^{\tau_2\lambda_{k_2}}\mathrm{e}^{\tau_1\lambda_{k_1}},
\label{eq:corr_Wick_sum}
\end{multline} 
where we have defined the vector of time differences $\bm{\tau} = ( t_n-t_{n-1},\cdots, t_2-t_1,t_1)$ and the coefficients
\begin{multline}
c^{(n)}_{\bm{\nu}}(k_1,\cdots,k_n) := \\ 
\bbra{\id_{\rm SE}}\mathbb{P}_{\nu_n}\kketbra{{k_{n}}}{{k_{n}}}\mathbb{P}_{\nu_{n-1}} \cdots \mathbb{P}_{\nu_1} \kket{{k_{1}}} \!\bbraket{{k_{1}}}{\rho^{\rm SE}_0}. \label{eq:residue_Laplace}
\end{multline}
Adapted from Ref.\ \cite{hubenerWicksTheoremMatrix2013}, we introduce 
\begin{definition}[$p$-number of a PCR]
\label{def:p_number}
Given an exactly compressed time-translation invariant cPT $\cPT$ in the process-canonical representation, the $p$-number of $\cPT$ is the minimum order $p$ such that one can find an eigenbasis of $\gen$ such that for any of its elements $k \in \{0, \dots, d_{\rm SE}^2 - 1\}$ there exists an integer $p_k' \leq p$, a sequence $\bm{\nu}_k = (\nu_1, \cdots, \nu_{p_k'})$, and a string $\bm{k} =(k_1,\cdots,k_{p_k'})$ that contains $k$ such that 
\begin{equation}
    \label{eq:p_number_condition}
     c^{(p_k')}_{\bm{\nu}_k}(\bm{k}) \neq 0.
\end{equation} 
If the minimum does not exist, we say that the $p$-number is infinite. 
\end{definition}
With the specification of an exactly compressed cPT, we exclude the trivial case where a part of the environment does not couple to the physical system at all, i.e., when $\gen = \mathbb{H}_{\rm SE_1}\otimes \id_{\rm E_2} + \id_{\rm SE_1} \otimes \mathbb{H}_{\rm E_2}$ and $\rho^{\rm SE}_0 = \rho^{\rm SE_1 }_0 \otimes \rho^{\rm E_2 }_0$. 
In this case, the $p$-number of the associated process will always be infinite, irrespective of $\mathbb{H}_{\rm SE_1}$ and $\rho^{\rm SE_1 }_0$, as can be directly verified from the definition.
We call a cPT exactly compressed if there is no basis such that $\gen$ and $\rho^{\rm SE}_0$ are of the above form. 

The key property of \cref{def:p_number} is that a process of finite $p$-number is completely determined by its marginals up to order $2p+1$. 
To see this, let us consider a cPT with $p$-number $p$ and let $\{\kket{k}\}_k$ be the eigenbasis that achieve the $p$-number value. 
We abbreviate the matrix elements of the jump matrices $\id_{\rm E}\otimes \mathbb{P}_\nu$ in this eigenbasis as $\bbra{{k_{2}}}\id_{\rm E}\otimes\mathbb{P}_{\nu_1}\kket{{k_{1}}} = P^{[1]}_{k_2,k_1}$. 
Note that the jump matrix acts as an identity on the environment. 
By \cref{def:p_number}, we can write for every index value $k$ of the eigenbasis the scalar identity
\begin{align}
    1^{(k)} &= \frac{ P^{[p'_k]}_{1,\ast}\cdots P^{[\ell]}_{\ast,k} P^{[\ell-1]}_{k,\ast} \cdots \bbraket{\ast}{\rho^{\rm SE}_0}}{c^{(p'_k)}_{\bm{\nu}_k}} \nonumber \\
    &= \frac{P^{[\ell-1]}_{k,\ast} \cdots \bbraket{\ast}{\rho^{\rm SE}_0} P^{[p'_k]}_{1,\ast}\cdots P^{[\ell]}_{\ast,k} }{c^{(p'_k)}_{\bm{\nu}_k}},
\end{align}
where $\ast$ denotes other, irrelevant indices and in the second line we have permuted the order of the $p+1$ scalar factors. 
Now, given a coefficient as in \cref{eq:residue_Laplace} for arbitrary $n$ and some $\bm{\nu}'$, we can insert resolutions of the scalar identity in terms of the appropriate index between every matrix element factor in $c^{(n)}_{\bm{\nu}'}(\bm{k})$ as
\begin{widetext}
    \begin{multline}
   c^{(n)}_{\bm{\nu}'}(\bm{k}) = P^{[n]}_{1,k_n}1^{(k_n)} P^{[n-1]}_{k_n,k_{n-1}}1^{(k_{n-1})}\cdots P^{[1]}_{k_2,k_1} 1^{(k_{1})} \bbraket{k_1}{\rho^{\rm SE}_0} \\
     = P^{[n]}_{1,k_n}\frac{P^{[\ell-1]}_{k_n,\ast} \cdots \bbraket{\ast}{\rho^{\rm SE}_0} P^{[p_k']}_{\id,\ast}\cdots P^{[\ell]}_{\ast,k_n} }{\left[c^{(p'_{k_{n}})}_{\bm{\nu}_{k_{n}}} \right]} P^{[n-1]}_{k_n,k_{n-1}} \cdots P^{[1]}_{k_2,k_1} 1^{(k_{1})} \bbraket{k_1}{\rho^{\rm SE}_0}. 
\end{multline}
\end{widetext}
By appropriately rearranging and regrouping the factors of matrix elements, we see that we can write
\begin{equation}
      c^{(n)}_{\bm{\nu}'}(\bm{k})  = \frac{c_{\bm{\nu}_n}^{(\ell_n)} c_{\bm{\nu}_{n-1}}^{(\ell_{n-1})}\cdots c_{\bm{\nu}_1}^{(\ell_1)}} {c_{\bm{\nu}_{k_{n}}}^{(p'_{k_{n}})}c_{\bm{\nu}_{k_{n-1}}}^{(p'_{k_{n-1}})}\cdots c_{\bm{\nu}_{k_{1}}}^{(p'_{k_{1}})}},
\end{equation}
where $\ell_j \leq 2(p+1 -1) +1 = 2p +1$ for all $j=1,2,\cdots, n$. We can summarise this result in the following
statement.

\begin{theorem}[Recovering multi-time correlation functions]
    For a cPT with $p$-number $p$, all $C^{(n)}_{\bm{\nu}}$ from \cref{eq:finite_particle_marginal_cPT} for any $\bm{\nu}$ and $n$ can be determined from the complete knowledge of $\{C^{(\ell)}_{\bm{\nu}}\}_{\bm{\nu}, \ell= 1,\cdots, 2p+1}$ in terms of the $\{c^{(\ell)}_{\bm{\nu}}\}_{\bm{\nu}, \ell= 1,\cdots, 2p+1}$ and $\{\lambda_i\}_i$. 
    \label{thm:p_number_Wick_thm}
\end{theorem}
In the case of non-degenerate eigenvalues of $\gen$ with bounded real part $\mathrm{Re}(\lambda_k)< \infty$ for all $k \in [d_{\rm SE}^2]$, the knowledge of $\{C^{(\ell)}_{\bm{\nu}}\}_{\bm{\nu}, \ell= 1,\cdots, 2p+1}$ alone (not given in terms of the coefficients $c_{\bm{\nu}}$ and $\gen$-eigenvalues $\lambda_k$) of a cPT with $p$-number $p$ is sufficient to construct $C^{(n)}_{\bm{\nu}}$ for any $n$ and $\bm{\nu}$.  
This follows from the argument in \cite{hubenerWicksTheoremMatrix2013}, which considers the Laplace transforms of $\{C^{(\ell)}_{\bm{\nu}}(\bm{\tau})\}_{\bm{\nu}, \ell= 1,\cdots, 2p+1}$ as functions of $\bm{\tau}$. 
Under the above conditions, the Laplace transform integral converges and allows extraction of the eigenvalues $\{\lambda_k\}_k$ and the associated coefficients $\{c_{\bm{\nu}_i}^{(\ell_i)}(k_1,\cdots,k_{\ell_i})\}_i$ from the $d^2_{\rm SE}$ distinct poles of the Laplace transform and its residues, respectively. 
From this point, the reconstruction of any $C^{(n)}_{\bm{\nu}}$ proceeds as above. 

We have thus shown that via \cref{thm:p_number_Wick_thm}, the $p$-number introduces a hierarchy of cPTs in the PCR. We can categorise PCRs based on the maximum particle number of the subspaces needed to fully determine the process. 
To build intuition about the $p$-number property, we now construct toy examples for processes with $p= 1,2$, and $\infty$ by choosing $\gen$ and $\rho^{\rm SE }_0$ appropriately. 
First, we construct an example of a process with $p=2$. 
By \cref{def:p_number}, there must exist a choice of $\nu, \nu'$ such that for all $k \in [d_{\rm SE}^2]$ there exists $k' \in [d_{\rm SE}^2]$ such that either $c^{(2)}_{\nu, \nu'}(k, k') \neq 0$ or $c^{(2)}_{\nu, \nu'}(k', k) \neq 0$, where
\begin{multline}
    c^{(2)}_{\nu, \nu'}(k, k') = \\ \llangle \id_{\rm SE} |\id_{\rm E}\otimes \mathbb{P}_{\nu} \kketbra{k}{k}\id_{\rm E}\otimes \mathbb{P}_{\nu'}\kket{k'} \!\bbraket{k'}{ \rho_0^{\rm SE}} .
\end{multline}
We target the condition $c^{(2)}_{\nu, \nu'}(k, k') \neq 0$ and suppose that we choose $k'$ such that $\llangle k' | = \llangle \id_{\rm SE} |$, which is always possible.
Suppose furthermore that $|k'\rrangle = |\id_{\rm SE}\rrangle$ and that the choice for which the term is non-zero is $\nu =  \nu' = (0,0)$.
Then we have
\begin{equation}
    c^{(2)}_{\nu, \nu'}(k, k') = \tr(S_k)^2,
\end{equation}
where $S_k \in L(\mathcal{H}_{\rm S}\otimes \mathcal{H}_{\rm E})$ is the matricised form of $| k \rrangle \equiv \kket{S_k}$.
Therefore, we require that for all $k$ the matrix $S_k$ has non-vanishing trace. 
One valid choice is, for example,
\begin{equation}
    | k \rrangle = \begin{cases}
        |\id_{\rm SE}\rrangle & k = 1\\
        |\id_{\rm SE}\rrangle + |P_k\rrangle & k \neq 1,
    \end{cases}
\end{equation}
where $P_k \neq \id_{\rm SE}$ are again the Pauli matrices in $L(\mathcal{H}_{\rm S}\otimes \mathcal{H}_{\rm E})$ for a system and environment each composed of qubits. 
Any generator 
\begin{equation}
   \gen = \sum_{k=1}^{d^2_{\rm SE}} \lambda_k \kketbra{k}{k} 
\end{equation}
with this eigenbasis and any initial state $\rho_0^{ \rm SE}$ defines a $p=2$ process if there exists an eigenstate $\kket{k}$ such that $\bbraket{k}{\rho_0^{ \rm SE}} = 0$.
This ensures that $p=2$ is indeed the minimal integer for which \cref{def:p_number} is satisfied. 
In the case that the initial state has overlap with all $\kket{k}$, we have constructed a $p=1$ process, as can be readily verified.  

We now construct an example of a $p = \infty$ process. Consider the generator 
\begin{equation}
    \gen = |\sigma_X \otimes \sigma_X\rrangle \!\llangle \sigma_X \otimes \sigma_X | ,
\end{equation}
where the system and the environment are each a single qubit and $\sigma_X$ is the Pauli-X operator. 
Let the initial state be the maximally mixed state $\rho_0^{\rm SE} = \frac{1}{d_{\rm SE}} \id_{\rm SE}$.
The eigenbasis of $\gen$ is $\{|\id_{\rm SE}\rrangle, \kket{\sigma_X \otimes \sigma_X}, \ldots\}$.
Due to the tracelessness of $\sigma_X$, there is no $c_{\bm{\nu}}^{(n)}(\bm{k})$ that is non-zero for some $\bm{k}$ containing $k=1$ with $\kket{1} = \kket{\sigma_X \otimes \sigma_X}$. 
Therefore, this process has an infinite $p$-number. 
Note that the choice of initial state is crucial for this behavior. 
Generalizing this argument, we can show that any process on qubit system and environment with a generator diagonal in the Pauli basis and a maximally mixed initial state has an infinite $p$-number. 

We conclude this subsection with some remarks. 
We observed that given the generator $\gen$, the choice of initial state can increase the $p$-number of the associated cPT by one if it does not have overlap with all eigenspaces of the generator, i.e., if there exists a $k$ such that $\bbraket{k}{\rho_0^{\rm SE}} =0$. 
Moreover, this implies that a Markovian process, corresponding to no environment being present ($\mathcal{H}_{\rm E} = \mathbb{C}$), will have $p$-number 1 or 2, depending on the initial state. 
Since we have shown in the above example that non-Markovian processes can also have $p$-number 1 or 2, the hierarchy induced by the $p$-number cannot always distinguish between Markovian and non-Markovian processes.
Finally, for infinite dimensional environments, we expect that in principle the $p$-number definition can be extended for reasonably behaving generators $\gen$. We leave the formalization of this for future work.

\section{Working with continuous process tensors in practice}\label{sec:apps}
Although this is principally a work to formalise process tensors in the continuum, we would like to conclude with a discussion around some potential applications of the framework. In particular, it is not on the face of it clear that one can even work with cPTs in practice. Here, we outline some methods in the important areas of \emph{simulation} and \emph{tomography} for which we expect there to be useful and efficient applications of cPTs. It is our intention that these expose some of the useful properties of cMPSs as applied to non-Markovian open quantum systems, as opposed to necessarily a comprehensive set of results. We will have more to say on this matter in forthcoming work.

\subsection{Simulation and compression}\label{ssec:compressibility}
% !Tex root = ./paper.tex

An appealing aspect of working with MPS and MPOs is that there exist well-established and well-studied algorithms from which one can perform bond-wise singular value truncations to obtain a compressed (and approximated) description of the vector. Each of these truncations is point-wise optimal in the 2-norm (and hence in 1-norm for MPSs). In process tensor literature, the basic strategy is to iteratively construct the process tensor Choi matrix as an MPO via Trotterisation, and to truncate whenever the internal bonds exceed a pre-determined threshold~\cite{PhysRevLett.126.200401,Cygorek-2022,Link_2024}.\par 
In this section, we would like to concretely consider the extension of these truncation strategies to cMPSs, and hence the question of \emph{compressed} continuous process tensors. That is, we will show how to implement a dynamical rank approximation for a given cPT based on the cMPS compression algorithm introduced by Park et al. in Ref.~\cite{parkTensorNetworkInfluence2024} for the purpose of simulating the dynamics of a many-body system. The nature of this approximation differs from that of discrete compression in that the projection onto the chosen dominant eigenspaces is encoded in the equations of motion of the process as opposed to applied separately after construction. As such, it can significantly reduce the impact of Trotter error, even when nominally solving a problem to the same order in time.
Note that we are not concerned with numerical optimality here, but rather in studying the \emph{existence} of nearby cPTs on smaller bond spaces. We truncate with an exact diagonalisation, but in principle it is clear that the procedure can be more efficient by employing the vast literature on dynamic rank approximation methods~\cite{koch2007dynamical,ceruti2022rank,lubich2014projector,Haegeman_2017}. Moreover, there is no obstacle in principle to truncating to a $\chi$ subspace while one iteratively constructs the environment, such that the full $D\times D$ object never be explicitly represented.

To be more specific, consider first a $\cPT$ in its process-canonical representation (which can be found to arbitrary precision, see~\cref{thrm:pcr density} and the companion work~\cite{fuksaDilationTheoremContinuum2026}), which is a cMPS with bond dimension $D = d_{\rm S}^2d_{\rm E}^2$. 
We wish to find $|\Upsilon_T'\rangle_\chi$, where we make explicit the bond dimension $\chi < D$. This new cPT should (i) fully define a new $Q(t)$ and $\{R_\nu(t)\}$ set (and hence a new cPT); (ii) preserve the instrument structure (physical legs) of the cPT; and (iii) (optionally) preserve the physicality of the cPT. We say `optionally' on this last point, since generic tensor network compression techniques tend to give up physicality of the representation in favour of computationally more efficient approximations of observables of the object. 
We will do this by performing a point-wise truncation of the singular values of $\cPT$, which is hence point-wise optimal. Importantly, in contrast to Trotter methods, this approach integrates the projection onto some dominant singular value space \emph{into} the equations of motion, and so their effects can be brought to genuinely arbitrarily low order in time. This mirrors the principles of the time-dependent variational principle (TDVP)~\cite{Haegeman_2016}.

We recall from~\cref{eq:cmps-gauges} that the cMPS representation is not unique and that its $Q$ and $R$ matrices may be gauge-transformed and still represent the same physical process. Analogous to MPS, certain canonical gauges are chosen for both numerical stability and because they reveal the continuous Schmidt information about the cMPS. These require defining the left and right environments of the cMPS, which we shall do in our cPT context. Let $l(0) \coloneqq \kket{\id_{\rm SE}}\!\bbra{\id_{\rm SE}}$ and $r(T)\coloneqq \kket{\rho_0^{\rm SE}}\!\bbra{\rho_0^{\rm SE}}$. These form the initial conditions of virtual density matrices $l(t)$ and $r(t)$, which are determined by the ODEs:
\begin{equation}
    \label{eq:env-eoms}
    \begin{split}
        \frac{\dd l(t)}{\dd t} &= l\cdot Q + Q^\dagger\cdot l + \sum_{\nu}\mathbb{P}_{\nu}^\dagger l\mathbb{P}_{\nu},\\
        \frac{\dd r(t)}{\dd t} &= -\left(Q\cdot r + r\cdot Q^\dagger + \sum_{\nu}\mathbb{P}_{\nu}r\mathbb{P}_{\nu}^\dagger\right).
    \end{split}
\end{equation}
These virtual density matrices are naturally also gauge-covariant objects. The \emph{left-orthonormal gauge} is the one such that $l(t) = \id~\forall~t$. This is achieved by choosing $g_L(t)\coloneqq [\sqrt{l(t)}]^{-1}$. Then the object $\sqrt{l(t)}\sqrt{r(t)} = \sqrt{r(t)}$ carries the field singular values at each point. There is a left-unitary freedom remaining in this gauge which preserves the left-orthonormal form: $g_L(t)\mapsto u(t)g_L(t)$. This unitary can be chosen such that its effective action on $r(t)$ brings it into its diagonal form. That is, take $u(t)$ such that 
\begin{equation}
    r(t) \mapsto \sigma(t) = u^\dagger(t)r(t)u(t),
\end{equation}
where $\sigma(t)$ is diagonal for all $t$ and in descending order. This is the diagonal form of the \emph{central canonical gauge}. For technical details about these points -- including how one finds the gauge transformations in practice -- we refer the reader to~\cref{app:gauges} for explicit details. The conceptual point is that a gauge transformation can be found to map $r(t)\mapsto \sigma(t)$.
After normalisation, this $\sigma(t)$ is the entanglement entropy of $\cPT$ at a given time $t$.

With this in mind, we propose the following algorithm based on the work of Ref.~\cite{parkTensorNetworkInfluence2024}. For simplicity, assume that the Schmidt values of the cPT remain non-degenerate (including eigenvalue crossings) for the entirety of the process. Given $\cPT$, suppose we transform $Q$ and $\{\mathbb{P}_\nu\}$ into this above central canonical form.
Let us denote the gauge transformed cMPS parameters (including the $u(t)$ transformation) by $Q^{(c)}(t)$ and $\{\mathbb P^{(c)}_{\nu}(t)\}$. Now, in this frame, the dynamics of $\sigma(t)$ are given by the ODE 
\begin{equation}
    \frac{\dd \sigma(t)}{\dd t} = -\!\left(\! Q^{(c)}(t)\sigma(t) + \sigma(t) Q^{{(c)}\dagger}(t) + \sum_\nu \mathbb{P}_\nu^{{(c)}}\sigma(t)\mathbb{P}_\nu^{{(c)}\dagger}\right).
\end{equation}
Let $P_\chi$ be the truncation projector onto the eigenspace spanned by the $\chi$ dominant eigenvalues as
\begin{equation}
    P_\chi := \begin{pmatrix}
        \id_{\chi\times \chi} & \bm{0} \\
        \bm{0} & \bm{0}
    \end{pmatrix}.
\end{equation}
Breaking $\sigma(t)$ up into these components, we have 
\begin{equation}
    \sigma(t) = \begin{pmatrix}
        \sigma_{(1, 1)}(t) & \sigma_{(1,2)}(t) \\
        \sigma_{(2,1)}(t) & \sigma_{(2,2)}(t)
    \end{pmatrix},
\end{equation}
with $\sigma_{(1,2)}(t) = \sigma_{(2,1)}(t) = \bm{0}$.
The matrices $Q^{(c)}(t)$ and $\{\mathbb{P}^{(c)}_\nu\}$ can be written in this block structure similarly. Now, the action of $P_\chi$ implicitly defines a new cMPS via the dynamics of $\sigma_{(1,1)}(t)\equiv \tilde{\sigma}(t)$
\begin{multline}
    \label{eq:eom-sigma}
        \frac{\dd \tilde{\sigma}}{\dd t}= -\left[Q^{(c)}_{(1,1)}\tilde{\sigma}(t) + \tilde{\sigma}(t)Q^{(c)\dagger}_{(1,1)} + \sum_\nu \mathbb{P}_{\nu,(1,1)}^{(c)}\tilde{\sigma} \mathbb{P}_{\nu,(1,1)}^{(c)\dagger} \right.\\
        + \left.\underbrace{\sum_\nu \mathbb{P}_{\nu,(1,2)}^{(c)}\sigma_{(2,2)}\mathbb{P}_{\nu,(1,2)}^{(c)\dagger}}_{\text{leakage term}}\right].
\end{multline}
Equation~\eqref{eq:eom-sigma} is exact. Now, the relevant approximation here we need for the construction of a compressed cMPS is to also zero out $\sigma_{(2,2)}$. If
\begin{equation}
    \sum_\nu \mathbb{P}_{\nu,(1,2)}^{(c)}\sigma_{(2,2)}\mathbb{P}_{\nu,(1,2)}^{(c)\dagger} \approx 0,
\end{equation}
then, this defines a new cMPS $|\Upsilon'_T\rangle_\chi$ such that 
\begin{equation}
    \cPT \approx \cpt{Q^{(c)}_{(1,1)}(t)}{\{R^{(c)}_{\nu,(1,1)}\}}{B^{(c)}_{(1,1)}}_\chi.
\end{equation}
We have written this new object as a cMPS, but not necessarily a cPT, since the projected operators may no longer exactly satisfy the positivity and causality conditions of cPTs as they are outlined in~\cref{ssec:physicality}. Nevertheless, the TDVP accommodates projections onto quite general manifolds~\cite{Hackl_2020}. This could be chosen to be constrained to a requirement of physicality as determined, for example, by a Lindblad or Hamiltonian generator and the basis constraints outlined in~\cref{ssec:physicality}. The existence of and difficulty in computing approximate cPTs for a given process remains a subject for future work. 

\subsection{Learning and tomography}
\label{sec:tomography}
% !Tex root = ./paper.tex

In addition to the task of simulation or compression, another application of central importance to the burgeoning field of quantum technologies is that of \emph{learning} non-Markovian processes. Non-Markovian tomography with the aid of process tensors has, in recent years, been formalised, experimentally demonstrated, and been made efficient~\cite{white2022NonMarkovianQuantumProcess,whiteNonMarkovianQuantumProcess2022,White-2025-Unifying-NM,Giarmatzi2025multitimequantum,Shrapnel_2018_supervised,PRXQuantum.2.040351,PhysRevA.104.022432,xiang2021quantifynonmarkovianprocessintervening,PhysRevLett.134.010803,Li_2024_NMGST}. But this tomography suffers from the same conceptual issues as the finite framework itself. Namely, that there is no place for the finite time of the instruments. 
This is particularly pressing since instruments can represent physically very different processes. In the common setting where single qubit gates, two qubit gates, and projective measurements can occur on three drastically different timescales with different drives, this introduces an inconsistency in how the tomographic reconstruction is ultimately performed.

Suppose a continuous quantum process $\cPT$, which we can re-run many times, and a parameter $\epsilon \ge 0$.
A cPT tomography consists of an experimental protocol and a post-processing pipeline.
The protocol runs the unknown process $n \times k$ times with various instruments, obtaining samples $\mathcal{X} = \{x^{(j)}_{i}\}_{i\in[k], j\in[n]}$, where the samples $x^{(j)}$ come from the probability distribution
\begin{equation}
    \Pr(X = x | \mathcal{I}_j) = \braket{\mathcal{I}_j(x) | \Upsilon_T} {\rm d}x,
\end{equation}
where $\{\bra{\mathcal{I}_j(x)}\}_x$ is a valid instrument.
Given the set $\mathcal{X}$, the classical post-processing pipeline of cPT tomography returns an estimate $\ket{\tilde \Psi}$, with the guarantee that (perhaps with high probability)
\begin{equation}
    d(\ket{\Psi}, \ket{\tilde \Psi}) \le \epsilon,
\end{equation}
for a meaningful notion of distance, see \cref{ssec:distance}.

As $\ket{\Psi}$ has an infinite number of degrees of freedom, performing tomography on it without any structural assumptions requires infinite $n\times k$, as well as infinite computational resources in the post-processing pipeline \cite{haahSampleOptimalTomographyQuantum2017,meleOptimalLearningQuantum2025,BenchmarkingReview}.
Therefore, in order to get a practical procedure we need to place assumptions on the structure of $\ket{\Psi}$, obtaining an ansatz class with a finite---and in practice reasonable---number of parameters.
This ansatz class should be well motivated on physical grounds, and is reasonable whenever correlations are short-ranged in a precise way \cite{AreaReview}.

Given our exposition of the cPT framework, a natural first step is to utilise the cMPS ansatz \cref{ssec:cMPS}, which is parametrised by the matrix functions $Q: [0, T] \rightarrow \mathbb{C}^{\chi d_{\rm S}^2 \times \chi d_{\rm S}^2}$, $R_\nu: [0, T] \rightarrow \mathbb{C}^{\chi d_{\rm S}^2 \times \chi d_{\rm S}^2}$ and the matrix $B \in \mathbb{C}^{\chi d_{\rm S}^2 \times \chi d_{\rm S}^2}$, subject to the physicality conditions of causality and positivity discussed in \cref{ssec:physical conditions}.
It may also be advantageous to restrict the cMPS parameters further to the process-canonical representation.
We leave the investigation of this for future work and note the apparent connection to self-consistency of tomographic methods, where we do not assume that we know exactly what instruments we are performing \cite{nielsen2021GateSetTomography,briegerCompressiveGateSet2023}.
In order to limit to a finite number of parameters, we need to make simplifying assumptions about the nature of the functions $Q(t)$ and $R_\nu(t)$.
The simplest case is to assume a time homogeneous process, in which case we have $Q(t) = Q$ and $R_\nu(t) = R_\nu$ and the number of parameters to be estimated reduces to $\chi^2 d_{\rm S}^4(d_{\rm S}^4 + 1)$, where $\chi$ is the dimension of the effective environment Liouville space, which can either be estimated from the tomography data, or assumed on physical grounds.

A possible approach to the tomography task, utilizing techniques from Ref.~\cite{steffensExperimentalQuantumfieldTomography2015} is through the measurement of low order correlation functions
\begin{equation}
    C^{(n)}_{\nu_1, \dots, \nu_n} (t_1, \dots, t_n) = \braket{\Omega | \psi_{\nu_n}(t_n) \dots \psi_{\nu_1}(t_1) | \Upsilon_T}.
\end{equation}
The work~\cite{steffensExperimentalQuantumfieldTomography2015} builds on the assumption of a finite $p$-number as encountered already in \cref{ssec:wick}.
Measuring $C^{(n)}$ corresponds precisely to the tomography of discrete process tensors, for which known tools exist \cite{whiteDemonstrationNonMarkovianProcess2020,whiteNonMarkovianQuantumProcess2022,white2022NonMarkovianQuantumProcess,whiteWhatCanUnitary2025}.
As we have already seen in \cref{eq:corr_Wick_sum} we have that  
\begin{equation}
\begin{split}
    &\hspace{-.3cm}C^{(n)}_{\nu_1, \dots, \nu_n}(t_1, \dots, t_n) =\\
                                       &=\tr[\e^{(T - t_n) Q}R_{\nu_n} \e^{(t_n - t_{n-1}) Q}R_{\nu_{n-1}} \dots R_{\nu_1} \e^{t_1 Q} B]\\
&= \sum_{k_{n+1} \dots k_1} \rho_{k_1 \dots k_{n+1}} \e^{(T-t_n) \lambda_{k_{n+1}}} \dots \e^{t_1 \lambda_{k_1}},
\end{split}
\end{equation}
where 
\begin{equation}
    \rho_{k_1 \dots k_{n+1}} \coloneqq \left(R_{\nu_n}\right)_{k_{n+1}, k_n} \dots \left(R_{\nu_1}\right)_{k_2, k_1} B_{k_1, k_{n+1}},
\end{equation}
where on the right hand side we wrote out the matrix elements $X_{k, k'} = \braket{k | X | k'}$, where $\{\ket{k'}\}$ and $\{\bra{k}\}$ are the (non-orthogonal) right and left eigenvectors of $Q$ and $\lambda_k$ are the eigenvalues of $Q$. 
In Ref.~\cite{steffensExperimentalQuantumfieldTomography2015} the authors suggest a method based on the Prony matrix pencil approach to recover the poles $\lambda_k$ and the matrices $R$ and $B$.
A limitation of this approach is the need to perform discrete process tensor tomography, which requires applying instruments on timescales much faster than the ambient dynamics of the process, potentially running into the quantum speed limit and, disappointingly, losing the advantage our framework provides in treating instruments as finite time processes.

To circumvent this issue, we could instead use genuine continuous time pulses as instruments $\{\bra{\mathcal{I}_j(i)}\}_i$.
To see the feasibility of such approach, note that from the results of \cite{janzingQuantumControlAccess2002} it follows that applying coherent control to the system, we can generate $\chi d_{\rm S}^2$ linearly independent states on the virtual space of a cMPS, if the cMPS is written in its compressed form.
Therefore, there exists a family of control operations $\mathcal{O} \coloneqq \{\bra{\mathcal{I}_{\ell, i, j}}\}_{i,j}$ such that
\begin{equation}
    y_{ij}[\ell] \coloneqq \braket{\mathcal{I}_{\ell, i, j} | \Upsilon_T} = \braket{\phi_i | e^{i\ell \Delta t Q} | \psi_j},
\end{equation}
where each $\bra{\mathcal{I}_{\ell, i, j}}$ performs a non-trivial control operation that transforms $\ket{\rho^{\rm SE}} \mapsto \ket{\psi_j}$, then does nothing for time $\ell \Delta t$ and finally performs a non-trivial operation that transforms $\bra{E} \mapsto \bra{\phi_i}$, where $\{\ket{\psi_j}\}_j$ and $\{\bra{\phi_i}\}_i$ are (non-orthogonal) bases for the virtual space of the cMPS and its dual.
Using the tensorESPRIT algorithm from \cite{hangleiterRobustlyLearningHamiltonian2024a} we can estimate the eigenvalues of $Q$ from the data $y$, as well as the matrix $Q$ in either the basis $\{\ket{\psi_j}\}_j$ or $\{\bra{\phi_i}\}_i$.

In the case when a time independent ansatz is not appropriate, we need to find a different parametrization of the cMPS ansatz with a finite number of parameters.
An approach that turned out to be useful in ground state calculations is to use a spline basis for the matrix valued functions $Q, R$, as proposed in Ref.~\cite{tuybensVariationalOptimizationContinuous2022}.
We leave the details of such a scheme to future work.

\section{Discussion and outlook }\label{sec:discussion}
Motivated by the physical principle of finite energy operations, in this work we have derived a genuine continuum limit of process tensors. 
We showed how continuous process tensors emerge as vectors in a bosonic non-relativistic field theory with different particle species, wherein the particle species are identified with a basis of non-trivial transformations of the system. This moreover imbues the process tensor framework with a more genuinely stochastic-process flavour, in that the coarse-graining marginalisation operation is uniquely associated with the field theory vacuum.
The cPT Fock space vector carries all possible finite-time marginals of the process encoded in its finite-particle amplitudes, and therefore the connection to the discrete families is made apparent. Notably, we have not made any assumptions on the environment and can freely accommodate both finite- and infinite-dimensional environments. \par 

Although we arrived in our derivation at what we term a `process-canonical representation' (PCR) as a cMPS, we are able to state the conditions of physicality on such processes purely operationally for general Fock space vectors, they are thus not tied to this explicit representation. In a companion work~\cite{fuksaDilationTheoremContinuum2026} we connect these two perspectives, showing that any cPT can be arbitrarily approximated by a PCR.
This also raises the natural question of whether other structured classes of field theory vectors, such as unnormalised Gaussian field theory states or broader continuous tensor network states~\cite{tilloyContinuousTensorNetwork2019,PhysRevLett.110.100402,PhysRevD.104.096007,PhysRevD.105.045016,RevModPhys.93.045003}, could represent physically-relevant cPTs, and moreover how positivity and causality conditions manifest in them. More generally, one would hope to formalise the cPT framework further by bridging it completely to rigorous continuum notions of quantum stochastic processes~\cite{accardiQuantumStochasticProcesses1982,Nurdin_2021}.
Connected to more general quantum stochastic processes, it would be also interesting to explore notions of process complexity~\cite{aloisio-complexity} in the context of these state spaces~\cite{Chapman_2018_complexity}, where one has an in-built notion of boundedness instead of arbitrary computation.

When applied to the problem of quantum non-Markovianity, our framework ameliorates some conceptual gaps in the literature.
Specifically, we demonstrated on the one hand how rigorous discrete measures of non-Markovianity are not compatible with the continuum (\cref{prop:vanishing_NM_measu}) -- and then on the other hand showed how cPTs can be applied to construct such measures which are indeed well-behaved.
We furthermore showed that our framework can be used to define a hierarchy of multi-time processes which are in the process-canonical representation. This is based on the minimally-weighted functions that fully determine the process (its $p$-number). As natural next steps, it will be essential to demonstrate their utility and insight by computing both these measures and the $p$-number for physically relevant models. Furthermore, to develop better intuition about the $p$-number of a PCR, it would be valuable to formalise its definition to the case of generators on infinite environments and investigate whether there is a systematic connection between the $p$-number and the environment correlation functions for bosonic environments. Thus, again, aiming to strengthen our understanding of the connection between open systems and multi-time physics. 
Our framework thus allows for an information-theoretic treatment of quantum processes that is in line with physical assumptions, most importantly the assumption of bounded-energy operations. It will be an interesting avenue of future investigation to explore applications of this to prototypical information-theoretic tasks such as, for example, the discrimination of quantum processes~\cite{gutoski_strategy_norm}.

Lastly, process tensors have enjoyed a great deal of success in the respective problems of characterising non-Markovian noise on quantum devices~\cite{White-2025-Unifying-NM,white2022NonMarkovianQuantumProcess, Giarmatzi2025multitimequantum,PhysRevLett.134.010803,Li_2024_NMGST}, and simulating open quantum systems with tensor networks~\cite{strathearn_efficient_2018,Cygorek-2022,Link_2024} -- the latter of which can furthermore be an important ingredient for, e.g., finding optimal control protocols in quantum devices~\cite{glaserTrainingSchrodingersCat2015,Fux_2021}. We outlined how to extend some of these methods to the continuum, and believe that the combination of a continuum model with state-of-the-art tensor network methods~\cite{ganahlContinuousMatrixProduct2017, Haegeman_2017} will be useful to address some ill-posed properties of these methods. We have left open the direct investigation of our proposals to future work, but have made clear that there is good reason to expect its utility. 

It is our broad contention that bridging process tensors -- which have been an immensely useful information-theoretic tool -- with the continuum will offer fruitful insights into the structure of non-Markovian open quantum systems and open new avenues for their analysis, classification, and simulation. As such, this work carves out a framework which carries conceptual insights and a mathematical structure, but furthermore represents a first step towards these more practically-oriented goals.

\textbf{Note added.}  At a late stage in our project, we became aware of the concurrent work by Costa and Yang \cite{costaContinuousOperationsNonMarkovian2025}, which independently investigates the continuum limit of process tensors (process matrices) for infinite-dimensional quantum systems. Our treatment focuses on finite-dimensional systems and thus we find that 
the two works provide a complementary perspective.

\section*{Acknowledgments}
We thank Kavan Modi and Valentin Link for fruitful conversations. 
This work has been supported by the BMFTR (DAQC, MuniQC-Atoms, QuSol, Hybrid++, PasQuops), Clusters of Excellence (ML4Q, MATH+), the Munich Quantum Valley, Berlin Quantum, the Quantum Flagship (Millenion, Pasquans2), the DFG (CRC 183, SPP 2514), the European Research Council (DebuQC), and the Alexander-von-Humboldt Foundation.

\bibliography{references_merged}

@book{wisemanQuantumMeasurementControl2014,
	address = {Cambridge},
	edition = {First paperback edition},
	title = {Quantum measurement and control},
	isbn = {978-0-521-80442-4 978-1-107-42415-9},
	publisher = {Cambridge University Press},
	author = {Wiseman, Howard M. and Milburn, Gerard J.},
	year = {2014},
}

@misc{guilminParametersEstimationFitting2024a,
  title         = {Parameters estimation by fitting correlation functions of continuous quantum measurement},
  author        = {Guilmin, Pierre and Rouchon, Pierre and Tilloy, Antoine},
  year          = 2024,
  eprint        = {2410.11955},
  publisher     = {arXiv},
  urldate       = {2026-01-02},
  archiveprefix = {arXiv}
}

@book{gardinerStochasticMethodsHandbook2009,
  title = {Stochastic methods: A handbook for the natural and social sciences},
  shorttitle = {Stochastic Methods},
  author = {Gardiner, Crispin W.},
  year = 2009,
  series = {Springer Series in Synergetics},
  edition = {4th ed},
  number = {13},
  publisher = {Springer},
  address = {Berlin Heidelberg},
  isbn = {978-3-642-08962-6 978-3-540-70712-7},
  langid = {english}
}

@article{vankampenItoStratonovich1981,
  title = {It\^o versus {{Stratonovich}}},
  author = {{van Kampen}, N. G.},
  year = 1981,
  month = jan,
  journal = {Journal of Statistical Physics},
  volume = {24},
  number = {1},
  pages = {175--187},
  issn = {1572-9613},
  doi = {10.1007/BF01007642},
  urldate = {2026-09-17},
  langid = {english}
}

@misc{rosal2025deterministicequationsfeedbackcontrol,
      title={Deterministic Equations for Feedback Control of Open Quantum Systems}, 
      author={Alberto J. B. Rosal and Patrick P. Potts and Gabriel T. Landi},
      year={2025},
      eprint={2507.01934},
      archivePrefix={arXiv},
      optoptprimaryClass={quant-ph}
}

@book{gardiner2004quantum,
  title={Quantum noise: a handbook of Markovian and non-Markovian quantum stochastic methods with applications to quantum optics},
  author={Gardiner, Crispin and Zoller, Peter},
  year={2004},
  publisher={Springer Science \& Business Media}
}

@article{whiteNonMarkovianQuantumProcess2022,
  title = {Non-{{Markovian quantum process tomography}}},
  author = {White, G.A.L. and Pollock, F.A. and Hollenberg, L.C.L. and Modi, K. and Hill, C.D.},
  year = 2022,
  month = may,
  journal = {PRX Quantum},
  volume = {3},
  number = {2},
  pages = {020344},
  publisher = {American Physical Society},
  doi = {10.1103/PRXQuantum.3.020344},
  urldate = {2026-01-05}
}

@article{white2022NonMarkovianQuantumProcess,
  title = {Non-{{Markovian quantum process tomography}}},
  author = {White, Gregory A. L. and Pollock, Felix A. and Hollenberg, Lloyd C. L. and Modi, Kavan and Hill, Charles D.},
  year = 2022,
  month = may,
  journal = {PRX Quantum},
  volume = {3},
  number = {2},
optoptprimaryclass = {quant-ph},
  pages = {020344},
  doi = {10.1103/PRXQuantum.3.020344},
  urldate = {2024-05-23},
  langid = {english}
}

@article{strathearn_efficient_2018,
  author     = {Strathearn, A. and Kirton, P. and Kilda, D. and Keeling, J. and Lovett, B. W.},
  copyright  = {2018 The Author(s)},
  doi        = {10.1038/s41467-018-05617-3},
  issn       = {2041-1723},
  journal    = {Nature Comm.},
  month      = aug,
  number     = {1},
  pages      = {3322},
  title      = {Efficient non-{Markovian} quantum dynamics using time-evolving matrix product operators},
  url        = {https://www.nature.com/articles/s41467-018-05617-3},
  volume     = {9},
  year       = {2018}
}

@article{rivas2014quantum,
  title={Quantum non-Markovianity: characterization, quantification and detection},
doi={10.1088/0034-4885/77/9/094001},
  author={Rivas, {\'A}ngel and Huelga, Susana F. and Plenio, Martin B},
  journal={Rep. Prog. Phys.},
  volume={77},
  number={9},
  pages={094001},
  year={2014},
  publisher={IOP Publishing}
}

@Article{Milz-2021-GMEinTime,
	title={Genuine Multipartite Entanglement in Time},
	author={Simon Milz and Cornelia Spee and Zhen-Peng Xu and Felix A. Pollock and Kavan Modi and Otfried Gühne},
	journal={SciPost Phys.},
	volume={10},
	issue={6},
	pages={141},
	year={2021},
	publisher={SciPost},
	doi={10.21468/SciPostPhys.10.6.141},
	url={https://scipost.org/10.21468/SciPostPhys.10.6.141}
}

@article{Cygorek-2022,
  author    = {Cygorek, Moritz and Cosacchi, Michael and Vagov, Alexei and Axt, Vollrath Martin and Lovett, Brendon W. and Keeling, Jonathan and Gauger, Erik M.},
  doi       = {10.1038/s41567-022-01544-9},
  pages=662,
volume=18,
  journal   = {Nature Phys.},
  publisher = {Springer US},
  title     = {{Simulation of open quantum systems by automated compression of arbitrary environments}},
  year      = {2022}
}

@article{Gribben2022usingenvironmentto,
  doi       = {10.22331/q-2022-10-25-847},
  url       = {https://doi.org/10.22331/q-2022-10-25-847},
  title     = {Using the {e}nvironment to {u}nderstand non-{M}arkovian {o}pen {q}uantum {s}ystems},
  author    = {Gribben, Dominic and Strathearn, Aidan and Fux, Gerald E. and Kirton, Peter and Lovett, Brendon W.},
  journal   = {{Quantum}},
  issn      = {2521-327X},
  publisher = {{Verein zur F{\"{o}}rderung des Open Access Publizierens in den Quantenwissenschaften}},
  volume    = {6},
  pages     = {847},
  month     = oct,
  year      = {2022}
}

@article{BruknerNoCausalOrder,
DOI={10.1038/ncomms2076},
author={O. Oreshkov and F. Costa and C. Brukner},
title={Quantum correlations with no causal order}, journal={Nature Comm.},
volume=3, pages=1092, year=2012}

@article{cotlerSuperdensityOperatorsSpacetime2018,
  title = {Superdensity Operators for Spacetime Quantum Mechanics},
  author = {Cotler, Jordan and Jian, Chao-Ming and Qi, Xiao-Liang and Wilczek, Frank},
  year = 2018,
  month = sep,
  journal = {JHEP},
  volume = {2018},
  number = {9},
  pages = {93},
  issn = {1029-8479},
  doi = {10.1007/JHEP09(2018)093},
  urldate = {2026-01-22},
  langid = {english}
}

@article{PhysRevLett.101.060401,
  title = {Quantum Circuit Architecture},
  author = {Chiribella, G. and D'Ariano, G. M. and Perinotti, P.},
  journal = {Phys. Rev. Lett.},
  volume = {101},
  issue = {6},
  pages = {060401},
  numpages = {4},
  year = {2008},
  month = {Aug},
  publisher = {American Physical Society},
  doi = {10.1103/PhysRevLett.101.060401},
  url = {https://link.aps.org/doi/10.1103/PhysRevLett.101.060401}
}

@article{Hardy_2012,
   title={The operator tensor formulation of quantum theory},
   volume={370},
   url={http://dx.doi.org/10.1098/rsta.2011.0326},
   DOI={10.1098/rsta.2011.0326},
   number={1971},
   journal={Phil. Trans. Roy. Soc. A},
   publisher={The Royal Society},
   author={Hardy, Lucien},
   year={2012},
   month=jul, pages={3385–3417} }

@article{osborneHolographicQuantumStates2010,
  title = {Holographic quantum states},
  author = {Osborne, Tobias J. and Eisert, Jens and Verstraete, Frank},
  date = {2010-12-20},
  year = {2010},
  journaltitle = {Phys. Rev. Lett.},
  journal = {Phys. Rev. Lett.},
  volume = {105},
  pages = {260401},
  issn = {0031-9007, 1079-7114},
  doi = {10.1103/PhysRevLett.105.260401},
  url = {http://arxiv.org/abs/1005.1268},
  urldate = {2025-01-22}
}

@article{Dowling-2024-Tree-PTs,
  title = {Capturing Long-Range Memory Structures with Tree-Geometry Process Tensors},
  author = {Dowling, Neil and Modi, Kavan and Mu\~noz, Roberto N. and Singh, Sukhbinder and White, Gregory A. L.},
  journal = {Phys. Rev. X},
  volume = {14},
  issue = {4},
  pages = {041018},
  numpages = {32},
  year = {2024},
  month = {Oct},
  publisher = {American Physical Society},
  doi = {10.1103/PhysRevX.14.041018},
  url = {https://link.aps.org/doi/10.1103/PhysRevX.14.041018}
}

@article{White-2025-Unifying-NM,
  title = {Unifying non-Markovian characterization with an efficient and self-consistent framework},
  author = {White, G. A. L. and Jurcevic, P. and Hill, C. D. and Modi, K.},
  journal = {Phys. Rev. X},
  volume = {15},
  issue = {2},
  pages = {021047},
  numpages = {43},
  year = {2025},
  month = {May},
  publisher = {American Physical Society},
  doi = {10.1103/PhysRevX.15.021047},
  url = {https://link.aps.org/doi/10.1103/PhysRevX.15.021047}
}

@article{milzQuantumStochasticProcesses2021,
  title = {Quantum {{stochastic processes}} and {{Quantum}} non-{{Markovian phenomena}}},
  author = {Milz, Simon and Modi, Kavan},
  year = 2021,
  month = jul,
  journal = {PRX Quantum},
  volume = {2},
  number = {3},
  pages = {030201},
  issn = {2691-3399},
  doi = {10.1103/PRXQuantum.2.030201},
  urldate = {2024-04-24},
  langid = {english}
}

@article{Milz_2021_QST_review,
   title={Quantum Stochastic Processes and Quantum non-Markovian Phenomena},
   volume={2},
   pages=030201,
   url={http://dx.doi.org/10.1103/PRXQuantum.2.030201},
   DOI={10.1103/prxquantum.2.030201},
   journal={PRX Quantum},
   publisher={American Physical Society (APS)},
   author={Milz, Simon and Modi, Kavan},
   year={2021},
   month=jul }

@article{accardiQuantumStochasticProcesses1982,
  title = {Quantum {{stochastic processes}}},
  author = {Accardi, Luigi and Frigerio, Alberto and Lewis, John T.},
  year = 1982,
  month = apr,
  journal = {Publ. Res. Inst. 
             Math. Sc.},
  volume = {18},
  pages = {97--133},
  issn = {0034-5318},
  doi = {10.2977/prims/1195184017},
  urldate = {2025-02-19},
  langid = {english}
}

@inproceedings{Nurdin_2021,
   title={From the {Heisenberg} to the {Schrödinger} picture: Quantum stochastic processes and process tensors},
   url={http://dx.doi.org/10.1109/CDC45484.2021.9683765},
   DOI={10.1109/cdc45484.2021.9683765},
   booktitle={2021 60th IEEE Conference on Decision and Control (CDC)},
   publisher={IEEE},
   author={Nurdin, Hendra I. and Gough, John},
   year={2021},
   month=dec, pages={4164–4169} 
}

@article{Pollock2018-Process-Tensors,
author = {Pollock, Felix A. and Rodr{\'{i}}guez-Rosario, C{\'{e}}sar and Frauenheim, Thomas and Paternostro, Mauro and Modi, Kavan},
doi = {10.1103/PhysRevA.97.012127},
url = {https://www.dx.doi.org/10.1103/PhysRevA.97.012127},
journal = {Phys. Rev. A},
number = {1},
pages = {012127},
title = {{Non-Markovian quantum processes: Complete framework and efficient characterization}},
volume = {97},
year = {2018}
}

@article{whiteDemonstrationNonMarkovianProcess2020,
  title = {Demonstration of non-{{Markovian}} process characterisation and control on a quantum processor},
  author = {White, G. A. L. and Hill, C. D. and Pollock, F. A. and Hollenberg, L. C. L. and Modi, K.},
  year = 2020,
  month = dec,
  journal = {Nature Communications},
  volume = {11},
  number = {1},
  pages = {6301},
  publisher = {Nature Publishing Group},
  issn = {2041-1723},
  doi = {10.1038/s41467-020-20113-3},
  urldate = {2026-01-05},
  copyright = {2020 The Author(s)},
  langid = {english}
}

@misc{tjoaContinuousMatrixProduct2025,
  title = {{Continuous matrix product operators for quantum fields}},
  author = {Tjoa, Erickson and Cirac, J. Ignacio},
  year = 2025,
  eprint = {2511.04545},
  optoptprimaryclass = {quant-ph},
  publisher = {arXiv},
  archiveprefix = {arXiv},
  langid = {english}
}

@article{verstraeteContinuousMatrixProduct2010,
  title         = {Continuous {{matrix product states}} for {{quantum fields}}},
  author        = {Verstraete, F. and Cirac, J. I.},
  journal = {Phys. Rev. Lett.},
  volume = {104},
  issue = {19},
  pages = {190405},
  numpages = {4},
  year = {2010},
  month = {May},
  publisher = {American Physical Society},
  doi = {10.1103/PhysRevLett.104.190405},
}

@article{tilloyContinuousTensorNetwork2019,
  title = {Continuous {{Tensor Network States}} for {{Quantum Fields}}},
  author = {Tilloy, Antoine and Cirac, J. Ignacio},
  year = 2019,
  month = may,
  journal = {Physical Review X},
  volume = {9},
  number = {2},
  pages = {021040},
  issn = {2160-3308},
  doi = {10.1103/PhysRevX.9.021040},
  urldate = {2025-11-19},
  langid = {english}
}

@article{steffensExperimentalQuantumfieldTomography2015,
  title = {Towards Experimental Quantum-Field Tomography with Ultracold Atoms},
  author = {Steffens, A. and Friesdorf, M. and Langen, T. and Rauer, B. and Schweigler, T. and H{\"u}bener, R. and Schmiedmayer, J. and Riofr{\'i}o, C. A. and Eisert, J.},
  year = 2015,
  month = jul,
  journal = {Nature Communications},
  volume = {6},
  number = {1},
  pages = {7663},
  publisher = {Nature Publishing Group},
  issn = {2041-1723},
  doi = {10.1038/ncomms8663},
  urldate = {2023-11-24},
  copyright = {2015 The Author(s)},
  langid = {english}
}

@article{haegemanCalculusContinuousMatrix2013,
  title = {Calculus of Continuous Matrix Product States},
  author = {Haegeman, Jutho and Cirac, J. Ignacio and Osborne, Tobias J. and Verstraete, Frank},
  year = 2013,
  month = aug,
  journal = {Physical Review B},
  volume = {88},
  number = {8},
  pages = {085118},
  issn = {1098-0121, 1550-235X},
  doi = {10.1103/PhysRevB.88.085118},
  urldate = {2022-11-10},
  langid = {english}
}

@article{milzKolmogorovExtensionTheorem2020,
  title = {Kolmogorov Extension Theorem for (Quantum) Causal Modelling and General Probabilistic Theories},
  author = {Milz, Simon and Sakuldee, Fattah and Pollock, Felix A. and Modi, Kavan},
  year = 2020,
  month = apr,
  journal = {Quantum},
  volume = {4},
  pages = {255},
  publisher = {Verein zur F\"orderung des Open Access Publizierens in den Quantenwissenschaften},
  doi = {10.22331/q-2020-04-20-255},
  urldate = {2024-12-04},
  langid = {british}
}

@article{Burgath_2015_TD_dilations,
  title = {Quantum simulations of dissipative dynamics: Time dependence instead of size},
  author = {Dive, Benjamin and Mintert, Florian and Burgarth, Daniel},
  journal = {Phys. Rev. A},
  volume = {92},
  issue = {3},
  pages = {032111},
  numpages = {7},
  year = {2015},
  month = {Sep},
  publisher = {American Physical Society},
  doi = {10.1103/PhysRevA.92.032111},
  url = {https://link.aps.org/doi/10.1103/PhysRevA.92.032111}
}

@misc{costaContinuousOperationsNonMarkovian2025,
  title = {Continuous operations on non-{{Markovian}} processes},
  author = {Costa, Fabio and Yang, Jing},
  year = 2025,
  month = dec,
  number = {arXiv:2512.05884},
  eprint = {2512.05884},
  optprimaryclass = {quant-ph},
  publisher = {arXiv},
  doi = {10.48550/arXiv.2512.05884},
  urldate = {2025-12-09},
  archiveprefix = {arXiv},
  langid = {english}
}

@article{Burgarth_2022_dilations,
   title={Control of Quantum Noise: On the Role of Dilations},
   volume={24},
   ISSN={1424-0661},
   url={http://dx.doi.org/10.1007/s00023-022-01211-y},
   DOI={10.1007/s00023-022-01211-y},
   number={1},
   journal={Annales Henri Poincaré},
   publisher={Springer Science and Business Media LLC},
   author={Burgarth, Daniel and Facchi, Paolo and Hillier, Robin},
   year={2022}, pages={325–347} }

@article{aloisio-complexity,
title={On the sampling complexity of open quantum systems},
author={Aloisio, Isobel A and White, Gregory A L and Hill, Charles D and Modi, Kavan},
eprint = {2209.10870},
year={2022},
doi={10.1103/PRXQuantum.4.020310},
journal={PRX Quantum},
volume={4},
pages={020310},
url={https://journals.aps.org/prxquantum/abstract/10.1103/PRXQuantum.4.020310}}

@article{wangResourceTheoryAsymmetric2019,
  title = {Resource Theory of Asymmetric Distinguishability for Quantum Channels},
  author = {Wang, Xin and Wilde, Mark M.},
  year = 2019,
  month = dec,
  journal = {Phys. Rev. Res.},
  volume = {1},
  number = {3},
  pages = {033169},
  publisher = {American Physical Society},
  doi = {10.1103/PhysRevResearch.1.033169},
  urldate = {2026-01-08}
}

@article{zambonProcessTensorDistinguishability2024,
  title = {Process Tensor Distinguishability Measures},
  author = {Zambon, Guilherme},
  year = 2024,
  month = oct,
  journal = {Phys. Rev. A},
  volume = {110},
  number = {4},
  pages = {042210},
  issn = {2469-9926, 2469-9934},
  doi = {10.1103/PhysRevA.110.042210},
  urldate = {2025-11-13},
  langid = {english}
}

@misc{ganahlContinuousMatrixProduct2017,
  title = {Continuous {{matrix product states}} for {{inhomogeneous quantum field theories}}: A {{basis-spline approach}}},
  shorttitle = {Continuous {{Matrix Product States}} for {{Inhomogeneous Quantum Field Theories}}},
  author = {Ganahl, Martin},
  year = 2017,
  month = dec,
  number = {arXiv:1712.01260},
  eprint = {1712.01260},
  publisher = {arXiv},
  urldate = {2025-02-13},
  archiveprefix = {arXiv}
}

@Article{AreaReview,
  title                     = {Area laws for the entanglement entropy},
  Author                   = {J. Eisert and M. Cramer and M. B. Plenio},
  doi={10.1103/RevModPhys.82.277},
  Journal                  = {Rev. Mod. Phys.},
  Year                     = {2010},
  Pages                    = {277},
  Volume                   = {82}
}

@article{BenchmarkingReview,
title={Quantum certification and benchmarking},
doi={10.1038/s42254-020-0186-4},
Author={J. Eisert and D. Hangleiter and N. Walk and I. Roth and D. Markham and R. Parekh and U. Chabaud and E. Kashefi},
journal={Nature Rev. Phys.},
volume=2, pages={382-390},
year={2020}
}

@article{tuybensVariationalOptimizationContinuous2022,
  title = {Variational optimization of continuous matrix product states},
  author = {Tuybens, Beno{\^i}t and Nardis, Jacopo De and Haegeman, Jutho and Verstraete, Frank},
  year = 2022,
  journal = {Phys. Rev. Lett.},
  volume = {128},
  pages = {020501},
  issn = {0031-9007, 1079-7114},
  doi = {10.1103/PhysRevLett.128.020501},
  urldate = {2025-02-13},
  archiveprefix = {arXiv},
  langid = {english}
}

@article{janzingQuantumControlAccess2002,
  title = {Quantum Control without Access to the Controlling Interaction},
  author = {Janzing, D. and Armknecht, F. and Zeier, R. and Beth, {\relax Th}.},
  year = 2002,
  month = jan,
  journal = {Phys. Rev. A},
  volume = {65},
  number = {2},
  pages = {022104},
  publisher = {American Physical Society},
  doi = {10.1103/PhysRevA.65.022104},
  urldate = {2025-12-30}
}

@article{whiteWhatCanUnitary2025,
  title = {What Can Unitary Sequences Tell Us about Multi-Time Physics?},
  author = {White, Gregory A. L. and Pollock, Felix A. and Hollenberg, Lloyd C. L. and Hill, Charles D. and Modi, Kavan},
  year = 2025,
  month = apr,
  journal = {Quantum},
  volume = {9},
  pages = {1695},
  publisher = {Verein zur F\"orderung des Open Access Publizierens in den Quantenwissenschaften},
  doi = {10.22331/q-2025-04-08-1695},
  urldate = {2026-01-05},
  langid = {british}
}

@misc{meleOptimalLearningQuantum2025,
  title = {Optimal Learning of Quantum Channels in Diamond Distance},
  author = {Mele, Antonio Anna and Bittel, Lennart},
  year = 2025,
  eprint = {2512.10214},
  optoptprimaryclass = {quant-ph},
  publisher = {arXiv},
  urldate = {2025-12-18},
  archiveprefix = {arXiv},
  langid = {english}
}

@article{haahSampleOptimalTomographyQuantum2017,
  title = {Sample-{{Optimal Tomography}} of {{Quantum States}}},
  author = {Haah, Jeongwan and Harrow, Aram W. and Ji, Zhengfeng and Wu, Xiaodi and Yu, Nengkun},
  year = 2017,
  month = sep,
  journal = {IEEE Trans. Inf. Th.},
  volume = {63},
  number = {9},
  pages = {5628--5641},
  issn = {1557-9654},
  doi = {10.1109/TIT.2017.2719044},
  urldate = {2025-01-09}
}

@article{parkTensorNetworkInfluence2024,
  title = {Tensor Network Influence Functionals in the Continuous-Time Limit: {{Connections}} to Quantum Embedding, Bath Discretization, and Higher-Order Time Propagation},
  shorttitle = {Tensor Network Influence Functionals in the Continuous-Time Limit},
  author = {Park, Gunhee and Ng, Nathan and Reichman, David R. and Chan, Garnet Kin-Lic},
  date = {2024-07-02},
  year = {2024},
  journal = {Phys. Rev. B},
  shortjournal = {Phys. Rev. B},
  volume = {110},
  number = {4},
  pages = {045104},
  issn = {2469-9950, 2469-9969},
  doi = {10.1103/PhysRevB.110.045104},
  url = {https://link.aps.org/doi/10.1103/PhysRevB.110.045104},
  urldate = {2024-12-10},
  langid = {english}
}

@article{sonnerInfluenceFunctionalManybody2021,
  title = {Influence Functional of Many-Body Systems: Temporal Entanglement and Matrix-Product State Representation},
  shorttitle = {Influence Functional of Many-Body Systems},
  author = {Sonner, Michael and Lerose, Alessio and Abanin, Dmitry A.},
  date = {2021-12},
  year = {2021},
  journal = {Ann. Phys.},
  shortjournal = {Ann. Phys.},
  volume = {435},
  eprint = {2103.13741},
  eprinttype = {arXiv},
  eprintclass = {quant-ph},
  pages = {168677},
  issn = {00034916},
  doi = {10.1016/j.aop.2021.168677},
  url = {http://arxiv.org/abs/2103.13741},
  urldate = {2024-12-10},
  langid = {english}
}

@article{hangleiterRobustlyLearningHamiltonian2024a,
  title = {Robustly Learning the {{Hamiltonian}} Dynamics of a Superconducting Quantum Processor},
  author = {Hangleiter, Dominik and Roth, Ingo and Fuksa, Jonáš and Eisert, Jens and Roushan, Pedram},
  date = {2024-11-06},
  year = {2024},
  journal = {Nature Comm.},
  shortjournal = {Nat Commun},
  volume = {15},
  number = {1},
  pages = {9595},
  publisher = {Nature Publishing Group},
  issn = {2041-1723},
  doi = {10.1038/s41467-024-52629-3},
  url = {https://www.nature.com/articles/s41467-024-52629-3},
  urldate = {2025-03-07},
  langid = {english}
}

@article{wisemanCompleteParameterizationInvariance2001,
	title = {Complete parameterization, and invariance, of diffusive quantum trajectories for {Markovian} open systems},
	volume = {268},
	issn = {03010104},
	url = {http://arxiv.org/abs/quant-ph/0012016},
	doi = {10.1016/S0301-0104(01)00296-8},
	number = {1-3},
	urldate = {2025-03-19},
	journal = {Chem. Phys.},
	author = {Wiseman, H. M. and Diosi, L.},
	month = jun,
	year = {2001},
	pages = {91--104},
}

@article{hubenerWicksTheoremMatrix2013,
	title = {Wick's theorem for matrix product states},
	volume = {110},
	issn = {0031-9007, 1079-7114},
	url = {http://arxiv.org/abs/1207.6537},
	doi = {10.1103/PhysRevLett.110.040401},
	urldate = {2025-08-20},
	journal = {Phys. Rev. Lett.},
	author = {Hübener, R. and Mari, A. and Eisert, J.},
	year = {2013},
	pages = {040401},
}

@article{PhysRevA.80.022339,
  title = {Theoretical framework for quantum networks},
  author = {Chiribella, Giulio and D'Ariano, Giacomo Mauro and Perinotti, Paolo},
  journal = {Phys. Rev. A},
  volume = {80},
  issue = {2},
  pages = {022339},
  numpages = {20},
  year = {2009},
  month = {Aug},
  publisher = {American Physical Society},
  doi = {10.1103/PhysRevA.80.022339},
  url = {https://link.aps.org/doi/10.1103/PhysRevA.80.022339}
}

@article{nielsen2021GateSetTomography,
  title = {Gate {{Set Tomography}}},
  author = {Nielsen, Erik and Gamble, John King and Rudinger, Kenneth and Scholten, Travis and Young, Kevin and {Blume-Kohout}, Robin},
  year = 2021,
  month = oct,
  journal = {Quantum},
  volume = {5},
  pages = {557},
  doi = {10.22331/q-2021-10-05-557},
  urldate = {2024-05-23},
  langid = {english}
}

@article{briegerCompressiveGateSet2023,
  title = {Compressive Gate Set Tomography},
  author = {Brieger, Raphael and Roth, Ingo and Kliesch, Martin},
  year = 2023,
  journal = {PRX Quantum},
  volume = {4},
  number = {1},
  optprimaryclass = {quant-ph},
  pages = {010325},
  issn = {2691-3399},
  doi = {10.1103/PRXQuantum.4.010325},
  urldate = {2025-04-16},
  langid = {english}
}

@article{Fux_2021,
   title={Efficient Exploration of Hamiltonian Parameter Space for Optimal Control of Non-Markovian Open Quantum Systems},
   volume={126},
   DOI={10.1103/physrevlett.126.200401},
   journal={Phys. Rev. Lett.},
   pages=200401,
   publisher={American Physical Society (APS)},
   author={Fux, Gerald E. and Butler, Eoin P. and Eastham, Paul R. and Lovett, Brendon W. and Keeling, Jonathan},
   year={2021},
   month=may }

@article{PhysRevLett.126.200401,
   title={Efficient Exploration of Hamiltonian Parameter Space for Optimal Control of Non-Markovian Open Quantum Systems},
   volume={126},
   ISSN={1079-7114},
   url={http://dx.doi.org/10.1103/PhysRevLett.126.200401},
   DOI={10.1103/physrevlett.126.200401},
   number={20},
   journal={Phys. Rev. Lett.},
   publisher={American Physical Society (APS)},
   author={Fux, Gerald E. and Butler, Eoin P. and Eastham, Paul R. and Lovett, Brendon W. and Keeling, Jonathan},
   year={2021}}

@article{PhysRevLett.110.100402,
  title = {Entanglement Renormalization for Quantum Fields in Real Space},
  author = {Haegeman, Jutho and Osborne, Tobias J. and Verschelde, Henri and Verstraete, Frank},
  journal = {Phys. Rev. Lett.},
  volume = {110},
  issue = {10},
  pages = {100402},
  numpages = {5},
  year = {2013},
  month = {Mar},
  publisher = {American Physical Society},
  doi = {10.1103/PhysRevLett.110.100402} 
}

@article{PhysRevD.104.096007,
  title = {Relativistic continuous matrix product states for quantum fields without cutoff},
  author = {Tilloy, Antoine},
  journal = {Phys. Rev. D},
  volume = {104},
  issue = {9},
  pages = {096007},
  numpages = {14},
  year = {2021},
  month = {Nov},
  publisher = {American Physical Society},
  doi = {10.1103/PhysRevD.104.096007}
}

@article{PhysRevD.105.045016,
  title = {Approximating relativistic quantum field theories with continuous tensor networks},
  author = {Shachar, Tom and Zohar, Erez},
  journal = {Phys. Rev. D},
  volume = {105},
  issue = {4},
  pages = {045016},
  numpages = {14},
  year = {2022},
  month = {Feb},
  publisher = {American Physical Society},
  doi = {10.1103/PhysRevD.105.045016}
}

@article{RevModPhys.93.045003,
  title = {Matrix product states and projected entangled pair states: Concepts, symmetries, theorems},
  author = {Cirac, J. Ignacio and P\'erez-Garc\'{\i}a, David and Schuch, Norbert and Verstraete, Frank},
  journal = {Rev. Mod. Phys.},
  volume = {93},
  issue = {4},
  pages = {045003},
  numpages = {65},
  year = {2021},
  publisher = {American Physical Society},
  doi = {10.1103/RevModPhys.93.045003}
}

@book{serafiniQuantumContinuousVariables2017,
  title = {Quantum Continuous Variables: A Primer of Theoretical Methods},
  shorttitle = {Quantum Continuous Variables},
  author = {Serafini, Alessio},
  year = 2017,
  publisher = {CRC Press, Taylor \& Francis Group, CRC Press is an imprint of the Taylor \& Francis Group, an informa business},
  address = {Boca Raton},
  isbn = {978-1-4822-4634-6},
  lccn = {QA76.889 .S47 2017}
}

@misc{tilloy2021notes,
  title        = {Notes on Continuous Tensor Network States},
  author       = {Tilloy, Antoine},
  year         = {2021},
  howpublished = {\url{https://atilloy.com/wp-content/uploads/2021/09/lecture_tensor21.pdf}},
  note         = {Lecture notes}
}

@article{le2008entanglement,
  title={Entanglement entropy, decoherence, and quantum phase transitions of a dissipative two-level system},
  author={Le Hur, Karyn},
  journal={Ann. Phys.},
  volume={323},
  number={9},
  pages={2208--2240},
  year={2008},
  publisher={Elsevier},
  doi = {https://doi.org/10.1016/j.aop.2007.12.003}
}

@Inbook{LeHurrBook2010,
author="Le Hur, Karyn",
editor="Carr, Lincoln D.",
title="Quantum Phase Transitions in Spin-Boson Systems: Dissipation and Light Phenomena",
bookTitle="Understanding quantum phase transitions",
year="2010",
publisher="CRC Press",
address="New York",
pages="217-240",
isbn="9781439802519",
doi="10.1201/b10273"
}

@article{Link_2024,
   title={Open Quantum System Dynamics from Infinite Tensor Network Contraction},
   volume={132},
   pages=200403,
   DOI={10.1103/physrevlett.132.200403},
   journal={Phys. Rev. Lett.},
   publisher={American Physical Society (APS)},
   author={Link, Valentin and Tu, Hong-Hao and Strunz, Walter T.},
   year={2024},
   month=may }

@article{Pollock_2018_measure,
   title={Operational Markov Condition for Quantum Processes},
   volume={120},
   DOI={10.1103/physrevlett.120.040405},
   pages=040405,
   journal={Phys. Rev. Lett.},
   publisher={American Physical Society (APS)},
   author={Pollock, Felix A. and Rodríguez-Rosario, César and Frauenheim, Thomas and Paternostro, Mauro and Modi, Kavan},
   year={2018}}

@misc{khatri2024principlesquantumcommunicationtheory,
      title={Principles of Quantum Communication Theory: A Modern Approach}, 
      author={Sumeet Khatri and Mark M. Wilde},
      year={2024},
      eprint={2011.04672},
      archivePrefix={arXiv}
}

@article{Giarmatzi2025multitimequantum,
  doi = {10.22331/q-2025-12-18-1952},
  url = {https://doi.org/10.22331/q-2025-12-18-1952},
  title = {Multi-time quantum process tomography on a superconducting qubit},
  author = {Giarmatzi, Christina and Jones, Tyler and Gilchrist, Alexei and Pakkiam, Prasanna and Fedorov, Arkady and Costa, Fabio},
  journal = {{Quantum}},
  issn = {2521-327X},
  publisher = {{Verein zur F{\"{o}}rderung des Open Access Publizierens in den Quantenwissenschaften}},
  volume = {9},
  pages = {1952},
  month = dec,
  year = {2025}
}

@article{PhysRevLett.103.210401,
  title = {Measure for the Degree of Non-Markovian Behavior of Quantum Processes in Open Systems},
  author = {Breuer, Heinz-Peter and Laine, Elsi-Mari and Piilo, Jyrki},
  journal = {Phys. Rev. Lett.},
  volume = {103},
  issue = {21},
  pages = {210401},
  numpages = {4},
  year = {2009},
  month = {Nov},
  publisher = {American Physical Society},
  doi = {10.1103/PhysRevLett.103.210401},
  url = {https://link.aps.org/doi/10.1103/PhysRevLett.103.210401}
}

@article{PhysRevA.83.052128,
  title = {Measures of non-Markovianity: Divisibility versus backflow of information},
  author = {Chru\ifmmode \acute{s}\else \'{s}\fi{}ci\ifmmode \acute{n}\else \'{n}\fi{}ski, Dariusz and Kossakowski, Andrzej and Rivas, \'Angel},
  journal = {Phys. Rev. A},
  volume = {83},
  issue = {5},
  pages = {052128},
  numpages = {6},
  year = {2011},
  month = {May},
  publisher = {American Physical Society},
  doi = {10.1103/PhysRevA.83.052128},
  url = {https://link.aps.org/doi/10.1103/PhysRevA.83.052128}
}

@article{Shrapnel_2018_supervised,
   title={Quantum Markovianity as a supervised learning task},
   volume={16},
   ISSN={1793-6918},
   url={http://dx.doi.org/10.1142/S0219749918400105},
   DOI={10.1142/s0219749918400105},
   number={08},
   journal={Int. J.Quant. Inf.},
   publisher={World Scientific Pub Co Pte Ltd},
   author={Shrapnel, Sally and Costa, Fabio and Milburn, Gerard},
   year={2018},
   month=dec, pages={1840010} }

@article{PRXQuantum.2.040351,
  title = {Randomized Benchmarking for Non-Markovian Noise},
  author = {Figueroa-Romero, Pedro and Modi, Kavan and Harris, Robert J. and Stace, Thomas M. and Hsieh, Min-Hsiu},
  journal = {PRX Quantum},
  volume = {2},
  issue = {4},
  pages = {040351},
  numpages = {28},
  year = {2021},
  month = {Dec},
  publisher = {American Physical Society},
  doi = {10.1103/PRXQuantum.2.040351},
  url = {https://link.aps.org/doi/10.1103/PRXQuantum.2.040351}
}

@article{PhysRevA.104.022432,
  title = {Experimental characterization of a non-Markovian quantum process},
  author = {Goswami, K. and Giarmatzi, C. and Monterola, C. and Shrapnel, S. and Romero, J. and Costa, F.},
  journal = {Phys. Rev. A},
  volume = {104},
  issue = {2},
  pages = {022432},
  numpages = {7},
  year = {2021},
  month = {Aug},
  publisher = {American Physical Society},
  doi = {10.1103/PhysRevA.104.022432},
  url = {https://link.aps.org/doi/10.1103/PhysRevA.104.022432}
}

@article{ObservationNonMarkovian,
  title                     = {Observation of non-Markovian micro-mechanical Brownian motion},
  Author                   = {S. Groeblacher and A. Trubarov and N. Prigge and M. Aspelmeyer and J. Eisert},
journal={Nature Comm.}, volume=6, pages=7606, year=2015, doi={10.1038/ncomms8606}
}

@misc{xiang2021quantifynonmarkovianprocessintervening,
      title={{Quantify the Non-Markovian process with intervening projections in a superconducting processor}}, 
      author={Liang Xiang and Zhiwen Zong and Ze Zhan and Ying Fei and Chongxin Run and Yaozu Wu and Wenyan Jin and Zhilong Jia and Peng Duan and Jianlan Wu and Yi Yin and Guoping Guo},
      year={2021},
      eprint={2105.03333},
      archivePrefix={arXiv}
}

@article{PhysRevLett.134.010803,
  title = {Quantum Comb Tomography via Learning Isometries on Stiefel Manifold},
  author = {Li, Ze-Tong and He, Xin-Lin and Zheng, Cong-Cong and Dong, Yu-Qian and Luan, Tian and Yu, Xu-Tao and Zhang, Zai-Chen},
  journal = {Phys. Rev. Lett.},
  volume = {134},
  issue = {1},
  pages = {010803},
  numpages = {7},
  year = {2025},
  month = {Jan},
  publisher = {American Physical Society},
  doi = {10.1103/PhysRevLett.134.010803},
  url = {https://link.aps.org/doi/10.1103/PhysRevLett.134.010803}
}

@article{Li_2024_NMGST,
   title={Non-Markovian quantum gate set tomography},
   volume={9},
   DOI={10.1088/2058-9565/ad3d80},
   number={3},
   journal={Quant. Sc. Tech.},
   publisher={IOP Publishing},
   author={Li, Ze-Tong and Zheng, Cong-Cong and Meng, Fan-Xu and Zeng, Han and Luan, Tian and Zhang, Zai-Chen and Yu, Xu-Tao},
   year={2024},
   month=may, pages={035027} }

@article{Milz_2019_div,
   title={Completely Positive Divisibility Does Not Mean Markovianity},
   volume={123},
   url={http://dx.doi.org/10.1103/PhysRevLett.123.040401},
   DOI={10.1103/physrevlett.123.040401},
   pages=040401,
   journal={Phys. Rev. Lett.},
   publisher={American Physical Society (APS)},
   author={Milz, Simon and Kim, M. S. and Pollock, Felix A. and Modi, Kavan},
   year={2019},
   month=jul }

@article{Arenz_2015_decoupling,
   title={Distinguishing decoherence from alternative quantum theories by dynamical decoupling},
   volume={92},
   pages=022102,
   url={http://dx.doi.org/10.1103/PhysRevA.92.022102},
   DOI={10.1103/physreva.92.022102},
   number={2},
   journal={Phys. Rev. A},
   publisher={American Physical Society (APS)},
   author={Arenz, Christian and Hillier, Robin and Fraas, Martin and Burgarth, Daniel},
   year={2015},
   month=aug }

@article{Milz_2018_restricted,
   title={Reconstructing non-Markovian quantum dynamics with limited control},
   volume={98},
pages=012108,
   DOI={10.1103/physreva.98.012108},
   journal={Phys. Rev. A},
   publisher={American Physical Society (APS)},
   author={Milz, Simon and Pollock, Felix A. and Modi, Kavan},
   year={2018},
   month=jul }

@inproceedings{Gutoski_2007, series={STOC07},
   title={Toward a general theory of quantum games},
   url={http://dx.doi.org/10.1145/1250790.1250873},
   DOI={10.1145/1250790.1250873},
   booktitle={Proceedings of the thirty-ninth annual ACM symposium on Theory of computing},
   publisher={ACM},
   author={Gutoski, Gus and Watrous, John},
   year={2007},
   month=jun, pages={565–574},
   collection={STOC07} }

@misc{fuksaDilationTheoremContinuum2026,
  title = {Dilation Theorem for Continuum Quantum Stochastic Processes},
  author = {Fuksa, Jon{\'a}{\v s} and Wassner, Clara and Eisert, Jens and White, Gregory A. L.},
  year = 2026,
  month = sep,
  number = {arXiv:2609.35403},
  eprint = {2609.35403},
  primaryclass = {quant-ph},
  publisher = {arXiv},
  doi = {10.48550/arXiv.2609.35403},
  urldate = {2026-09-29},
  archiveprefix = {arXiv},
  pages = {2609.35403},
  journaltitle = {}
}

@article{Dowling_2024_chaos,
   title={Operational Metric for Quantum Chaos and the Corresponding Spatiotemporal-Entanglement Structure},
   volume={5},
   pages=010314,
   DOI={10.1103/prxquantum.5.010314},
   number={1},
   journal={PRX Quantum},
   publisher={American Physical Society (APS)},
   author={Dowling, Neil and Modi, Kavan},
   year={2024},
   month=feb }

@misc{tingley2023notesfockspace,
      title={Notes on Fock space}, 
      author={Peter Tingley},
      year={2023},
      eprint={2211.12463},
      archivePrefix={arXiv},
url={https://arxiv.org/abs/2211.12463}, 
}

@misc{wolf2012quantum,
  title={Quantum channels and operations-guided tour},
  author={Wolf, Michael M},
note={lecture notes},
url={https://mediatum.ub.tum.de/node?id=1701036},
  year={2012}
}

@Article{MarkovCirac,
  title                     = {Assessing non-Markovian quantum dynamics},
  Author                   = {M. M. Wolf and J. Eisert and T. S. Cubitt and J. I. Cirac},
  doi={10.1103/PhysRevLett.101.150402},
  Journal                  = {Phys. Rev. Lett.},
  Year                     = {2008},
  Pages                    = {150402},
  Volume                   = {101}
}

@article{pechukas_reduced_1994-1,
	title = {Reduced {Dynamics} {Need} {Not} {Be} {Completely} {Positive}},
	volume = {73},
	issn = {0031-9007},
	doi = {10.1103/PhysRevLett.73.1060},
	number = {8},
	urldate = {2023-09-11},
	journal = {Phys. Rev. Lett.},
	author = {Pechukas, Philip},
	month = aug,
	year = {1994},
	pages = {1060--1062},
}

@article{Abanin_2021-IF,
  title = {Influence Functional of Many-Body Systems: {{Temporal}} Entanglement and Matrix-Product State Representation},
  shorttitle = {Influence Functional of Many-Body Systems},
  author = {Sonner, Michael and Lerose, Alessio and Abanin, Dmitry A.},
  year = 2021,
  month = dec,
  journal = {Ann. Phys.},
  series = {Special Issue on {{Philip W}}. {{Anderson}}},
  volume = {435},
  pages = {168677},
  issn = {0003-4916},
  doi = {10.1016/j.aop.2021.168677},
  urldate = {2026-01-22}
}

@article{koch2007dynamical,
  title={Dynamical low-rank approximation},
  author={Koch, Othmar and Lubich, Christian},
  journal={SIAM J. Mat. Ana. Appl.},
url={http://www.othmar-koch.org/papers/papers/lowrank.pdf},
  volume={29},
  pages={434--454},
  year={2007},
  publisher={SIAM}
}

@article{ceruti2022rank,
  title={A rank-adaptive robust integrator for dynamical low-rank approximation},
doi={10.48550/arXiv.2104.05247},
  author={Ceruti, Gianluca and Kusch, Jonas and Lubich, Christian},
  journal={BIT Num. Math.},
  volume={62},
  number={4},
  pages={1149--1174},
  year={2022},
  publisher={Springer}
}

@article{lubich2014projector,
  title={A projector-splitting integrator for dynamical low-rank approximation},
  author={Lubich, Christian and Oseledets, Ivan V.},
  journal={BIT Num. Math.},
  volume={54},
doi={10.48550/arXiv.1301.1058},
  pages={171--188},
  year={2014},
  publisher={Springer}
}

@article{Haegeman_2017,
   title={Quantum Gross-Pitaevskii Equation},
doi={10.21468/SciPostPhys.3.1.006},
   volume={3},
   number={1},
pages=006,
   journal={SciPost Phys.},
   publisher={Stichting SciPost},
   author={Haegeman, Jutho and Draxler, Damian and Stojevic, Vid and Cirac, Ignacio and Osborne, Tobias and Verstraete, Frank},
   year={2017},
   month=jul }

@article{Haegeman_2016,
   title={Unifying time evolution and optimization with matrix product states},
pages=165116,
   volume={94},
   DOI={10.1103/physrevb.94.165116},
   number={16},
   journal={Phys. Rev. B},
   publisher={American Physical Society (APS)},
   author={Haegeman, Jutho and Lubich, Christian and Oseledets, Ivan and Vandereycken, Bart and Verstraete, Frank},
   year={2016},
   month=oct }

@article{glaserTrainingSchrodingersCat2015,
  title = {Training {{Schr\"odinger}}'s Cat: Quantum Optimal Control},
  shorttitle = {Training {{Schr\"odinger}}'s Cat},
  author = {Glaser, Steffen J. and Boscain, Ugo and Calarco, Tommaso and Koch, Christiane P. and K{\"o}ckenberger, Walter and Kosloff, Ronnie and Kuprov, Ilya and Luy, Burkhard and Schirmer, Sophie and {Schulte-Herbr{\"u}ggen}, Thomas and Sugny, Dominique and Wilhelm, Frank K.},
  year = 2015,
  month = dec,
  journal = {Europ. Phys. J. D},
  volume = {69},
  number = {12},
  pages = {279},
  issn = {1434-6079},
  doi = {10.1140/epjd/e2015-60464-1},
  urldate = {2026-01-22},
  langid = {english}
}

@article{hudson1984quantum,
  title={Quantum Ito's formula and stochastic evolutions},
  author={Hudson, Robin L and Parthasarathy, Kalyanapuram R},
  journal={Commun. Math. Phys.},
  DOI={doi.org/10.1007/BF01258530},
  volume={93},
  number={3},
  pages={301--323},
  year={1984},
  publisher={Springer}
}

@misc{parthasarathy2014quantumstochasticcalculusquantum,
      title={{Quantum stochastic calculus and quantum Gaussian processes}}, 
      author={K. R. Parthasarathy},
      year={2014},
      eprint={1408.5686},
      archivePrefix={arXiv},
      optprimaryclass={math-ph}
}

@article{Hackl_2020,
   title={Geometry of variational methods: dynamics of closed quantum systems},
   volume={9},
   DOI={10.21468/scipostphys.9.4.048},
pages={048},
   journal={SciPost Phys.},
   publisher={Stichting SciPost},
   author={Hackl, Lucas and Guaita, Tommaso and Shi, Tao and Haegeman, Jutho and Demler, Eugene and Cirac, Ignacio},
   year={2020},
   month=oct }

@article{Chapman_2018_complexity,
   title={Toward a Definition of Complexity for Quantum Field Theory States},
   volume={120},
   DOI={10.1103/physrevlett.120.121602},
   pages=121602,
   journal={Phys. Rev. Lett.},
   publisher={American Physical Society (APS)},
   author={Chapman, Shira and Heller, Michal P. and Marrochio, Hugo and Pastawski, Fernando},
   year={2018},
   month=mar }

@article{audenaertSharpContinuityEstimate2007,
  title = {A Sharp Continuity Estimate for the von {{Neumann}} Entropy},
  author = {Audenaert, Koenraad M R},
  year = 2007,
  month = jun,
  journal = {J. Phys. A},
  volume = {40},
  number = {28},
  pages = {8127},
  issn = {1751-8121},
  doi = {10.1088/1751-8113/40/28/S18},
  urldate = {2026-04-29},
  langid = {english}
}

@article{gutoski_strategy_norm,
    author = {Gutoski, Gus},
    title = {On a measure of distance for quantum strategies},
    journal = {J. Math. Phys.},
    volume = {53},
    number = {3},
    pages = {032202},
    year = {2012},
    issn = {0022-2488},
    doi = {10.1063/1.3693621},
    url = {https://doi.org/10.1063/1.3693621}
}

@article{hastingsConnectingEntanglementTime2015,
  title = {Connecting Entanglement in Time and Space: {{Improving}} the Folding Algorithm},
  shorttitle = {Connecting Entanglement in Time and Space},
  author = {Hastings, M. B. and Mahajan, R.},
  year = 2015,
  month = mar,
  journal = {Phys. Rev. A},
  volume = {91},
  number = {3},
  pages = {032306},
  publisher = {American Physical Society},
  doi = {10.1103/PhysRevA.91.032306},
  urldate = {2025-01-08}
}

\clearpage
\appendix
\onecolumngrid
% !Tex root = ./paper.tex
\crefalias{section}{appendix}
\section{Summary of notation}
\label{app:list_notation}

{\renewcommand{\arraystretch}{1.3}
\setlength{\tabcolsep}{12pt}
\noindent
\begin{tabular}{@{}rl}
\toprule
        \textbf{Symbol} & \textbf{Meaning} \\  
        $\mathcal B(\mathcal H)$ & Bounded linear operator on Hilbert space $\mathcal H$ \\
        $[k]$ & $\coloneqq \{1, \dots, k\}$ for integer $k$ \\
        $A^\ast$ & (Entry-wise) complex conjugation of $A$  \\
        $\{R_\nu\}_\nu$, $\{R_\nu\}$ & Set of objects $R_\nu$ indexed by $\nu$, where index set is clear from context \\
        $\id$ & identity operator \\
        $\mathcal T$& Time-ordering operator \\
        $\kket{\,\cdot\, }$ & Liouville space vector \\
        $\mathcal A$ &   Trace non-increasing CP-map  \\
        $\{P_\mu\}_{\mu \in [d_{\rm S}^2]}$ & Basis of normalised Pauli matrices on the system Hilbert space\\
        $\{\mathbb{P}_\nu\}_{\nu \in [d_{\rm S}^4]}$ & Basis of normalised Pauli matrices on the system Liouville space\\
        $\PT$ & Discrete process tensor on $k$ times represented as an abstract map \\     
        $ \mathcal A^{(x)}$ & Instrument associated to the outcome $x$ \\
        $\{M^x\}_{x\in \mathcal X}$ & General POVM with outcome set $\mathcal X$ \\
         $\mathbf{A}_{k:0}^{(\mathbf{x})}$ & Sequence of instruments associated to the outcome sequence $\mathbf x = (x_0,\cdots,x_k)$ \\
         $\hat \Upsilon_{k:0}$/$\hat{\mathbf{A}}_{k:0}$ & Choi state of process tensor/ instrument \\
         $\mathfrak{o}_j / \mathfrak{i}_{j+1}$ & Output space of the process at time $t_j$ / input at time $t_j$\\
        $\Gamma(\mathcal H)$ & Fock space to Hilbert space $\mathcal H$  \\ 
        $\ket{\Omega}$ & Vacuum state of a field theory \\ 
        $\Gamma_T^{(d)}$ & Continuum bosonic Fock space on the interval $[0,T]$ with 
        $d$ particle species \\
        $\int_0^T \mathrm{D}^{n}t $ & Abbreviation for $\int_{0\leq t_1 \leq \cdots \leq t_n \leq T} \mathrm{d}t_1 \cdots\mathrm{d}t_n $ \\
        $\cmps{Q}{\{R_\nu\}}{B}$ & Continuous matrix product state as in \cref{def:cMPS} \\ 
        $\cPT$ & Continuous process tensor field theory state on interval $[0,T]$. \\
        $\ket{\mathcal{I}(x)}$ & Field theory state representing an instrument element to the outcome $x$  \\
         \bottomrule
\end{tabular}%
}

\section{Details on the derivation of the process tensor continuum limit}
\label{app:derivation}

In the main text, the Fock space underlying the second-quantised process tensor is denoted
\(\Gamma(\mathcal{H}_j^\nu)\), and its vacuum is identified with the linear functional 
\(\bra{\mathrm{vac}}_j\). We now make this identification precise.
Let \(\mathcal{H}\) be a single-particle Hilbert space with an orthogonal decomposition
\[
\mathcal{H} \;=\; \mathcal{H}_1 \oplus \mathcal{H}_2 .
\]
The (bosonic) Fock space over \(\mathcal{H}\) is defined as
\[
\Gamma(\mathcal{H}) \;:=\; \bigoplus_{n=0}^{\infty} \mathrm{Sym}^n(\mathcal{H}),
\]
where \(\mathrm{Sym}^n(\mathcal{H})\) denotes the \(n\)-fold symmetric tensor power.
From this, one can then identify the unitary isomorphism
\begin{equation}
\Gamma(\mathcal{H}_1 \oplus \mathcal{H}_2)
\;\cong\;
\Gamma(\mathcal{H}_1)\otimes \Gamma(\mathcal{H}_2),
\end{equation}
which follows from the decomposition
\begin{equation}
\mathrm{Sym}^n(\mathcal{H}_1 \oplus \mathcal{H}_2)
\;\cong\;
\bigoplus_{k=0}^n
\mathrm{Sym}^k(\mathcal{H}_1)\otimes \mathrm{Sym}^{n-k}(\mathcal{H}_2),
\end{equation}
and is compatible with the grading by total particle number.

Now consider a local single-particle Hilbert space \(\mathcal{H}_j\) with a distinguished
vacuum mode,
\[
\mathcal{H}_j \;=\; \mathcal{H}_j^{(\mathrm{vac})} \oplus \mathcal{H}_j^\nu,
\qquad
\mathcal{H}_j^{(\mathrm{vac})} \cong \mathbb{C}\,|\mathrm{vac}\rangle\!\rangle_j,
\]
where \(\mathcal{H}_j^\nu\) is the excited subspace. Applying the above construction yields
the canonical factorisation
\begin{equation}
\Gamma(\mathcal{H}_j)
\;\cong\;
\Gamma(\mathcal{H}_j^{(\mathrm{vac})}) \otimes \Gamma(\mathcal{H}_j^\nu).
\end{equation}
Under this isomorphism, the global Fock vacuum
\(\ket{\Omega}_j\in\Gamma(\mathcal{H}_j)\) is mapped to
\[
\ket{\Omega}_j
\;\longmapsto\;
\ket{\Omega^{(\mathrm{vac})}} \otimes \ket{\mathrm{vac}}_j,
\]
where \(\ket{\Omega^{(\mathrm{vac})}}\) is the vacuum of
\(\Gamma(\mathcal{H}_j^{(\mathrm{vac})})\) and \(\ket{\mathrm{vac}}_j\) is the vacuum of
\(\Gamma(\mathcal{H}_j^\nu)\).

Since \(\mathcal{H}_j^{(\mathrm{vac})}\) is one-dimensional,
\(\Gamma(\mathcal{H}_j^{(\mathrm{vac})})\) carries no nontrivial excitations and can play the role of a reference sector. We may therefore identify the effective Fock space of excitations relative to the local vacuum as
\begin{equation}
\tilde{\Gamma}(\mathcal{H}_j^\nu)
\;\equiv\;
\Gamma(\mathcal{H}_j^\nu),
\end{equation}
with the understanding that the vacuum vector
\(\ket{\mathrm{vac}}_j \in \Gamma(\mathcal{H}_j^\nu)\) is identified with the vector space
\[
\Gamma(\mathbb{C}\kket{\rm vac}_j)\otimes \ket{\mathrm{vac}}_j
\;\in\;
\Gamma(\mathcal{H}_j)
\]
embedded in $\Gamma(\mathcal{H}_j)$.
In this sense, \(\Gamma(\mathcal{H}_j^\nu)\) describes excitations relative to the
reference vacuum mode \(|\mathrm{vac}\rangle_j\).
This makes precise the sense in which particles in \(\Gamma(\mathcal{H}_j^\nu)\) are excitations relative to a chosen subset of modes in the larger Fock space.
The physical meaning of this in our process tensor formalism should be taken to be the fact that at all orders in the power expansion of a generator $\mathbb{G}$, the identity component cannot contribute anything more than an unobservable global phase to the transformation of the system.

\section{Continuos matrix product operators (cMPO) and their use in the formalism}\label{app:cmpo}

In the formulation of our framework we make use of cMPOs, which have been recently introduced in Ref.~\cite{tjoaContinuousMatrixProduct2025}.
Concretely, we use them define the process associated to a continuous instrument in~\cref{ssec:born rule}, and in ~\cref{ssec:choi duality} we claim that they parametrise Choi operators of cPTs.
In this appendix we provide more details to both points. 

First, we need to generalise the definition of a cMPO from \cite{tjoaContinuousMatrixProduct2025} to the case of multi-species Fock spaces with field operators $\{\psi_\mu(t)\}$ and vacuum $\ket{\Omega}$. 
Consider the following supermaps acting on operators on the associated Fock space
\begin{align}
l_{\mu}(t)[\bullet] &:= \psi_\mu^\dagger(t)\,\bullet, \\
r_{\mu'}(t)[\bullet] &:= \bullet\,\psi_{\mu'}(t), \\
\mathrm{Ad}_{\mu,\mu'}(t)[\bullet] &:= \psi_\mu^\dagger(t)\,\bullet\,\psi_{\mu'}(t), \\
\id[\bullet] &:= \bullet, 
\end{align}
let $Q(t), L_\mu(t), R_{\mu'}(t), T_{\mu,\mu'}(t) \in M_D(\mathbb{C})$ be matrix-valued
functions on the interval $[0,T]$, and let $B\in M_D(\mathbb{C})$ be a boundary matrix.
A multi-species cMPO with bond dimension $D$ is defined as the operator
\begin{equation}
O = \mathrm{Tr}_D \!\left(
B\,\mathcal{T}\exp\!\left[\int_I dt\,\mathcal{L}_t\right]
\right)\bigl(|\Omega\rangle\langle\Omega|\bigr),
\end{equation}
where the local generator $\mathcal{L}_t$ is given by
\begin{equation}
\mathcal{L}_t
=
Q(t)\otimes \id
+ \sum_{\mu} L_\mu(t)\otimes l_{\mu}(t)
+ \sum_{\mu'} R_{\mu'}(t)\otimes r_{\mu'}(t)
+ \sum_{\mu,\mu'} T_{\mu,\mu'}(t)\otimes \mathrm{Ad}_{\mu,\mu'}(t) .
\end{equation}
Now we can immediately see that the Choi operator of a PCR with Liouvillian $\mathbb{H}_{\rm SE}$ and initial state $\rho^{\rm SE}_0$ is a cMPO on the Fock space $\Gamma_T^{d_{\rm S}^2-1}$ with the parameters $Q = \mathbb{H}_{\rm SE}$, $R_\mu = \mathbb{P}_{\mu0}$, $L_\mu=\mathbb{P}_{0\mu}$, $T_{\mu\mu'} = 0$ and $B = \kket{\rho^{\rm SE}_0}\!\bbra{\id_{\rm SE}}$.
Because of how the cMPO parameters factorise, it is possible to split the Liouville space into its Hilbert space constituents and obtain~\cref{eq:cmpo choi state}, the Kraus operator-sum representation
of the process.

The operator $\mathbb{T}$ defined in \cref{eq:T operator} to obtain the cPT associated to a continuous instrument has precisely this structure.
To verify its correct action on the coherent control instrument~\cref{eq:inst_coherent_control_A}, we need to compute the product of two multi-species cMPOs, building on the single species results from~\cite{tjoaContinuousMatrixProduct2025}.
Let $O_1$ and $O_2$ be two multi-species cMPOs with bond dimensions $D_1$ and $D_2$,
specified by the parameters
\begin{equation}
(B^{(i)}, Q^{(i)}, L^{(i)}_\mu, R^{(i)}_{\mu'}, T^{(i)}_{\mu,\mu'}), \qquad i=1,2 .
\end{equation}
Then their product $O = O_2 O_1$ is again a multi-species cMPO with bond dimension
$D \le D_1 D_2$, boundary matrix
\begin{equation}
B = B^{(1)} \otimes B^{(2)},
\end{equation}
and local tensors
\begin{align}
Q &= Q^{(1)} \otimes \id_{D_2}
    + \id_{D_1} \otimes Q^{(2)}
    + \sum_{\nu} L^{(1)}_{\nu} \otimes R^{(2)}_{\nu}, \\[0.5em]
    L_{\mu} &= \id_{D_1} \otimes L^{(2)}_{\mu} 
    + \sum_{\nu} L^{(1)}_{\nu} \otimes T^{(2)}_{\mu,\nu} , \\[0.5em]
    R_{\mu'} &= R^{(1)}_{\mu'} \otimes \id_{D_2} 
    + \sum_{\nu} T^{(1)}_{\nu,\mu'} \otimes R^{(2)}_{\nu}, \\[0.5em]
T_{\mu,\mu'} &= \sum_{\nu} T^{(1)}_{\nu, \mu'} \otimes T^{(2)}_{\mu,\nu} .
\end{align}
Since we can write any cMPS $\ket{\Phi[Q, \{R_\mu\}, B]}$ as $O\vac$, where $O$ is a cMPO with local tensors $Q, R_\mu, 0, 0$ and boundary matrix $B$, the above calculation shows that the action of a cMPO with local matrices $Q^{(2)}, R_\mu^{(2)}, L_\mu^{(2)}, T_{\mu, \mu'}^{(2)}$ and boundary matrix $B^{(2)}$ on a cMPS $\ket{\Phi[Q^{(1)}, \{R_\mu^{(1)}\}, B^{(1)}]}$ produces a cMPS $\ket{\Phi[Q, \{R_\mu\}, B]}$ with
\begin{align}
    Q &= Q^{(1)} \otimes \id + \id \otimes Q^{(2)} + \sum_{\mu} R_\mu^{(1)} \otimes R_\mu^{(2)}\\
    R_\mu &= \id \otimes L_\mu^{(2)} + 
    \sum_{\mu'} R_{\mu'}^{(1)} \otimes T_{\mu, \mu'}^{(2)} \\
    B &= B^{(1)} \otimes B^{(2)}.
\end{align}
Now, to see that $\mathbb{T}$  correctly splits and connects the indices as explained intuitively in~\cref{ssec:born rule}, consider its action on the correlated coherent instrument
\begin{equation}
    \ket{\mathcal{I}[(\mathbb{L}^{\rm A})^*, \{c_\nu\}_\nu, \kket{(\rho^{\rm A})^*}\!\bbra{\id_A}]},
\end{equation}
where $c_\nu = \tr_{\mathrm{S}}[\mathbb{L}^{\mathrm{SA}}\mathbb{P}_\nu]^*$ and $\mathbb{L}^{\rm A} = \tr_{\mathrm{S}}[\mathbb{L}^{\rm SA}]$.
We get
\begin{equation}
    \bbra{A}\mathbb{T}\kket{\rho}\ket{\mathcal{I}[(\mathbb{L}^{\rm A})^*, \{c_\nu\}_\nu, \kket{(\rho^A)^*}\!\bbra{\id_A}]} = \ket{\Phi[(\mathbb{L}^{\rm{SA}})^*, \{\mathbb{P}_\nu\}, \kket{\rho}\!\bbra{A}\otimes\kket{\rho^{(\rm A)^*}}\!\bbra{\id_A}]},
\end{equation}
which is precisely what the process corresponding to the coherent control instrument should be -- a cMPS with the control Lindbladian as its drift matrix, the PCR jump matrices $\{\mathbb{P}_\nu\}$ and boundary matrix $\kket{\rho}\!\bbra{A}$.

\section{cMPS key quantitites and gauge transformations}
\label{app:gauges}

In this appendix section, we briefly recap the central objects of utility in cMPS, the cMPS gauge freedom, as well as the corresponding gauge equations of motion which we employ throughout this work in relation to cPTs. For convenient reference, we list the principal objects and their properties in Table~\ref{tab:cMPS_objects}.

\begin{table}[ht]
\centering
\begin{tabular}{@{}lllll@{}}
\toprule
\textbf{Object }
& \textbf{Definition }
& \textbf{Equation / Construction}
& \textbf{Notable properties }
& \textbf{Gauge transformation} \\ 
\midrule

$Q(t)$ 
& cPT variational parameter
& -- 
& Generates joint $\rm{SE}$ dynamics
& $Q \mapsto g^{-1}Qg + g^{-1}\dot g$ \\

$R(t)$ 
& cPT variational parameter
& -- 
& System jump / emission operator
& $R \mapsto g^{-1}Rg$ \\

$B$ 
& Boundary matrix
& Appears in $\mathrm{Tr}[B\,\mathcal{T}e^{\int Q + R\psi^\dagger}]$
& Encodes boundary conditions
& $B \mapsto g(0)^{-1} B g(T)$ \\

\midrule

$\mathbb{W}(t)$ 
& Transfer generator
& $Q\otimes\id + \id\otimes Q^\ast + R\otimes R^\ast$
& Lindblad-type generator
& $W \mapsto (g^{-1}\!\otimes g^{-1{\ast}})\, W\, (g\!\otimes g^\ast)$ \\

$\mathcal{W}(t_1,t_2)$ 
& Transfer propagator
& $\mathcal{T}\exp\!\int_{t_1}^{t_2} \mathbb{W}(t)\,\dd t$
& CPTP
& Similarity transform \\

\midrule

$r(t)$ 
& Right density matrix
& $\dot r = Q r + r Q^\dagger + R r R^\dagger$
& Positive, gauge covariant
& $r \mapsto g^{-1} r g^{-1{\dagger}}$ \\

$l(t)$ 
& Left density matrix
& $-\dot l = Q^\dagger l + l Q + R^\dagger l R$
& Positive, gauge covariant
& $l \mapsto g^{-1 \dagger} l g$ \\

\midrule

$r_\mathrm{fp}$ 
& Right fixed point
& $\mathcal{W}(0,T)[r] = r$
& Defines right canonical gauge
& Covariant \\

$l_\mathrm{fp}$ 
& Left fixed point
& $[l]\,\mathcal{W}(0,T) = l$
& Defines left canonical gauge
& Covariant \\

\midrule

$C(t)$ 
& Schmidt map
& $C=\sqrt{l}\sqrt{r}$
& Central canonical form
& Gauge invariant \\

\bottomrule
\end{tabular}
\caption{Key cMPS objects used in this work. 
}
\label{tab:cMPS_objects}
\end{table}

In matrix product states (MPS), there is a well-known gauge freedom for which, at the $i$th bond, one may left-multiply $A_i$ by an invertible matrix $G$, and right-multiply $A_{i+1}$ by $G^{-1}$ without changing the physical state. The analogous transformation for cMPSs is a position-dependent gauge matrix $g(t)$ on the interval $[0,T]$ which transforms
\begin{equation}
    \begin{split}
        Q(t)&\mapsto g(t)^{-1}Q(t)g(t) + g(t)^{-1}\frac{\dd g}{\dd t}(t),\\
        R(t)&\mapsto g^{-1}(t)R(t)g(t),\\
        B&\mapsto g(0)^{-1}B g(T).
    \end{split}
\end{equation}
Note that for ease of notation we will denote the jump matrices as constituting a single species, but the general case is the same.

Following standard prescriptions, we will list the appropriate gauge transformations which put a cMPS into its canonical forms, for further information, see, e.g., Ref.~\cite{ganahlContinuousMatrixProduct2017}. All objects are defined in~\cref{tab:cMPS_objects}. The cMPS transfer operator $\mathbb{W}(t)$ acts as the generator of the dynamics of $r(x)$ and $l(x)$ as
\begin{align}
    \frac{\dd r(t)}{\dd t} &= \mathbb{W}^{(r)}[r(t)] = Q(t)r(t) + r(t)Q^\dagger(t) + R(t)r(t)R^\dagger(t),\\
    \frac{\dd l(t)}{\dd t} &= [l(t)]\mathbb{W}^{(l)} = l(t)Q(t) + Q^\dagger l(t) + R^\dagger l(t) R(t).
\end{align}
When the cMPS has closed boundary conditions on a finite interval with $B = |l'\rangle\!\langle r'|$ then $\bbra{l(0)} = \langle l'|\otimes \langle l'^\ast|$ and $\kket{r(t)} = |r'\rangle\otimes |r'^\ast\rangle$; otherwise for open boundaries on an infinite interval they are the dominant left and right eigenvectors of the transfer propagator $\mathcal{W}(0,T)$. 
The matrices $Q(t)$ and $R(t)$ can be gauge transformed to either $(Q^{(l)}(t),R^{(l)}(t))$ or $(Q^{(r)}(t),R^{(r)}(t))$. Respectively, these are said to be in left or right orthogonal form at position $t$ if they satisfy
\begin{align}
    [\id]\mathbb{W}^{(l)}(t) &= Q^{(l)}(t) + Q^{(l)\dagger}(t) + R^{(l)\dagger}(t)R^{(l)}(t) = 0, \\
    \mathbb{W}^{(r)}(t)[\id] &= Q^{(r)}(t) + Q^{(r)\dagger} + R^{(r)}(t)R^{(r)\dagger}(t) = 0.
\end{align}
The requisite transformations on $Q(t)$ and $R(t)$ are given as 
\begin{equation}
    \label{appeq:left-orth}
    \begin{split}
        Q^{(l)}(t) &= \sqrt{l(t)}Q(t)\left[\sqrt{l(t)}\right]^{-1} - \frac{\dd \sqrt{l(t)}}{\dd t}\left[\sqrt{l(t)}\right]^{-1},\\
        R^{(l)}(t) &= \sqrt{l(t)} R(t)\left[\sqrt{l(t)}\right]^{-1},
    \end{split}
\end{equation}
and
\begin{equation}
    \label{appeq:right-orth}
    \begin{split}
        Q^{(r)}(t) &= \left[\sqrt{r(t)}\right]^{-1}Q(t)\sqrt{r(t)} - \sqrt{r(t)}\frac{\dd \sqrt{r(t)}}{\dd t},\\
        R^{(l)}(t) &= \left[\sqrt{r(t)}\right]^{-1} R(t)\sqrt{r(t)}.
    \end{split}
\end{equation}

Finally, in contrast to the discrete case, cMPS does not have a `mixed' canonical form in the sense of a simultaneous left and right orthogonal form. Since the transformation of $Q(t)$ involves a differential, these aforementioned global conditions cannot generically be satisfied locally at the same $t$. Instead, the \emph{central} canonical form of a cMPS carries its entanglement structure.
This is given by matrix functions $\Lambda_Q(t),\Lambda_R(t),C(t)$, and $\frac{\dd C(t)}{\dd t}$ such that the left and right orthogonal matrices can be given by
\begin{equation}
    \begin{split}
        Q^{(l)}(t) &= C(t)\Lambda_Q(t),\\
        R^{(l)}(t) &= C(t)\Lambda_R(t),\\
        Q^{(r)}(t) &= \Lambda_Q(t)C(t) + \left[C(t)\right]^{-1}\frac{\dd C}{\dd t},\\
        R^{(r)}(t) &= \Lambda_R(t)C(t),
    \end{split}
\end{equation}
and where
\begin{equation}
    \begin{split}
        \Lambda_Q(t) &= \frac{1}{\sqrt{r(t)}}Q(t)\frac{1}{\sqrt{l(t)}} - \frac{1}{\sqrt{r(t)}}\frac{1}{\sqrt{l(t)}}\frac{\dd \sqrt{l(t)}}{\dd t}\frac{1}{\sqrt{l(t)}}\\
        \Lambda_R(t) &= \frac{1}{\sqrt{r(t)}}R(t)\frac{1}{\sqrt{l(t)}}\\
        C(t) &= \sqrt{l(t)}\sqrt{r(t)}.
    \end{split}
\end{equation}
Finally, the central canonical matrices $Q^{(c)}(t)$ and $R^{(c)}(t)$ are related to the left and right orthogonal matrices via
\begin{equation}
    \begin{split}
        Q^{(c)}(t) &= Q^{(l)}(t)C(t) = C(t)Q^{(r)}(t) - \frac{\dd C}{\dd t}\\
        R^{(c)}(t) &= R^{(l)}(t)C(t) = C(t)R^{(r)}(t).
    \end{split}
\end{equation}
We call the matrix $C(t)$ the \emph{Schmidt map}, as it connects the left and right auxiliary spaces. Explicitly, it is the coefficient matrix with respect to some orthogonal bases on $\mathsf{P}$ and $\mathsf{F}$, where these are the intervals $[0,t)$ and $(t,T]$, respectively:
\begin{equation}
    \cPT = \sum_{a,b}C(t)_{a,b}|\Psi_a\rangle_{\mathsf{P}}\otimes |\Psi_b'\rangle_{\mathsf{F}}.
\end{equation}
Note that $C(t)$ is not manifestly diagonal here, but can be chosen to be. In particular, there still exists a unitary freedom in which $g(t)\mapsto g(t)u(t)$ preserving the left orthogonality condition which can be used to pointwise diagonalise $r(t)$. However, this transformation can be cumbersome and introduce non-essential geometric and non-smooth structure. In this case, we have the additional transformation
\begin{equation}
    r(t)\mapsto \tilde{r}(t) = U(t)\sigma(t)U^\dagger(t),
\end{equation}
(which is the defining property of $U(t)$), giving
\begin{equation}
    \begin{split}
        \tilde{Q}^{(c)}(t) &= U^\dagger (t) Q^{(c)}(t)U(t) + U^\dagger \frac{\dd U}{\dd t},\\
        \tilde{R}^{(c)}(t) &= U^\dagger(t) R^{(c)}(t) U(t).
    \end{split}
\end{equation}
Thus, letting $C(t)$ be written in its diagonal form, and absorbing the unitary transformation along with the eigenvalues $\sqrt{\sigma_i(t)}$ into $|\Psi_a\rangle_{\mathsf{P}}$ and $|\Psi_b'\rangle_{\mathsf{F}}$, we can construct the instantaneous Schmidt decomposition of the cPT at time $t$:
\begin{equation}
    \cPT = \sum_i \ket{\widetilde{\Psi}_i}_{\mathsf{P}}\otimes|\widetilde{\Psi}'_i\rangle_{\mathsf{F}},
\end{equation}
with $\{\ket{\widetilde{\Psi}_i}_{\mathsf{P}}\}_i$ and $\{\ket{\widetilde{\Psi}'_i}_{\mathsf{F}}\}_i$ the orthonormal left and right Schmidt vectors, respectively. 

\section{Distance measures}\label{ssec:distance}

%!Tex root=./paper.tex

Here, we briefly discuss meaningful notions of distance between continuous process tensors.
The meaningfulness of a notion of a distance is given by its operational interpretation:
it should measure how difficult it is to distinguish two continuous processes $\ket{\Upsilon^{(1)}_T}$ and $\ket{\Upsilon_T^{(2)}}$ in the lab.
The outcomes of experiments are samples from probability distributions given by the generalised Born rule defined in \cref{ssec:born rule}.
The sample complexity of distinguishing two probability from samples is determined by their total variation (TV) distance
\begin{equation}
    \operatorname{TV}\left(p^{(1)}, p^{(2)}\right) \coloneqq \sup_{s \in \mathcal{X}} |p^{(1)}(x) - p^{(2)}(x)|.
\end{equation}
Therefore, to get the theoretical limit on the distinguishability of processes, we need to design the experiment (instrument) that maximises the TV distance between the resulting probability distributions. 
Hence, we obtain an operationally meaningful notions of distance
\begin{equation}\label{eq:operational hard distance}
    d_\text{TV}(\ket{\Upsilon^{(1)}_T}, \ket{\Upsilon^{(2)}_T}) \coloneqq
    \sup_{\{\bra{\mathcal{I}(x)}\}_{x \in \mathcal{X}}} \operatorname{TV}\left(\braket{\mathcal{I}(x) | \Upsilon_T^{(1)}}, \ \braket{\mathcal{I}(x) | \Upsilon_T^{(2)}}\right),
\end{equation}
where the supremum is taken over all valid instruments, see \cref{ssec:born rule}.
This distance can be understood as a continuous generalization of the generalised comb divergence \cite{wangResourceTheoryAsymmetric2019,zambonProcessTensorDistinguishability2024} and it quantifies the sample complexity of distinguishing the processes under the best possible instrument.
The operational advantage of the generalised comb divergences is that they satisfy the data processing inequality under causality and positivity preserving linear transformations of process tensors \cite{zambonProcessTensorDistinguishability2024}.

To get a more practical measure, which potentially takes into account practical limitations of a laboratory setup, we can restrict the supremum to a subset of all valid instruments.
One meaningful option is to restrict to coherent control instruments with a POVM measurement at time $t \in [0, T]$, which have the form 
\begin{equation}
    \sum_s M^s_\nu \bra{\mathcal{I}[c_0, \{c_{\nu'}\}, 1]} \psi_\nu(t),
\end{equation}
where $M^s_\nu \coloneqq \tr[\kket{\id}\!\bbra{M^s} \mathbb{P}_\nu]$ with $\{M^s\}_s$ a POVM, and $c_\nu$ for $\nu \in\{0, \dots, d_{\rm S}^4-1\}$ are real numbers such that $\sum_\nu c_\nu^* \mathbb{P}_\nu$ is a vectorised Lindbladian.
Note that this is stricly speaking not an instrument, as its effects are not normalisable.
Hence, it can only act in a rigorous way on continuous processes, as presented in the companion paper \cite{fuksaDilationTheoremContinuum2026}.
For a given coherent control, the supremum of the TV distance over the POVM elements is given by the trace distance of the system density matrix at $t$, so that we obtain
\begin{equation}\label{eq:operational simplified distance}
    d_\text{coherent}(\ket{\Upsilon^{(1)}_T}, \ket{\Upsilon^{(2)}_T}) \coloneqq
    \sup_{\substack{c_\nu \in \mathbb{R}: \ \text{Lindblad}\\t \in [0, T]}} \frac12 \Big\|\sum_\nu M^{(i,j)}_\nu \bra{\mathcal{I}[c_0, \{c_{\nu'}\}, 1]} \psi_\nu(t)
    \left(\ket{ \Upsilon_T^{(1)}} - \ket{\Upsilon_T^{(2)}}\right)\Big\|_1,
\end{equation}
where $M^{(i,j)}_\nu \coloneqq \tr[\kket{\id_{\rm S}}\!\bra{i}\!\bra{j} \mathbb{P}_\nu]$, $c_\nu$ are restricted such that $\sum_\nu c_\nu^* \mathbb{P}_\nu$ is a vectorised Lindbladian and the trace distance is taken over the matrix indexed by $i,j$.

Meaningful simpler measures on continuous processes are the so-called Choi divergences, which are state distinguishabiliy measures applied to the Choi state (\cref{ssec:choi duality}) of the continuous process tensor.
For instance, we have
\begin{equation}
    \|\hat \Upsilon_T^{(1)} - \hat \Upsilon_T^{(2)}\|_1 \ , \quad D(\hat \Upsilon_T^{(1)} \| \hat \Upsilon_T^{(2)}),
\end{equation}
where $\|\cdot \|_1$ is the trace distance and $D(\cdot \|\cdot )$ the relative entropy.
Choi divergences have recently been shown not to satisfy the data processing inequality, although they upper bound the more operationally motivated generalised comb divergences \cite{zambonProcessTensorDistinguishability2024}.

To compute Choi divergences of process tensors in process-canonical representation with a Liouvillian drift matrix, we can utilise the decomposition \cref{eq:kraus form}.
The $\ket{u_{i,j}}$ define a basis for the image of the Choi matrix, allowing us to compute the Choi divergences efficiently.
In this case we have that
\begin{equation}
\hat\Upsilon_T^{(i)} = \sum_{j,k=1}^{d_{\rm S} d_{\rm E}^{(i)}} \lambda_{j}\ket{u^{(i)}_{j,k}}\!\bra{u^{(i)}_{j,k}}
\end{equation}
where
\begin{equation}
    \ket{u^{(i)}_{j,k}} \coloneqq \ket{\Phi[H^{(i)}_{\rm SE}, \{\id_{\rm E} \otimes P_\mu\}, \ket{j}\!\bra{k}]}.
\end{equation}
The vectors $\mathcal{B} = \{\ket{u_j}\} \coloneqq \{\ket{u^{(1)}_{j,k}}\} \cup \{\ket{u^{(2)}_{j,k}}\}$ define an overcomplete basis.
Evaluating cMPS overlaps we can compute the corresponding Gram matrix $G_{i,j} = \braket{u_i | u_j}$ and compute the eigendecomposition $G = U\Lambda U^\dag$, where we exclude all the zero eigenvalues, so that $U$ becomes an isometry and $\Lambda = \diag(\{\lambda_j\})$ with $\lambda > 0$.
We can now define an orthonormal basis $\{\ket{i}\}_{i\in[r]}$, where 
\begin{equation}
r = \dim(\Span(\mathcal{B})) \le d_{\rm S}^2 \left((d_{\rm E}^{(1)})^2 + (d_{\rm E}^{(2)})^2\right),
\end{equation}
in which
\begin{equation}
    \ket{u_j} = \sum_k \sqrt{\lambda_k} U_{k,j}^\dag \ket{k},
\end{equation}
where for negative $\lambda_k$ we choose some branch of the square root.
Now we can write $\hat\Upsilon^{(1)}_T$ and $\hat\Upsilon_T^{(2)}$ in this basis and compute any Choi divergence of interest.

Finally, we mention that a simple, but practical notion of distance between continuous process tensors is the $L^2$ norm
\begin{equation}
    \|\ket{\Upsilon_T^{(1)}} - \ket{\Upsilon_T^{(2)}}\|_2.
\end{equation}
This is the norm that is used to compress process tensors, see \cref{ssec:compressibility}.
The operational meaning of the $L^2$ norm can be obtained by 
\begin{equation}
    \int_\mathcal{X} {\rm d} x \ |\!\braket{\mathcal{I}(s)|\Upsilon_T^{(1)}} - \braket{\mathcal{I}(s)|\Upsilon_T^{(2)}} \!| \le
    \|\ket{\mathcal{I}(s)}\|_{\Gamma \to L^1(\mathcal{X})} \|\ket{\Upsilon_T^{(1)}} - \ket{\Upsilon_T^{(2)}}\|_2,
\end{equation}
where we define an operator norm for the instruments.
Hence, the $L^2$ norm bounds the distinguishability of processes under instruments with a bounded norm.
In the companion work \cite{fuksaDilationTheoremContinuum2026} we explore the instrument norm further, showing that it is related to how difficult (energy intensive) is the instrument to be performed in the lab.

\section{Details on the examples of continuous instruments}
\label{app:continuous_instruments}
In this appendix we present the details of how to arrive at the instruments corresponding to continuous measurements, which were presented in \cref{ssec:control_family}. 
The connection between the cMPS formalism (which is the process canonical representation of a cPT) and continuous measurements was established 
in Ref.~\cite{osborneHolographicQuantumStates2010}. This connection serves as the starting point for the following derivations in both the quantum jump and diffusive measurement cases.

\subsection{Quantum jump measurements}
 A quantum jump measurement defined by the stochastic differential equation \cref{eq:cont_measurement_SME}  and \cref{eq:jump_meas_increment}  can be understood as the application of the following effect matrices every ${\rm d} t$ on the system $\rm{S}$
 \begin{equation}
     \begin{aligned}
         E_0 &= \id - \tfrac12 \sum_{j \in [k]} L_j^\dagger  L_j(t) \, \mathrm dt , \\ 
         E_j &= L_j(t) \, \sqrt{\mathrm d t} \ , \quad \forall j \in [k], 
     \end{aligned}
 \end{equation}
 where $\mathrm d t$ is an infinitesimal time increment.
 This corresponds to the POVM 
 \begin{equation}\label{eq:jump_povm}
 \{\id - \sum_{j \in [k]} L_j^\dag L_j(t) \, \mathrm d t\} \cup \{L_j^\dagger L_j \mathrm d t\}_{j\in[k]}
 \end{equation}
 every ${\rm d}t$.
 This means that most of the time we will obtain the outcome ``0" whereas with an infinitesimal probability we will record a ``click", corresponding to one of the outcomes $j \in [k]$.

Consider first that $\mathrm d t \mapsto \shortT$ is a finite time step, keeping in mind that we will have to take $\shortT \rightarrow 0$ to recover the infinitely weak measurement.
Suppose a spinful one-dimensional bosonic lattice theory on the interval $[0,T]$ with $\frac{T}{\shortT}$ sites and bosonic lowering operators $a_j[\ell]$, where the index $j \in [k]$ labels the different spin values that the boson at the $\ell$-th site for $\ell \in \left[{T}/{\shortT}\right]$ can have.
We will encode the outcomes from measuring the POVM in \cref{eq:jump_povm}  into a state of the lattice theory.
In particular, if we get a click $j \in [k]$ at time $\ell \shortT$ we will create a particle with spin $j$ at the $\ell$-th site.
As will become apparent, and as follows 
from Ref.~\cite{osborneHolographicQuantumStates2010}, the lattice theory becomes a kind of quantum memory for the measurement if we couple it with the system through the Hamiltonian
\begin{equation}
    H(t) = \sqrt{\shortT} \sum_{\ell \in \left[\frac{T}{\shortT}\right]} \delta(t - \ell \shortT) \sum_{j\in [k]} \left( i L_j(\ell \shortT) \otimes a^\dag_j[\ell] - i L_j^\dag(\ell \shortT) \otimes a_j[\ell]\right).
   \label{eq:interaction_Ham_meter_system}
\end{equation}
If the initial state of the system is $\ket{\psi}$ and the initial state of the lattice theory is the vacuum $\ket{\Omega}$, then in the limit $\shortT \rightarrow 0$ the state at time $T$ can be written as 
\begin{equation}\label{eq:weak measurement time evolution}
    \lim_{\shortT \rightarrow 0} \prod_{\ell \in \left[\frac{T}{\shortT}\right]} \exp\left[\sqrt{\shortT} \sum_{j\in[k]} \left(L_j(\ell\shortT) \otimes a^\dag_j[\ell] - L_j^\dag(\ell\shortT) \otimes a_j[\ell]\right)\right]\ket{\psi}\ket{\Omega}.
\end{equation}
As observed in
Ref.~\cite{osborneHolographicQuantumStates2010}, we can use the BCH formula
\begin{equation}
    e^{\epsilon X} e^{\epsilon Y} = e^{\epsilon X + \epsilon Y + \frac12 \epsilon^2 [X, \, Y] + \mathcal O (\epsilon^3)}
\end{equation}
to obtain 
\begin{equation}
    \begin{split}
         e^{\sqrt{\shortT}(L \otimes a^\dag - L^\dag \otimes a)} \ket{\psi}\ket{\Omega} &= e^{\sqrt{\shortT} \, L\otimes a^\dag + \mathcal{O} (\shortT^{1.5})} e^{\frac12\shortT [L\otimes a^\dag, \, L^\dag \otimes a]} e^{- \sqrt{\shortT} L^\dag \otimes a} \ket{\psi}\ket{\Omega} \label{eq:BCH_on_state} \\
     &= e^{\sqrt{\shortT} \, L\otimes a^\dag + \mathcal{O} (\shortT^{1.5})} e^{-\frac12 \shortT \, L^\dag L \otimes  a a^\dag} \ket{\psi}\ket{\Omega}\\    &= e^{\sqrt{\shortT} \, L \otimes  a^\dag + \mathcal{O} (\shortT^{1.5})} e^{-\frac12 \shortT \, L^\dag L \otimes  [a, \, a^\dag]}e^{-\frac12 \shortT \, L^\dag L \otimes  a^\dag a} \ket{\psi}\ket{\Omega}\\             &= e^{\sqrt{\shortT} \, L \otimes  a^\dag + \mathcal{O} (\shortT^{1.5})} e^{-\frac12 \shortT \, L^\dag L \otimes  \id} \ket{\psi}\ket{\Omega}\\
    &= e^{- \frac12 \shortT \, L^\dag L \otimes  \id + \sqrt{\shortT} \, L \otimes  a^\dag + \mathcal{O}(\shortT^{1.5})} \ket{\psi}\ket{\Omega},
    \end{split}
\end{equation}
where we have used the bosonic commutation relationships and the definition of the vacuum state vector $a \ket{\Omega} = 0$.
Now we can rewrite the time evolution \eqref{eq:weak measurement time evolution} as
\begin{align}
    &\lim_{\shortT \rightarrow 0} \prod_{\ell \in \left[\frac{T}{\shortT}\right]} \exp\left[\sum_{j \in [k]} -\tfrac12 \shortT \, L_j^\dag L_j(\ell\shortT) \otimes \id + \sqrt{\shortT} L_j(\ell\shortT) \otimes a^\dag_j [\ell]\right]\ket{\psi}\ket{\Omega}=\label{eq:which clicks}\\
    &\hspace{4cm}= \mathcal{T} \exp\left[\int_0^T \mathrm d t \, \sum_{j\in[k]}-\tfrac12 L_j^\dag L_j(t) \otimes \id + L_j(t) \otimes \psi_j^\dag(t)\right]\ket{\psi}\ket{\Omega}\label{eq:almost cmps},
\end{align}
where $\psi_j(t) \coloneqq \lim_{\shortT \rightarrow 0} \frac1{\sqrt{\shortT}} a_j[\lfloor \frac{t}{\shortT}\rfloor]$ is the field operator.
Equation \eqref{eq:almost cmps} is reminiscent of the cMPS state with $Q = \sum_{j \in [k]} -\tfrac12 L^\dag_j L_j$ and $R_j = L_j$ for $j \in [k]$.
Furthermore, \eqref{eq:which clicks} shows that the field theory indeed can be considered as a meter weakly measuring the system under the POVM \eqref{eq:jump_povm}: Each of the factors in \eqref{eq:which clicks} can be expanded as
\begin{align}
    \exp\Big[\sum_{j \in [k]} -\tfrac12 \shortT \, L_j^\dag L_j(\ell \shortT) \otimes \id &+ \sqrt{\shortT} \, L_j(\ell\shortT) \otimes a^\dag_j[\ell]\Big] =\\
 &=\id + \sum_{j\in[k]} - \tfrac12 \shortT \, L_j^\dag L_j(\ell\shortT) \otimes \id + \sqrt{\shortT} \, L_j(\ell\shortT) \otimes a^\dag[\ell]
\end{align}
and, 
therefore, we create a particle with spin $j$ at the site $\ell$ iff we apply the effect operator corresponding to the click $j$ on the system, otherwise we apply the effect operator $0$ corresponding to no click.
Furthermore, the probability of observing an excitation of type $j$ at site $\ell$ is to first order in $\shortT$ equal to the probability of measuring the click $j$ with the finite time POVM 
\begin{equation}
    \{\id - \sum_{j\in[k]} L_j^\dag L_j \shortT\} \cup \{L_j^\dag L_j \shortT\}_{j\in[k]}
\end{equation}
at time $\ell\shortT$.
As we take the limit $\shortT \to 0$ the measurement becomes simultaneously weaker and more frequent.
The scaling is chosen such that probability of observing a finite number of jumps in the interval remains finite throughout.

Suppose that we now wish to compute the probability of getting a click $x \in [k]$ at time $\tau \in [0,T]$.
By the Born rule and through the commutation relations along with the definition of the vacuum state $\ket{\Omega}$
\begin{align}
    \begin{split}
        p(x, \tau) &= \lim_{\shortT \rightarrow 0} \sum_{i \in [d_{\rm S}]} \bigl|\bra{i} \bra{\Omega} \id \otimes a_x[\tfrac{\tau}{\shortT}] \prod_{\ell \in [T/\shortT]} \big(\id + \sum_{j \in [k]}- \tfrac12 \shortT \, L_j^\dag L_j(\ell\shortT) \otimes \id + \nonumber\\
                   &\hspace{6cm}+ \sqrt{\shortT} \, L_j(\ell\shortT) \otimes a^\dag[\ell]\big) \ket{\psi}\ket{\Omega}\bigr|^2
    \end{split}\nonumber \\
    \begin{split}
    &= \lim_{\shortT \rightarrow 0} \sum_{i \in [d_{\rm S}]} \bigl|\bra{i} \bra{\Omega} \prod_{\ell > (\tau /\shortT)} (\id - \tfrac12 \shortT \sum_{j \in [k]} L_j^\dag L_j \otimes \id) \left(\sqrt{\shortT} \, c_x \otimes \id \right)\times   \\
    &\hspace{6cm}\times\prod_{\ell < \tau/\shortT} (\id - \tfrac12 \shortT \sum_{\nu\in[k]} L_j^\dag L_j \otimes \id) \ket{\psi}\ket{\Omega}\bigr|^2
    \end{split}  \\
    &= \mathrm{d}\tau \sum_{i\in[d_{\rm S}]} \bigl|\bra{i} \mathcal{T} \e^{-\frac12\sum_{j\in[k]} \int_\tau^T L_j^\dag L_j\mathrm{d}t} \, c_x \, \e^{-\frac12\sum_{j\in[k]} \int_0^\tau L_j^\dag L_j \mathrm{d}t} \ket{\psi}\bigr|^2 \nonumber  \\ 
      &= \mathrm{d}\tau \, \llangle\id_{\rm S}| \mathcal{T} \e^{-\frac12\sum_{j\in[k]} \int_\tau^T L_j^\dag L_j\otimes \id + \id \otimes L_j^T L_j^* \mathrm{d}t} \, c_x \otimes c_x^* \, \e^{-\frac12\sum_{j\in[k]} \int_0^\tau L_j^\dag L_j \otimes \id + \id \otimes L_j^T L_j^* \mathrm{d}t} |\rho\rrangle. \nonumber
\end{align}
We now need to rewrite this equation as an inner product between the continuous process tensor in process canonical representation $\cPT = \cpt{\gen}{\{\id_{\rm E} \otimes \mathbb P_\nu\}_{\nu}}{B}$ and some other field theory state. 
As in the main text we will write $\mathbb P_\nu$ instead of $\id_{\rm E} \otimes \mathbb P_\nu $ for brevity.
We will make use of the following identity valid for general cMPS \cite{haegemanCalculusContinuousMatrix2013}
\begin{equation}
        \psi_{\nu}(\tau) \cpt{\gen}{\{ \mathbb P_{\nu'}\}_{\nu'}}{B} =
       \tr\left[B \mathcal{T}\left(\e^{\int_\tau^T dt \,\gen + \sum_{\nu'}( \mathbb P_{\nu'})\otimes \psi_{\nu'}^\dag}\right)  \mathbb P_{\nu'} \mathcal{T}\left(\e^{\int_0^\tau dt \,\gen + \sum_{\nu'} \mathbb P_{\nu'}\otimes \psi_{\nu'}^\dag}\right) \right] \ket{\Omega}.
\end{equation}
This way we are able to ``bring down'' the basis element $\mathbb P_\nu$ from the exponential at the right time.
We now write the state above in the particle number basis, to get 
\begin{equation}
    \begin{split}
        \psi_{\nu}&(\tau)\cpt{\gen}{\{ \mathbb P_{\nu'}\}_{\nu'}}{B} =\\
        &\sum_{N, N'=0}^\infty \sum_{\nu_1, \nu'_1, \dots, \nu_N, \nu'_{N'}}\int_{T\ge t_1 \ge \dots \ge t_N \ge\tau} \mathrm{d}t_1\dots \mathrm{d}t_N \int_{\tau\ge t'_1 \ge \dots \ge t'_{N'} \ge 0} \mathrm{d}t'_1 \dots \mathrm{d}t'_{N'}  
        \tr\big[B M_{\gen}(T, t_1) \mathbb P_{\nu_1} M_{\gen}(t_1, t_2) \mathbb P_{\nu_2} \dots \\
        & \dots \mathbb P_{\nu_N} M_{\gen}(t_N, \tau) \mathbb P_{\nu} M_{\gen}(\tau, t'_1) \mathbb P_{\nu'_{1}} M_{\gen}(t'_1, t'_2) \dots P_{\nu'_{N'}} M_{\gen}(t'_{N'}, 0)\big] \psi^\dag_{\nu_1}(t_1)\dots\psi^\dag_{\nu_N}(t_N)\psi^\dag_{\nu'_1}(t'_1)\dots\psi^\dag_{\nu'_{N'}}(t'_{N'})\ket{\Omega},
    \end{split}
\end{equation}
where $M_{\gen}(x,y) = \mathcal{T} \exp[\int_x^y \gen \mathrm{d}z]$. When contracting the vector above with some arbitrary cMPS, all the involved integration variables have a fixed ordering. Therefore using
\begin{equation}
    \begin{split}
        \bra{\Omega} \psi_{\nu_1}(y_1) \dots \psi_{\nu_M}(y_M) \psi^\dag_{\mu_1}(x_1) &\dots \psi^\dag_{\mu_N}(x_N) \ket{\Omega} =\\  &\delta_{MN}\delta_{\mu_1\nu_1}\dots\delta_{\mu_N\nu_N}\delta(x_1-y_1) \dots \delta(x_N - y_N)
    \end{split}
\end{equation}
we find that
\begin{multline}
       \bra{Q, \, \{R_{\nu'}\}_{\nu'}, \, B'} \psi_{\nu}(\tau) \cpt{\gen}{\{ \mathbb P_{\nu'}\}_{\nu'}}{B}= \\  = \tr \left[ B\otimes (B')^* \mathcal{T} \e^{\int_\tau^T \gen \otimes \id + \id \otimes Q^* + \sum_{\nu'} \mathbb P_{\nu'} \otimes R^*_{\nu'}} \mathbb P_\nu  \mathcal{T}\e^{\int_0^\tau \gen \otimes \id + \id \otimes Q^* + \sum_{\nu'} \mathbb P_{\nu'} \otimes R^*_{\nu'}}\right].
\end{multline}
This result now allows us to finally express the weak measurement probability as
\begin{equation}
    p(x, \tau) =  \sum_{\nu = 0}^{d_S^4 - 1} \chi^x_{\nu} \bra{\Phi[c_0, \, \{c_{\nu'}\}_{\nu'\in[d_{\rm S}^4-1]}, \, 1]} \xi_{\nu}(\tau) \cPT \mathrm{d}\tau,
\end{equation}
where $c_{\nu}$ and $\chi^x_{\nu}$ are complex numbers given by
\begin{equation}
\begin{split}
    c_\nu &\coloneqq - \tfrac{1}{2} \tr\Bigl[\mathbb{P}_\nu\bigl(\sum_{j\in[k]} L_j^\dag L_j \otimes \id + \id \otimes L_j^T L_j^*\bigr)\Bigr]\\
    \chi^x_\nu &\coloneqq \tr\left[L_x \otimes L_x^* \mathbb{P}_\nu\right]^*.
    \end{split}
\end{equation}
and $\chi_\nu$ is defined as in \cref{ssec:discrete marginals}.

\subsection{Diffusive measurement}

We now turn to the description of diffusive measurements in our formalism. A diffusive measurement of the operators $\{L_j\}_{j \in [k]}$ on the system during the time interval $[0,T]$ produces as outcomes a set of $k$ continuous, complex-valued integrated \textit{records} $ Y_j: [0,T] \longrightarrow \mathbb C$, $j \in [k]$. Their derivatives in a distributional sense are the measurement ``currents'' $J_j := \mathrm d Y_j/ \mathrm d t$.  Due to the randomness in the quantum measurement the currents are apriori stochastic quantities described by the complex random variables
\begin{equation}
     \bm{J}_j(t)\mathrm dt =  \langle \sum_{j'} u_{j,j'}L^\dagger_{j'} + L_j \rangle_{\psi(t)}\mathrm dt + \mathrm d\bm Z_j(t),   \label{eq:gener_dyne_currents}
\end{equation}
where $\psi(t)$ is the current state of the system. 
The $\mathrm{d}\bm{Z}_j$ are increments of complex Wiener processes which in the most general form of a diffusive measurement satisfy \cite{wisemanCompleteParameterizationInvariance2001}
\begin{align}
\label{eq:general_wiener_incr_complex}
    \mathrm d\bm Z_j(t) \mathrm d\bm Z^\ast_{j'}(t) &= \delta_{j,j'} \mathrm d t
    ,\\
    \mathrm d\bm Z_j(t) \mathrm d\bm Z_{j'}(t) &= u_{j,j'} \mathrm d t, \label{eq:general_wiener_incr_complex_var}
\end{align}
where $u_{j,j'}$ are arbitrary complex numbers with the only constraint that the spectral norm of the matrix $ u$ is bounded: $\|u\|^2 \leq 1$. 
As we will show in the following such a diffusive measurement can be modelled (similarly to the quantum jump case) via a coupling of the system to bosonic meter modes via the Hamiltonian in~\cref{eq:interaction_Ham_meter_system}  which are then subject to a so-called general-dyne measurement \cite{serafiniQuantumContinuousVariables2017}, a generalization of homodyne and heterodyne detection. The POVM corresponding to the general-dyne measurement derives from the following resolution of the identity on the Hilbert space of $k$ bosonic modes
\begin{equation}
\label{eq:gen_dyne_POVM}
    \id = \frac{1}{\pi^k} \int_{\mathbb C^k}\mathrm d\vec{\alpha} S D (\vec{\alpha}) \ket{0}\bra{0} D^\dagger(\vec{\alpha}) S^\dagger.
\end{equation}
 Here, $D$ is a displacement operator on $k$ bosonic modes, $\vec{\alpha}$ is a vector of $k$ complex displacements, and $S$ is any $k$-mode squeezing operation. We use the abbreviation $\mathrm d\vec{\alpha} \equiv \mathrm d\alpha_1 \mathrm d\alpha_2 \cdots \mathrm d\alpha_k $. 
 The case of homodyne detection which produces real outcome currents (as presented in the main text) corresponds to the limiting POVM where the squeezing strength of $S$ is send to infinite such that we are effectively measuring in the eigenbasis of a quadrature operator. Here, we will restrict to the case where $S=\id$ for simplicity of presentation. This corresponds to the case of $u_{j,j'} \equiv 0$ in \cref{eq:general_wiener_incr_complex_var} and hence the complex infinitesimal Wiener (Ito) increments now satisfy 
\begin{equation}
\begin{split}
      \mathrm d\bm Z_j(t) \mathrm d\bm Z^\ast_{j'}(t) &= \delta_{j,j'} \mathrm d t, \\
    \mathrm d\bm Z_j(t) \mathrm d\bm Z_{j'}(t) &= 0.
\end{split}
\label{eq:wiener_complex_incr}
\end{equation}
This is heterodyne detection of multiple modes. The complex measurement currents decompose into two real currents $J^R_j(t), J_j^I(t)$ as
\begin{equation}
\label{eq:complex_outcome_diff}
    J_j(t)\mathrm d t = \frac{1}{\sqrt{2}}\bigl(J^R_j(t)+ i J^I_j(t)\bigr)\mathrm d t.
\end{equation}
The $J^R_j, J^I_j$ are the records of two conjugate quadratures and the factor $\tfrac{1}{\sqrt{2}}$ ensures that the two real currents carry white noise of unite strength, given the standard normalization of the complex Wiener increments in \cref{eq:general_wiener_incr_complex_var}. This is the same convention that relates a complex coherent amplitude to its quadratures: $\alpha= \tfrac{1}{\sqrt{2}}(q + ip)$ \cite{wisemanQuantumMeasurementControl2014,serafiniQuantumContinuousVariables2017}. 

As mentioned, the system and meters are again coupled via the Hamiltonian in Eq.~\eqref{eq:interaction_Ham_meter_system} and we first consider a small but finite time step $\shortT$. The set of outcomes is a finite sequence of complex vectors $(\vec{\alpha}[\ell])_{\ell \in [\frac{T}{\shortT}]}$.  We focus now on one single small timestep $[\ell\shortT, (\ell +1)\shortT]$ of the evolution where a general-dyne measurement of the meter modes at time $\ell \shortT$ yields the outcome $  \vec{\alpha}[\ell]$. This corresponds to the projection of the meter modes onto the POVM element $\bigl(D(\vec{\alpha}[\ell])\ket{0}\bra{0}D^\dagger(\vec{\alpha}[\ell])\bigr)/\pi^k$ and yields as the output state
vector of the system
\begin{equation}
    \begin{split}
         \ket{\psi[(\ell +1 )\shortT]}_{\vec{\alpha}[\ell]}
   &=  \id \otimes \pi^{-k/2} \bra{0}D^\dagger(\vec{\alpha}[\ell])  \left[\id   -\frac {\shortT}{2} \, L_j^\dag L_j[\ell] \otimes \id + \sqrt{\shortT} L_j[\ell] \otimes a^\dag_j [\ell]\right]\ket{\psi[\ell\shortT]}\ket{0} \\
     & = \frac{\bra{0}D^\dagger(\vec{\alpha}[\ell])\ket{0}}{\pi^{k/2}}   \bigl(\id - \frac {\shortT}{2} \, L_j^\dag L_j[\ell] \bigr)\ket{\psi[\ell]}  +   \frac{\sqrt{\shortT}}{\pi^{k/2}}L_j(\ell\shortT)\ket{\psi[\ell]}  \bra{0}D^\dagger(\vec{\alpha}[\ell])  a^\dag_j  [\ell]\ket{0}.
    \end{split}
\end{equation}
where we are implicitly summing over $j \in [k]$. Recalling that $D(\vec{\alpha}) = \prod_j D(\alpha_j)$, the last term further gives
\begin{equation}
    \begin{split}
          & \sqrt{\shortT} L_j(\ell\shortT)\ket{\psi[\ell]}  \bra{0}\prod_{j'} D^\dagger(\alpha_{j'}[\ell]) a^\dag_j  [\ell]\ket{0} \\
    =&  \sqrt{\shortT} L_j(\ell\shortT)\ket{\psi[\ell]}\bra{0} D^\dagger(\alpha_{j}) a^\dag_j [\ell]\ket{0} \prod_{j' \neq j}\bra{0} D^\dagger(\alpha_{j'}) \ket{0} \\
    =& \sqrt{\shortT}\,  L_j(\ell\shortT)\ket{\psi[\ell]}\alpha^\ast_{j} \bra{0}D^\dagger(\alpha_{j})\ket{0} \prod_{j' \neq j}\bra{0} D^\dagger(\alpha_{j'}) \ket{0} \\
    =& \sqrt{\shortT}  \,L_j(\ell\shortT)\ket{\psi[\ell]}\alpha^\ast_{j} \bra{0} D^\dagger(\vec{\alpha}[\ell]) \ket{0}.
    \end{split}
\end{equation}
In the second to third line we used the definition of a coherent state and that it is an eigenstate of the annihilation operator. Moreover, the vacuum expectation value of the displacement operator is
\begin{equation}
    \bra{0} D(\vec{\alpha}[\ell]) \ket{0} = \prod_{j'} \e^{- \frac12|\alpha_{j'}[\ell]|^2}.
\end{equation}
Overall, we then get that
\begin{align}
   \ket{\psi[(\ell +1 )\shortT]}_{\vec{\alpha}[\ell]} =  \frac{\prod_{j'} \e^{- \frac12|\alpha_{j'}[\ell]|^2}}{\pi^{k/2}}   \bigl(\id - \frac {\shortT}{2} \, L_j^\dag L_j(\ell\shortT) +\sqrt{\shortT} \,\alpha_j^\ast L_j(\ell \shortT) \bigr)\ket{\psi[\ell]}. 
\end{align}
 We now interpret
 \begin{equation}
     M_{\vec{\alpha}[\ell]} = \id  - \frac12 \shortT \, L_j^\dag L_j(\ell\shortT) + \sqrt{\shortT} L_j(\ell\shortT)\alpha^\ast_{j}
 \end{equation}
  as the effect operator of the measurement on the system corresponding to the outcome ${\vec{\alpha}[\ell]}$.  
 So the probability of obtaining the outcome ${\vec{\alpha}[\ell]}$ in the small time interval $[\ell \shortT,(\ell+1)\shortT]$ is 
 \begin{equation}
     \begin{split}
  p({\vec{\alpha}[\ell]})\mathrm d\vec{\alpha}_\ell  &= \langle \psi[(\ell +1 )\shortT]\ket{\psi[(\ell +1 )\shortT]}_{\vec{\alpha}[\ell]} \mathrm d\vec{\alpha}_\ell  \\ 
  &=  \bra{\psi[\ell]}M_{_{\vec{\alpha}[\ell]}}^\dagger M_{_{\vec{\alpha}[\ell]}}\ket{\psi[\ell]}\frac{\prod_{j'} \e^{- |\alpha_{j'}[\ell]|^2}}{\pi^{k}} \mathrm d\vec{\alpha}_\ell 
  \\
  & = \frac{\prod_{j'} \e^{- |\alpha_{j'}[\ell]|^2}}{\pi^{k}} \bigl(1  + \sqrt{\shortT}(\text{Re}(\alpha_j)\underbrace{ \langle  L_{j}[\ell] +L^\dagger_{j}[\ell]\rangle_{\psi[\ell]}}_{\langle x_j [\ell] \rangle} +  \text{Im}(\alpha_j) \underbrace{i\langle  L^\dagger_{j}[\ell] -L_{j}[\ell]\rangle_{\psi[\ell]}}_{\langle y_j [\ell] \rangle})  + \mathcal O(\shortT) \bigr) \mathrm d\vec{\alpha}_\ell  \\
  &= \frac{\prod_{j'} \e^{- |\alpha_{j'}[\ell]|^2}}{\pi^{k}} \bigl( (1  + \sum_j \sqrt{\shortT}\text{Re}(\alpha_j)\langle x_j [\ell]\rangle)(1  + \sum_j \sqrt{\shortT}\text{Im}(\alpha_j) \langle y_j [\ell] \rangle)  + \mathcal O(\shortT)  \bigr) \mathrm d\vec{\alpha}_\ell
     \end{split}
 \end{equation}
Remember that we are implicitly summing over $j \in [k]$. To proceed further we notice that
 \begin{equation}
 \label{eq:diff_compl_square_discret}
     \begin{split}
           \prod_j  \e^{-(\text{Re}(\alpha_j)-\langle x_j \rangle \sqrt{\shortT}/2)^2} &=  \prod_j \sum_{m=0}^\infty \frac{1}{m!} (-1)^m \sum_{n=0}^{2m}\binom{2m}{n} \text{Re}(\alpha_j)^{2m-n}(-\frac{\langle x_j \rangle \sqrt{\shortT}}{2})^{n} 
   \\
   &= \prod_j \sum_{m=0}^\infty \frac{1}{m!} (-1)^m \bigl(\text{Re}(\alpha_j)^{2m} - m\text{Re}(\alpha_j)^{2m-1} \langle x_j \rangle \sqrt{\shortT})+ \mathcal O(\shortT )\bigr) \\
   &=   \prod_j  \bigl(e^{-\text{Re}(\alpha_j)^2} +  \sum_{m=0}^\infty \frac{(-1)^{m}}{m!}  \text{Re}(\alpha_j)^{2m+1} \langle x_j \rangle \sqrt{\shortT})+ \mathcal O(\shortT )\bigr) \\
   &= \bigl(\prod_{j'}  e^{-\text{Re}(\alpha_{j '})^2} \bigr)\bigl(1 + \sum_{j}  \text{Re}(\alpha_j) \langle x_j \rangle \sqrt{\shortT}+ \mathcal O(\shortT )\bigr).
     \end{split}
 \end{equation}
  Using this we see that to the leading order we have
  \begin{equation}
  \label{eq:diff_inst_chang_measure_disc}
      \begin{split}
           p({\vec{\alpha}[\ell]})\mathrm d\vec{\alpha}_\ell 
     &= \Bigl( \prod_j  \frac{e^{-(\text{Re}(\alpha_j)-\langle x_j \rangle \sqrt{\shortT}/2)^2}}{\sqrt{\pi}}\frac{e^{-(\text{Im}(\alpha_j)-\langle y_j \rangle \sqrt{\shortT}/2)^2}}{\sqrt{\pi}}  + \mathcal O(\shortT) \Bigr) \mathrm d\vec{\alpha}_\ell  \\
     &= \Bigl(  \prod_j  \frac{\sqrt{\shortT}e^{-\shortT(\frac{\text{Re}(\alpha_j)}{\sqrt{\shortT}}-\frac{\langle x_j \rangle}{2})^2}}{\sqrt{\pi}}\frac{\sqrt{\shortT}e^{-\shortT(\frac{\text{Im}(\alpha_j)}{\sqrt{\shortT}}-\frac{\langle y_j \rangle}{2} )^2}}{\sqrt{\pi}}  + \mathcal O(\shortT) \Bigr)\mathrm d\bigl(\frac{\vec{\alpha}_\ell}{\shortT^k}\bigr)
      \end{split}
  \end{equation}
 We see that the real and imaginary part of the quantities $\bm{\alpha}_j /\sqrt{\shortT}$ are independent and follow the statistics of a Gaussian distribution with mean $\langle x_j[\ell]\rangle/2$ and $\langle y_j [\ell] \rangle /2 $, respectively,  and variance $1/(2\shortT)$. Using the rules how a 
 Gaussian random variable transforms under linear transformations we can  write 
 \begin{align}
  \frac{\text{Re}(\bm \alpha_j)[\ell]}{\sqrt{\shortT}}+ i \, \frac{\text{Im}(\bm{\alpha}_j) [\ell]}{\sqrt{\shortT}} = \frac{1}{2}  \langle x_j[\ell]\rangle+i\frac{1}{2}  \langle y_j[\ell]\rangle   + \frac{\delta\bm Z_j}{\shortT} = \langle L_j \rangle +\frac{\delta\bm Z_j}{\shortT}
  \label{eq:discretised_diff_outcome}
 \end{align}
 where $\delta\bm Z_j$ satisfies \cref{eq:wiener_complex_incr} at the level of the discretised increments. 
  In the limit of infinitely many but infinitely small time-steps, i.e., $\frac{T}{\shortT} \rightarrow \infty$  and $\shortT \rightarrow \mathrm d t$, $\delta\bm Z_j$ becomes a complex valued Wiener increment $\mathrm d \bm Z_j$ as in \cref{eq:wiener_complex_incr} and the $(\bm{\vec{\alpha}}[\ell]/\sqrt{\shortT})_\ell $ become, after summation, the continuous integrated records 
 \begin{equation}
 \label{eq:diff_limit_intrecord}
    \bm{ Y}_j(t) :=  \lim_{\shortT \to 0} \shortT \sum_{\ell=0}^{\lfloor t/\shortT \rfloor} \frac{\bm{\alpha}_j[\ell]}{\sqrt{\shortT}} = \int_{0}^t \braket{L_j}_{\psi(s)}\mathrm ds + \bm{Z}_j(t).
 \end{equation}
 In this limit the currents are recovered as $\bm{J}_j(t) = \mathrm d \bm Y_j(t)/\mathrm d t$ with statistics as in Eq.~\eqref{eq:gener_dyne_currents}. So we see that continuously monitoring the meter modes under heterodyne measurements reproduces the statistics of a diffusive measurement on the system.
  
 Let us consider the limit in more detail in order to obtain now an expression for the instrument density of a diffusive measurement. The following equivalences hold up to and including order $\shortT$. For the discrete set of ordered outcomes $(\vec{\alpha}[\ell])_{\ell \in \left[\frac{T}{\shortT}\right]}$ we have 
 \begin{equation}
     \begin{split}
    p\bigl((\vec{\alpha}[\ell])_\ell\bigr)\mathrm{d}\vec{\alpha} &= 
     \| \prod_{\ell \in \left[\frac{T}{\shortT}\right]} \Bigl(\id  \otimes \bra{0_\ell}
      \frac{D^\dagger(\vec{\alpha}[\ell])}{\pi^{k/2}}   \left[\id   -\frac {\shortT}{2} \sum_{j} L_j^\dag L_j[\ell] \otimes \id + \sum_{j}\sqrt{\shortT} L_j[\ell] \otimes a^\dag_j [\ell]\right]\Bigr)\ket{\psi}\ket{0}^{\otimes \left[\frac{T}{\shortT}\right]}\|_2^2 \mathrm{d}\vec{\alpha} \\
      &=  \|\sum_{i=1}^{d_{\rm S}} \ket{i}\bra{i} \otimes  \bra{0}^{\otimes \left[\frac{T}{\shortT}\right]}\bigotimes_{\ell \in [\frac{T}{\shortT}]} 
      \frac{D^\dagger(\vec{\alpha}[\ell])}{\pi^{k/2}} \prod_{\ell \in \left[\frac{T}{\shortT}\right]} \e^{\sum_{j \in [k]} -\frac12 \shortT \, L_j^\dag L_j[\ell]\otimes \id + \sqrt{\shortT} L_j[\ell] \otimes a^\dag_j [\ell]}\ket{\psi}\ket{0}^{\otimes \left[\frac{T}{\shortT}\right]}\|_2^2 \mathrm{d}\vec{\alpha}.
     \end{split}
 \end{equation}
 From the properties of the displacement operator we get
 \begin{equation}
     \begin{split}
            \bigotimes_{\ell \in [\frac{T}{\shortT}]} D(\vec{\alpha}[\ell]) \ket{0} &=    \bigotimes_{\ell} \e^{\sum_j \alpha_j[\ell]\cdot a_j^\dagger[\ell] - \alpha_j^\ast[\ell] \cdot a_j[\ell]}\ket{0} \\
  &= \Bigl(\prod_{\ell,j'} e^{- \frac12|\alpha_{j'}[\ell]|^2}\Bigr) \bigotimes_{\ell} \e^{\sum_j \alpha_j[\ell]\cdot a_j^\dagger[\ell]} \e^{- \alpha_j^\ast[\ell] \cdot a_j[\ell]}\ket{0} \\
  &= \Bigl(\prod_{\ell,j'} e^{- \frac12|\alpha_{j'}[\ell]|^2}\Bigr) \bigotimes_{\ell} \e^{\sum_j \alpha_j[\ell]\cdot a_j^\dagger[\ell]}\ket{0},
     \end{split}
 \end{equation}
 such that the probability becomes 
 \begin{multline}
    p\bigl((\vec{\alpha}[\ell])_\ell\bigr)\mathrm{d}\vec{\alpha} =\\   \sum_{i=1}^{d_{\rm S}}\bigl |\Bigl( \bigotimes_{\ell}\bra{0}\e^{\sum_j \alpha^\ast_j[\ell]\cdot a_j[\ell]}\Bigr)\Bigl(\bra{i}\prod_{\ell \in \left[\frac{T}{\shortT}\right]} e^{\sum_{j \in [k]} -\frac12 \shortT \, L_j^\dag L_j[\ell]\otimes \id + \sqrt{\shortT} L_j[\ell] \otimes a^\dag_j [\ell]}\ket{\psi}\ket{0}^{\otimes \left[\frac{T}{\shortT}\right]}\Bigr)\bigr|^2 \Bigl(\prod_{\ell} \frac{e^{-\sum_{j'}|\alpha_{j'}[\ell]|^2}}{\pi}\, \mathrm{d}\vec{\alpha}_{\ell}\Bigr).
    \label{eq:prob_diff_meas_finite}
 \end{multline}
 In the limit of $\shortT \to 0$ while keeping $T$ constant, 
the evolution state on the right becomes a cMPS with bond dimension $d_{\mathrm{S}}$ on the Fock space built from the single-particle Hilbert space $L^2([0,T], \mathbb C^k)$.
 The left hand side in the inner product, carrying the information about the discretised outcome currents, is formally of cMPS-type too with bond dimension 1, $Q=0$ and $R_j=J_j$: 
 \begin{equation}
 \label{eq:fake_cmps_diffusive_inst}
     \begin{split}
   \lim_{\shortT \to 0}  \bigotimes_{\ell} \e^{\sum_j \alpha_j[\ell]\cdot a_j^\dagger[\ell]}\ket{0} 
   &= \lim_{\shortT \to 0}  \bigotimes_{\ell} \e^{\shortT \sum_\ell\sum_j \tfrac{\alpha_j[\ell]}{\sqrt{\shortT}}\cdot \tfrac{a_j^\dagger[\ell]}{\sqrt{\shortT}}}\ket{0}   \\
  &=  \exp\Bigl[\sum_j \int_0^T \mathrm dt \, J_j(t)\psi_j^\dagger(t) \Bigr]\ket{\Omega} \\
  &\hat{=} \cmps{0}{\{J_j(t)\}_j}{1}.
     \end{split}
 \end{equation}
 This is \textit{not} a cMPS in the strict sense: the currents $J_j$ are not square-integrable so the norm of  $ \cmps{0}{\{J_j(t)\}_j}{1}$ diverges and the limit does not exist in the Fock space. What does have a limit is the product of this state with the Gaussian weight in \cref{eq:prob_diff_meas_finite}. This measure part becomes the product of $k$ standard complex Wiener path integral measures $\mathrm d W$
 \begin{equation}
     \lim_{\shortT \to 0} \prod_{\ell} \frac{e^{-\sum_{j'}|\alpha_{j'}[\ell]|^2}}{\pi}\, \mathrm{d}\vec{\alpha}_{\ell} = \mathrm d W^{\otimes k }(\vec{J})  =: \mathrm d \mu_W(\vec J).
 \end{equation}
The last line in \cref{eq:fake_cmps_diffusive_inst} is therefore to be read as a formal expression, defined by its pairing which produce a probability density with respect to the measure $\mathrm d \mu_W(\vec J)$
\begin{equation}
\label{eq:density_diff_limit_prestep}
      p(\vec{J}) = \sum_{i \in d_{\rm S}} \bigl | \bra{\Phi[0, \{J_j(t)\}_{j},1 ]} \Phi[\sum_{j \in[k]} -\frac12 L^\dag_j L_j, \, \{L_j(t)\}_{j\in[k]}, \, \ket{\psi}\!\bra{i}] \rangle \bigr|^2.
\end{equation}
We note that \cref{eq:diff_limit_intrecord} and \cref{eq:density_diff_limit_prestep} capture the same distribution law with respect to two different reference measures. Before \cref{eq:prob_diff_meas_finite}, the non-zero mean $\braket{L_j}$ is absorbed into the Gaussian by completing the square (see \cref{eq:diff_compl_square_discret}) so that the $\bm{Y}_j$ appear as Wiener processes with non-zero mean. In Eqs.~(\ref{eq:prob_diff_meas_finite})-(\ref{eq:density_diff_limit_prestep}) this step is not performed: the zero-centered Gaussian of the coherent state resolution of the identity is kept as the reference measure $\mu_W$ under which the currents have zero mean. The non-zero mean instead appears through the probability density in \cref{eq:density_diff_limit_prestep}. Consequently, from now the $J_j(t)$ are centred, jointly Gaussian, and obey 
\begin{equation}
    \label{eq:white_noise_prop}
    \mathbb E[J_j(t)J_i^\ast(s)]= \delta_{ij}\delta(t-s), \qquad  \mathbb E[J_j^\ast(t)J_i^\ast(s)] =  \mathbb E[J_j(t)J_i(s)]= 0.
\end{equation}
To bring the density expression further into the form of our framework we repeat
\begin{equation}
\label{eq:prob_diff_step_1}
    \begin{split}
          p(\vec{J}) 
    &= \sum_{i \in d_{\rm S}} \bra{i} \mathcal{T} \e^{\int_0^T \mathrm d t \sum_{j \in [k]} -\frac12 L_j^\dag L_j + J^\ast_j(t)L_{j}(t)}\ket{\psi}\bra{\psi} \mathcal{T} e^{\int_0^T \mathrm d t \sum_{j \in [k]} -\frac12 L_j^\dagger L_j + J_j(t)L^\dagger_{j}(t)}\ket{i} \\
     &= \llangle \id_{\rm S}| \mathcal{T} \e^{\int_0^T \mathrm d t \sum_{j \in [k]} -\frac12 L_j^\dag L_j + J^\ast_j(t)L_{j}(t)} \otimes \mathcal{T} \e^{\int_0^T \mathrm d t \sum_{j \in [k]} -\frac12 L_j^T L_j^* + J_j(t)L^\ast_{j}(t)}  |\rho\rrangle  \\
       &= \tr \left[ |\rho\rrangle \llangle \id_{\rm S}| \mathcal{T} e^{\int_0^T \mathrm d t \sum_{j \in [k]} - \frac12 L_j^\dag L_j \otimes \id - \frac12 \id \otimes L_j^T L_j^* +  J^\ast_j(t)L_{j}(t) \otimes \id + \id \otimes J_j(t)L^\ast_{j}(t)}\right]
    \end{split}
\end{equation}
where from the second to third line we expressed everything in Liouville space which just amounts to applying the vectorization map. The initial state $\rho_0$ on the system can now also be mixed.
It is important to note that in \cref{eq:prob_diff_step_1} the stochastic integrals are of Stratonovich-type \cite{gardinerStochasticMethodsHandbook2009, vankampenItoStratonovich1981}. This in particular amounts to setting 
\begin{equation}
    \begin{split}
        \int_t^{t^*} \mathrm d s \; \delta (t^*-s)= \int^t_{t^*} \mathrm d s \;\delta (t^*-s)= \tfrac 1 2 .
    \end{split}
\end{equation}
Since calculating with this rule is in general more cumbersome than with an Itô interpretation of stochastic integrals, we want to proceed by converting \cref{eq:prob_diff_step_1} to involve the Itô version of stochastic integrals. This is possible because \cref{eq:prob_diff_step_1} can be interpreted as the solution of the Stratonovich stochastic differential equation
\begin{equation}
    \kket{\mathrm d \bm \rho} =  \bigl(\sum_{j \in [k]} - \tfrac12 (L_j^\dag L_j \otimes \id + \id \otimes L_j^T L_j^*)\mathrm d t +  L_{j} \otimes \id \mathrm d\bm Z_j^\ast(t) + \id \otimes L^\ast_{j} \mathrm d\bm Z_j\bigr)\kket{\bm \rho}.
\end{equation}
Converting it into an Itô stochastic differential equation adds a term $\sum_{j,i}\tfrac{1}{2}(L_{j} \otimes \id \mathrm d\bm Z_j^\ast(t) + \id \otimes L^\ast_{j} \mathrm d\bm Z_j)(L_{i} \otimes \id \mathrm d\bm Z_i^\ast(t) + \id \otimes L^\ast_{i} \mathrm d\bm Z_i)\kket{\bm \rho}$ where the Ito rule \cref{eq:wiener_complex_incr} has to be applied \cite{gardinerStochasticMethodsHandbook2009}.
Then we get
\begin{equation}
     \kket{\mathrm d \bm \rho} =  \bigl(\sum_{j \in [k]}  (- \tfrac12L_j^\dag L_j \otimes \id - \tfrac12 \id \otimes L_j^T L_j^*+ L_j\otimes L_j^\ast)\mathrm d t +  L_{j} \otimes \id \mathrm d\bm Z_j^\ast(t) + \id \otimes L^\ast_{j} \mathrm d\bm Z_j\bigr)\kket{\bm \rho},
\end{equation}
which should be compared with \cref{eq:cont_measurement_SME}.
Consequently, the probability, being the solution of this stochastic differential equation, is expressible as
\begin{equation}
    p(\vec{J}) = \tr \left[ |\rho\rrangle \llangle \id_{\rm S}| \mathcal{T}_{\text{ito}} e^{\int_0^T \mathrm d t \sum_{j \in [k]} - \frac12 L_j^\dag L_j \otimes \id - \frac12 \id \otimes L_j^T L_j^* +L_j\otimes L_j^\ast+  J^\ast_j(t)L_{j}(t) \otimes \id + \id \otimes J_j(t)L^\ast_{j}(t)}\right],
\end{equation}
where now the integrands are considered as \textit{non-anticipating} which in particular amounts to employing
\begin{equation}
    \begin{split}
        \int_t^{t^*} \mathrm d s \; \delta (t^*-s) &= 0 \\
         \int^t_{t^*} \mathrm d s \;\delta (t^*-s) &= 1
    \end{split}
    \label{eq:ito_def_delta_limit}
\end{equation}
when taking expectation values. Expressing now the involved operators again in the Pauli basis $\{\mathbb P_\nu\}_\nu$ with coefficients 
\begin{equation}
    \begin{split}
     c_\nu &\coloneqq - \tfrac{1}{2} \tr\Bigl[\mathbb{P}_\nu\bigl(\sum_{j\in[k]} L_j^\dag L_j \otimes \id + \id \otimes L_j^T L_j^*\bigr)\Bigr]^\ast\\ 
     d_\nu &\coloneqq \sum_{j \in [k]} \tr[\mathbb P_\nu (L_j\otimes L_j^\ast)]^\ast \\
        \chi^{\vec{J}}_\nu(t) &\coloneqq \tr\left[\mathbb{P}_\nu \sum_{j\in[k]} \bigl(J^\ast_j(t) L_j(t) \otimes \id + \id \otimes J_j(t)L^\ast_{j}(t)\bigr)\right]^\ast
    \end{split}
\end{equation}
we get 
\begin{align}
    p(\vec{J}) &= \braket{\Phi[c_0 + d_0+\chi_0^{\vec{J}},\{c_{\nu} + d_\nu+(\chi^{\vec{J}}_{\nu})\}_{\nu},1]|\Phi[ 0, \ \{\mathbb P_\nu\}_{\nu}, \ |\rho_0\rrangle\!\llangle \id_{\rm S}|]}.
    \label{eq:final_prob_diff_measurement}
\end{align}
So the probability density of a diffusive measurement is again expressible as the contraction of a cPT in process canonical representation with an object in the form of a cMPS with bond dimension 1 which carries all the control information.  \cref{eq:final_prob_diff_measurement} allows us to directly read off the instrument corresponding to a diffusive measurement.
We emphasise that this object is not a proper vector in the dual Fock space $(\Gamma_T^{d_{\rm S}^4-1}\bigr)^*$ because its wave functions contain linear functionals of white noise.

\end{document}